\documentclass{JHEP3}
\usepackage[centertags]{amsmath}
\usepackage{amssymb,amsbsy}
\usepackage{amsthm}
\usepackage[dvips]{graphicx}
\usepackage{subfigure}
\usepackage{mathrsfs}
\usepackage{amsfonts}
\usepackage{amsmath}
\usepackage{cite}
\usepackage{epsfig}
\usepackage{longtable}

\graphicspath{{Figures/}}

\def\be{\begin{equation}}
\def\ee{\end{equation}}
\def\baray{\begin{eqnarray}}
\def\earay{\end{eqnarray}}
\def\ba{\begin{eqnarray}}
\def\ea{\end{eqnarray}}

\theoremstyle{definition}

\theoremstyle{definition}

\theoremstyle{definition}

\theoremstyle{definition}

\numberwithin{equation}{section}

\newcommand{\nn}{\nonumber}

\newcommand{\half}{\frac{1}{2}}

\newcommand{\al}{\alpha}

\newcommand{\pd}{\partial}

\newcommand{\Asc}{\mathscr{A}}

\newcommand{\Bsc}{\mathscr{B}}

\title{Conifolds and Tunneling in the String Landscape}

\author{Pontus Ahlqvist$^{a}$, Brian R.~Greene$^{a}$, David Kagan$^{a}$, Eugene A.~Lim$^{a,b,c}$, \newline Saswat Sarangi$^{a}$, and I-Sheng Yang$^{a}$ \\
$^{a}$Institute of Strings, Cosmology and Astroparticle Physics \\
Department of Physics \\
Columbia University, New York, NY 10027, USA \\ \\ 
$^{b}$Department of Applied Mathematics and Theoretical Physics \\
Cambridge University, Wilberforce Road, CB3 0WA, United Kingdom\\ \\ 
$^{c}$Universit\'e de Fondwa, UNIF2004, Tom-Gato, Haiti
}
\date{\today}
\abstract{We investigate flux vacua on a variety of one-parameter Calabi-Yau compactifications, and find many examples that are connected through continuous monodromy transformations. For these, we undertake a detailed analysis of the tunneling dynamics and find that tunneling trajectories typically graze the conifold point---particular 3-cycles are forced to contract during such vacuum transitions. Physically, these transitions arise from the competing effects of minimizing the energy for brane nucleation (facilitating a change in flux), versus the energy cost associated with dynamical changes in the periods of certain Calabi-Yau 3-cycles. We find that tunneling occurs only when warping due to back-reaction from the flux through the shrinking cycle is properly taken into account.}

                                         
\preprint{}

\begin{document}

\section{Introduction}

For some time, the landscape of vacua has been a dominant theme in string theory research. Yet, explicit and detailed investigations of this multidimensional terrain remain mathematically challenging. Even so, broad-brush outlines of a landscape-based cosmological scenario have been put forward \cite{Linde:1993xx,Bousso:2000xa,Susskind}.  By melding eternal inflation with the string landscape, a multiverse consisting of bubble universes is generated, each bubble realizing one or another of the locally-stable minima in the string landscape. Going beyond this broad-brush picture requires analytic control of many details, including the topography of the string landscape (the location and nature of the locally-stable minima) and the process of bubble nucleation (Coleman-DeLuccia tunneling in the string landscape). With our current level of understanding, and with the mathematical tools we've so far developed, gaining such control over the entire landscape is well beyond reach. An alternative strategy, then, is to glean insights from a thorough study of portions of the landscape that are sufficiently restricted to be mathematically tractable while sufficiently representative to reveal general physical properties. In this paper, we take a modest step in this direction through the study of flux compactifications on explicit Calabi-Yau manifolds.

Specifically, we focus on flux compactifications of type IIB string theory which, as is well known, generate flux potentials that exhibit a huge landscape of lower-dimensional vacua \cite{Bousso:2000xa}. In the context of eternal inflation, we make the standard assumption that tunneling transitions are the dominant processes that nucleate bubbles of different vacua, but recognize that it is essential to understand the details of their dynamics. The study of such cosmological tunneling process predates the string theory landscape, of course. In recent years, though, researchers have realized that string theory leads to novel effects in this context \cite{Sarangi:2007jb, BrownSashShlaerWeltman, BlancoPillado}. Most manifest is the fact that, typically, string compactifications introduce hundreds of degrees of freedom: the moduli that describe the fluctuating geometry of the manifold on which the theory is compactified. Based on recent studies \cite{Yan09,AJL,BroDah10}, extra degrees of freedom are expected to play an important role in explicit models of vacuum.

An important early work focusing on the ``topography" of the Calabi-Yau string landscape \cite{Danielsson} established that vacua corresponding to different flux configurations can be smoothly connected via a multi-sheet potential. This work focused primarily on the vacuum structure of the mirror quintic compactification \cite{GreenePlesser}, and was developed further by \cite{Larfors} in which the authors estimated tunneling rates between mirror quintic vacua endowed with different fluxes. These works provide an important backdrop to the current paper. The mirror quintic is but one of a number of one-parameter Calabi-Yau compactifications, so a natural question---taken up in the first sections of this paper---is the degree to which the observations of \cite{Danielsson} extend to the full class of such mathematically tractable examples. We will find that, for the most part, they do. Next, the tunneling trajectories in \cite{Larfors} were estimated based on qualitative features evident in the relevant flux potentials, so a natural question---taken up on the later sections of the paper---is the degree to which these estimates are borne out by explicit calculation. We will find that they aren't; the estimated tunneling paths \cite{Larfors} turn out not to capture the key dynamical features of stringy tunneling transitions.

In this paper, we carefully address both of these issues numerically and analytically. We find that tunneling solutions exhibit a form that we call ``conifunneling": they are driven into the vicinity of the conifold point, and the geometry of the Calabi-Yau becomes almost singular and strongly warped.  Similar to existing examples of multi-field tunneling, the additional fields play a crucial role and can take extreme values during transitions.  This provides a detailed new picture of string landscape tunneling, as the extreme situations are often under better analytical control.  For example, we are able give an analytical upper bound on how distant vacua can be and still be connected by a conifunneling transition.

This paper is structured as follows. In section~\ref{sec-MQ} we describe flux compactifications for one-parameter Calabi-Yau manifolds. These manifolds have been classified and form a set of fourteen models, which are organized in four distinct families. In section \ref{sec-VacuumHunt} we apply the techniques of \cite{Danielsson,Larfors} to a cross-section of these 14 models to investigate similarities and differences in the vacuum structure of their flux potentials. We tabulate some of the non-supersymmetric and supersymmetric vacua found in these models, while also developing a new procedure to rapidly locate minima of the flux potential. Section \ref{sec-CDL} describes methods for investigating multi-field vacuum transitions while section \ref{sec-conifunneling} applies numerical techniques to the specific problem of multi-field tunneling in a stringy flux landscape. The result is the conifunneling phenomena described above. In section~\ref{sec-analysis} we provide analytical arguments and other supporting evidence for conifunneling as a general effect of transitions in the vicinity of special points in a string-like potential landscape.

\section{One-Parameter Calabi-Yaus and Flux Compactification}
 \label{sec-MQ}

Given a compact Calabi-Yau manifold $\cal M$, one can describe its moduli space of complex structures in terms of the periods of the holomorphic $3$-form. The period integrals are
\baray
\Pi_I = \int_{C_I} \Omega = \int_{\cal M} C_I \wedge \Omega, 
\earay
where $\Omega$ is the holomorphic $3$-form and $C_I$ describes a basis of $H_3({\cal M})$. The index $I$ runs from $0$ to $2h^{1,2}+1$. The intersection matrix $Q$ is given by
\baray
Q_{IJ} = \int_{\cal M}C_I \wedge C_J.
\earay
The explicit form of $Q$ depends on the choice of the basis $C_I$. However it is always possible to choose a basis where $Q$ has the following symplectic form
\baray
\label{symp}
Q = \left(
\begin{array}{cccccc}
   &  &  & & & -1 \\
   &  &  &  & 1&    \\
   &  &  & \cdot &  &    \\
   &  & \cdot &  &  &  \\
   & -1 &    &  &  & \\
1 & & & & &  \end{array} 
\right).
\earay
We refer to this as the symplectic basis of periods. The intersection matrix is invariant under symplectic transformations. It is convenient to represent the periods using a vector
\baray
\Pi(z) = \left(
\begin{array}{c}
\Pi_N(z)\\
\Pi_{N-1}(z) \\
 \cdot \\
 \cdot \\
 \cdot\\
\Pi_0(z) \end{array} 
\right)
\earay
where $N \equiv 2h^{1,2}+1$ and $z$ is an $h^{1,2}$ dimensional complex coordinate on the moduli space. In this paper, we deal with Calabi-Yau manifolds with one complex modulus, $h^{1,2}=1$, so $z$ is a complex number. We use the symplectic basis above to generate the flux potentials for the various models.

In general, the periods are subject to monodromies. Going around non-trivial loops in the moduli space changes the periods. This change is given in terms of the monodromy matrices $T$
\baray
\Pi \to T \cdot \Pi.
\earay 
The monodromy matrices preserve the intersection matrix and are thus elements of $Sp(N,\mathbb{Z})$. The Calabi-Yaus we consider have three mondromy matrices for each of the three special points in the moduli space, the large complex structure point (LCS) at $z=0$, the conifold point at $z=1$, and the Landau-Ginzburg point at $z=\infty$.

The periods behave in characteristic ways around the special points. Near $z=0$, $\Pi_0 \sim 1$, while the other periods go as $\Pi_i \sim (\log z)^i$ with possible subleading log terms. Near the conifold point, the cycle whose period is $\Pi_3$ collapses (absent effects from warping due to fluxes or branes) while the dual cycle with period $\Pi_0$ is only defined up to a monodromy. The other two periods approach constant values near the conifold point. The behavior around the Landau-Ginzburg point depends on the type of Calabi-Yau---in particular, the 14 one-parameter models break up into four families. The first set of models have regular behavior around the Landau-Ginzburg point for an appropriately chosen complex structure coordinate $\psi$. The remaining families involve some combination of periods developing logarithmic behavior.

\subsection{The Meijer basis of periods} 

The periods of a given Calabi-Yau are solutions to a set of differential equations called the Picard-Fuchs equations. For one-parameter models, the equations reduce to a single ordinary differential equation, whose natural basis of solutions are conveniently expressed in terms of Meijer G-functions \cite{Greene}. The Meijer basis of periods, $U_j(z)$, $j = 0,1,2,3$, proves to be convenient for computing the monodromies around the special points in the moduli space. In general, Meijer G-functions are solutions to ODEs of the form
\baray
\label{hyp'}
\left[\delta \prod_{i=1}^q (\delta + \beta_i -1) - z \prod_{j=1}^p (\delta + \alpha_j)\right] u(z) =0,
\earay
where $\delta = z\,d/dz$. The class of Calabi-Yaus that we consider in this paper have $\beta_i = 0$. These are referred to as `the generic family of compact one-parameter models' in \cite{Greene}. The periods satisfy the following Picard-Fuchs equation
\baray
\label{meijer}
\left[\delta^4 - z~(\delta + \alpha_1) (\delta + \alpha_2) (\delta + \alpha_3) (\delta + \alpha_4)\right] u(z) =0,
\earay
where $\alpha_r$ are rational numbers. The monodromy matrices computed in appendix \ref{UniversalMonodromies} describe the effect on the Meijer periods upon going around the special points in the moduli space. If $T$ is a monodromy expressed in the Meijer basis, the monodromies in the symplectic basis are given by $L T L^{-1}$, where $L$ transforms from the Meijer to symplectic basis of periods, $\Pi = LU$.

We use the Meijer functions $U_j(z)$ to construct the symplectic basis $\Pi(z)$ for various Calabi-Yaus (i.e. for different choices of $\alpha_i$). From the periods and a choice of fluxes we calculate the ${\cal N}=1$ scalar potential for the complex structure modulus $z$. One needs some numerical aid to do a reasonably speedy calculation of the Meijer periods. To facilitate the computations in this paper, we used {\em Mathematica} to generate discretized Meijer functions on square lattices on the complex plane with a spacing of $0.05$ for each of the models we investigated. Appendix \ref{sec-VacuumNumerics} details precisely how the numerical Meijer functions are computed and outlines the procedure for computing the Kahler potential, superpotential, and flux potential built from these periods. Figure \ref{MeijerFunction} shows a portion of one of the Meijer functions for model 8 in appendix \ref{TheModels}.

\begin{figure}\label{MeijerFunction}
\begin{center}
\includegraphics[width=14cm]{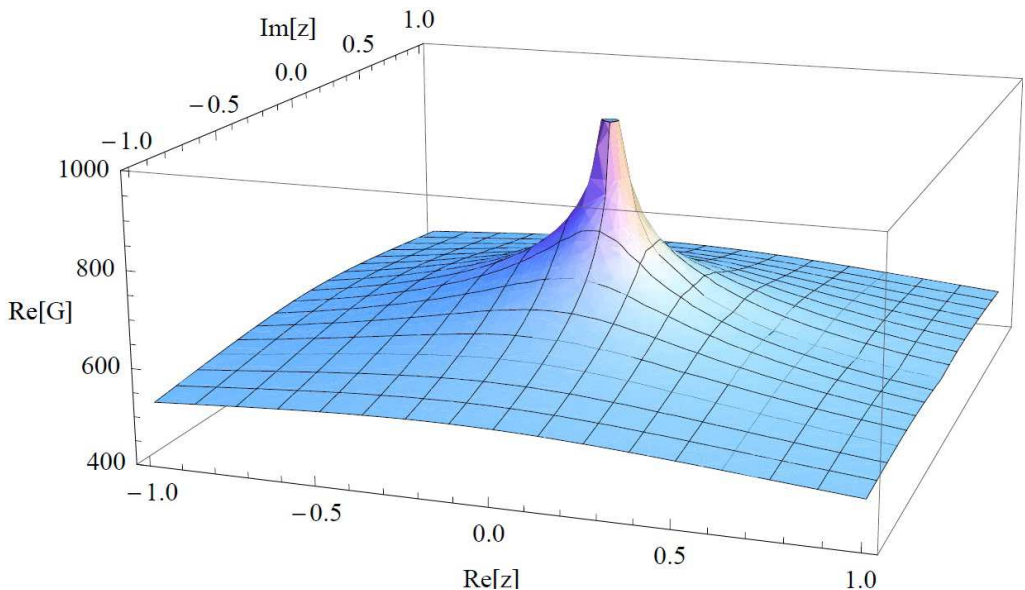}
\vspace{0.1in}
\caption{The Meijer function {\tt MeijerG[\{\{3/4,1/2,1/2,1/4\},\{\}\},\{\{0,0,0\},\{0\}\},-z]} in {\em Mathematica}'s notation for these functions.}
\end{center}
\end{figure}

\subsection{Flux compactification} \label{potential}

Flux compactifications of type IIB string theory on orientifolded Calabi-Yau manifolds were studied in \cite{GKP}. Wrapped $3$-form fluxes can stabilize the complex structure moduli and axio-dilaton by inducing a flux potential that may possesses both supersymmetric and non-supersymmetric minima (or perhaps neither). Non-perturbative and perturbative corrections to the tree level flux potential can stabilize Kahler moduli \cite{KKLT, BBCQ}. At tree level, the flux potentials do not fix the overall volume of the Calabi-Yau and are thus referred to as ``no-scale" models. We focus on the no-scale GKP compactifications for any given Calabi-Yau and thus, are only concerned with the dynamics of the axio-dilaton and complex structure moduli.

Wrapping fluxes around the different $3$-cycles of an orientifold of Calabi-Yau ${\cal M}$ induces the Gukov-Vafa-Witten potential:
\baray
W = \int_{\cal M} \Omega \wedge \left( F_{(3)}-\tau H_{(3)} \right) = F \cdot \Pi -\tau H 
\cdot \Pi \equiv A+B\tau,
\earay
where the axio-dilaton $\tau = C_{(0)} + i e^{-\phi}$, and $F$ and $H$ are 
the Ramond-Ramond (R-R) and Neveu-Schwarz Neveu-Schwarz (NS-NS) flux vectors, 
respectively. Since we are only considering the cases with $h^{1,2}=1$, the 
flux vectors $F$ and $H$ have $2 h^{1,2} + 2 = 4$ entries whose values give 
the strength of the fluxes piercing the relevant dual cycle; $F_0$ for 
example represents the flux threading the cycle whose period is $\Pi_3$. 

The Kahler potential is given by
\baray\label{KahlerPotential}
K = -\log \left(-i(\tau - \bar{\tau})\right) + K_{cs}\left(z,\bar{z}\right) - 3\log\left(-i(\rho-\bar{\rho})\right),
\earay 
where $\rho = \rho_R + i \rho_I$ is the volume modulus (also referred to as the universal Kahler modulus). The volume of the Calabi-Yau goes like $V_{CY} \sim \rho_I^{3/2}$. We will assume that some mechanism stabilizes this at a large value.

The Kahler potential for the complex structure depends on the periods of the $3$-cycles $\Pi$ and the intersection
matrix $Q$
\baray
\label{kcs}
K_{cs} = -\log \left( i \int_{\cal M} \Omega \wedge \overline{\Omega} \right) = -\log \left( i\Pi^{\dagger} Q^{-1}\Pi\right)
\earay
and has to be calculated individually for each Calabi-Yau. 

Note that the above expressions are valid when the effects of warping are essentially constant over the manifold. This assumption will suffice for constructing flux potentials and searching for minima. However, the effects of warping will need to be considered when we discuss tunneling between minima. A brief overview of dimensional reduction of type IIB supergravity on a warped Calabi-Yau is provided in appendix \ref{Reduction}.

The scalar potential for the complex structure 
moduli is given by the ${\cal N}= 1$ supergravity formula
\baray
\label{pot'}
V(z, \tau) = e^K\left( K^{z\bar{z}}D_zWD_{\bar{z}}\overline{W} + K^{\tau \bar{\tau}}D_\tau W D_{\bar{\tau}} \overline{W} 
+ K^{\rho \bar{\rho}}D_\rho W D_{\bar{\rho}} \overline{W} -3|W|^2 \right),
\earay
where $K_{i\bar{j}} = \pd_i \pd_{\bar{j}} K$ are the components of the Kahler metric. In no-scale models the last two terms in Eq.(\ref{pot'}) cancel as is easily checked using the Kahler potential (\ref{KahlerPotential}). The result is a flux potential given by
\baray
\label{pot}
V(z, \tau) = e^K\left( K^{z\bar{z}}D_zWD_{\bar{z}}\overline{W} + K^{\tau \bar{\tau}}D_\tau W D_{\bar{\tau}} \overline{W} \right).
\earay

The cancellation that yields the no-scale potential (\ref{pot}) means that $D_\rho W \sim W$. A supersymmetric vacuum should satisfy $D_z W = D_\tau W = 0$ and $W = 0$. However, we shall refer to vacua that satisfy the first two conditions as supersymmetric, regardless of whether the superpotential vanishes.

The number of $D3$ branes $N_{D3}$, the number of orientifold planes $N_{O3}$ and the fluxes 
are related by the tadpole cancellation condition
\baray
N_{D3} - \frac{1}{4}N_{O3} + \int_{\cal M} F_{(3)} \wedge H_{(3)} = 0,
\earay
which can be rewritten as a condition on the flux vectors:
\baray
F \cdot Q \cdot H = {1\over 4} N_{O3} - N_{D3}.
\earay

\subsection{The effect of monodromies}

The potential $V(z)$ is defined in terms of the periods, and so monodromy transformations will in general change the potential:
\baray
W = (F-\tau H)\cdot \Pi \to (F - \tau H) \cdot T_{\Pi} \cdot \Pi.
\earay
This suggests that another way to account for the monodromies is to keep the periods fixed and to change the fluxes:
\baray
\label{mon}
F \to F \cdot T_{\Pi}, \nonumber \\
H \to H \cdot T_{\Pi}.
\earay
Hence instead of going to a different sheet of the complex plane through the branch cut originating from the special point ($z=0,1,\infty$), one could just stay on the original sheet and change the fluxes according to (\ref{mon}). The crucial point is that the new set of vacua thus obtained by changing the fluxes are still continuously connected to the original set of vacua due to the monodromies. This is how the authors of \cite{Danielsson} generated multiple connected flux vacua for the mirror quintic.

\section{Finding Vacua}\label{sec-VacuumHunt}

Compact one-parameter models were classified in \cite{Doran:2005gu}. The mirror quintic (model 1 in appendix \ref{TheModels} below) is the most familiar of these 14 models. The periods of these Calabi-Yaus are solutions to a Picard-Fuchs equation specified by a set of four rational numbers $\alpha_r$ with $r = 1,2,3,4$ (the mirror quintic corresponds to the case where $\alpha_r = r/5$). We provide a table summarizing the various parameters that characterize all 14 models in appendix \ref{TheModels}.

Following \cite{Greene}, we organize the examples as follows: 
\begin{itemize}
\item Case 1: all $\alpha_r$'s different \\ 
($\alpha_1 \neq \alpha_2 \neq \alpha_3 \neq \alpha_4$; Models 1 -- 7).
\item Case 2: two of the $\alpha_r$'s equal to each other \\ 
($\alpha_1 \neq \alpha_2,\ \alpha_1 \neq \alpha_4,\ \alpha_2 \neq \alpha_4,\  \alpha_2 = \alpha_3$; Models 8 -- 10).
\item Case 3: two equal pairs \\ 
($\alpha_1 \neq \alpha_3$, $\alpha_1 =\alpha_2$, $\alpha_3 = \alpha_4$; Models 11 -- 13).
\item Case 4: All the $\alpha_r$'s equal \\ 
($\alpha_1 = \alpha_2 = \alpha_3 = \alpha_4$; Model 14).
\end{itemize}

In this section we extend the analyses of \cite{Danielsson} to new examples from each of these cases. Finding vacua by hand is not an easy task. Away from special points in the moduli space, one essentially must resort to trial-and-error methods \cite{Danielsson}. Fortunately, we find that there are useful tricks for finding analog vacua across the different compactifications, particularly for vacua that lie within the unit disk in the $z$-plane. Furthermore, we adapt techniques used to count vacua in the vicinity of special points in the moduli space \cite{KachruTaxonomy} to generate a potentially huge new number of vacua---many of which need {\em not} be near the original special point. This allows for a more thorough exploration of flux potential topography.

\subsection{Minimizing the axio-dilaton}
The potential given in Eq.(\ref{pot}) depends on both the complex structure $z$ and the axio-dilaton $\tau$. We would like to express the potential entirely in terms of $z$. To this end we minimize the potential with respect to $\tau$, i.e. impose the condition $\partial_\tau V(z,\tau) = 0$. This minimization leads to the following quadratic equation for the axio-dilaton
\baray
\alpha + \beta \bar{\tau} + \gamma \bar{\tau}^2 = 0 
\earay
where $\alpha$, $\beta$ and $\gamma$ are real valued functions of the fluxes and $z$.
\baray
\alpha &=& |A|^2 + K^{z\bar{z}}D_zAD_{\bar{z}}\bar{A}, \nonumber \\
\beta &=& \bar{A}B + A\bar{B} + K^{z\bar{z}}D_zAD_{\bar{z}}\bar{B}+ K^{\bar{z}z}D_{\bar{z}}\bar{A}D_zB, \nonumber \\
\gamma &=& |B|^2 + K^{z\bar{z}}D_zBD_{\bar{z}}\bar{B}.
\earay
The quadratic equation can be solved to express $\tau$ in terms of the fluxes and $z$
\baray
\tau(z) = -\frac{\beta}{2\gamma} + \sqrt{\frac{\beta^2}{4\gamma^2}-\frac{\alpha}{\gamma}}.
\earay 
The term under the square root is negative semidefinite. Since the string coupling is always positive we have kept the plus sign in front of the square root. 

The scalar potential can then be expressed as
\baray
\label{spot}
V(z) \equiv V(z,\tau(z)) = \frac{ie^{K_{cs}}}{\tau - \bar{\tau}} \left(\alpha +\tau(z)\beta_1 + \overline{\tau(z)} \beta_2 +\tau(z)\overline{\tau(z)}\gamma \right),
\earay
with
\baray
\beta_1 = A\bar{B} + K^{z\bar{z}}D_{\bar{z}}\bar{A}D_zB, \nonumber \\
\beta_2 = \bar{A}B + K^{\bar{z}z}D_zAD_{\bar{z}}\bar{B}.
\earay
Given the potential in this form, the minima we find in $z$ will automatically minimize in $\tau$ as well. This procedure is fine when searching for minima, but will have to be reconsidered when we turn to studying tunneling transitions between vacua.

\subsection{Flux vacuum distributions}

In general, choosing fluxes such that the resulting potential exhibits at least one minimum is a trial-and-error process. However, many vacua can be found by adapting the vacuum counting methods of \cite{KachruTaxonomy,ConlonQuevedo}. In principle, the following prescription can be performed for any special point in the moduli space where one has analytic expressions for the Calabi-Yau periods. For concreteness, we focus on generating vacua near the conifold point $z=1$.

Near the conifold the periods can be expanded to linear order
\begin{eqnarray}
\Pi_3 &\approx& \xi, \nn \\
\Pi_2 &\approx& c_0 + c_1 \xi, \nn \\
\Pi_1 &\approx& b_0 + b_1 \xi, \nn \\
\Pi_0 &\approx& {\xi \over 2\pi i} \log(-i\xi) + a_0 + a_1 \xi, \label{ncexpansion}
\end{eqnarray}
where $\xi = d_1 (z-1)$. The coefficients for any given model can be found by fitting the functions above to the numerically computed periods in the vicinity of the conifold point on a grid that goes from $\textrm{Re}(z) \in [0.9,\,1.1]$ and $\textrm{Im}(z) \in [-0.1,\,0.1]$ (see the end of appendix \ref{sec-NearConifoldPotential} for the coefficients relevant to the mirror quintic model).

The Kahler potential can be expressed as
\begin{equation}
K \approx -\log\left(\mu_0 + \mu_1 \xi + \overline{\mu}_1 \bar{\xi} + \mu_2 |\xi|^2 \log |\xi|^2 + \mu_3 |\xi|^2 + \mu_4 \xi^2 + \overline{\mu}_4 \bar{\xi}^2
+\cdots \right),
\end{equation}
where $\cdots$ indicates higher order terms in $\xi$. The coefficients that are relevant to our analysis are given by
\begin{eqnarray}
\mu_0 &=& i b_0 \bar{c}_0 - i \bar{b}_0 c_0, \nn \\
\mu_1 &=& i b_1 \bar{c}_0 - i \bar{b_0} c_1 + i \bar{a}_0, \nn \\
\mu_2 &=& {1 \over 2\pi}, \nn \\
\mu_3 &=& i b_1 \bar{c}_1 - i \bar{b}_1 c_1 - i a_1 + i \bar{a}_1.
\end{eqnarray}

We can now impose the conditions $D_\xi W = 0$ and $D_\tau W = 0$ and solve for $\log(-i\xi)$ and $\tau$ to leading order. The result is
\begin{equation*}
\tau = {F \cdot \Pi^\dag \over H \cdot \Pi^\dag} \approx {F_1 \bar{c}_0 + F_2 \bar{b}_0 + F_3 \bar{a}_0 \over H_1 \bar{c}_0 + H_2 \bar{b}_0 + H_3 \bar{a}_0},
\end{equation*}
and
\begin{eqnarray*}
{1\over 2\pi i}(\log(-i\xi) + 1) &\approx&  \\
& a_0 {\mu_1 \over \mu_0} & - a_1 
 - \left( {(F_2 - \tau H_2)(c_1 - c_0 {\mu_1 \over \mu_0}) + (F_1 - \tau H_1)(b_1 - b_0 {\mu_1 \over \mu_0}) - (F_0 - \tau H_0) \over F_3 - \tau H_3}
 \right).
\end{eqnarray*}
Given the above we can randomly choose flux vectors $F$ and $H$ and assemble a list of potential vacua. We drop candidate vacua whose $\xi$ is too far away from the conifold or whose Im$(\tau)$ too small (i.e. when the string coupling is large). By letting all entries of the flux vectors range freely between flux values from $-50$ to $50$ or so, and imposing the tadpole condition, we are able to reproduce the sorts of scatter plots of near conifold flux vacua found in \cite{KachruTaxonomy}. These generally show that the flux vacua become increasingly dense as one nears the conifold point as predicted in \cite{Douglas:2003um}. 

\begin{figure}[t]
\subfigure[]
{\includegraphics[width=.5\textwidth]{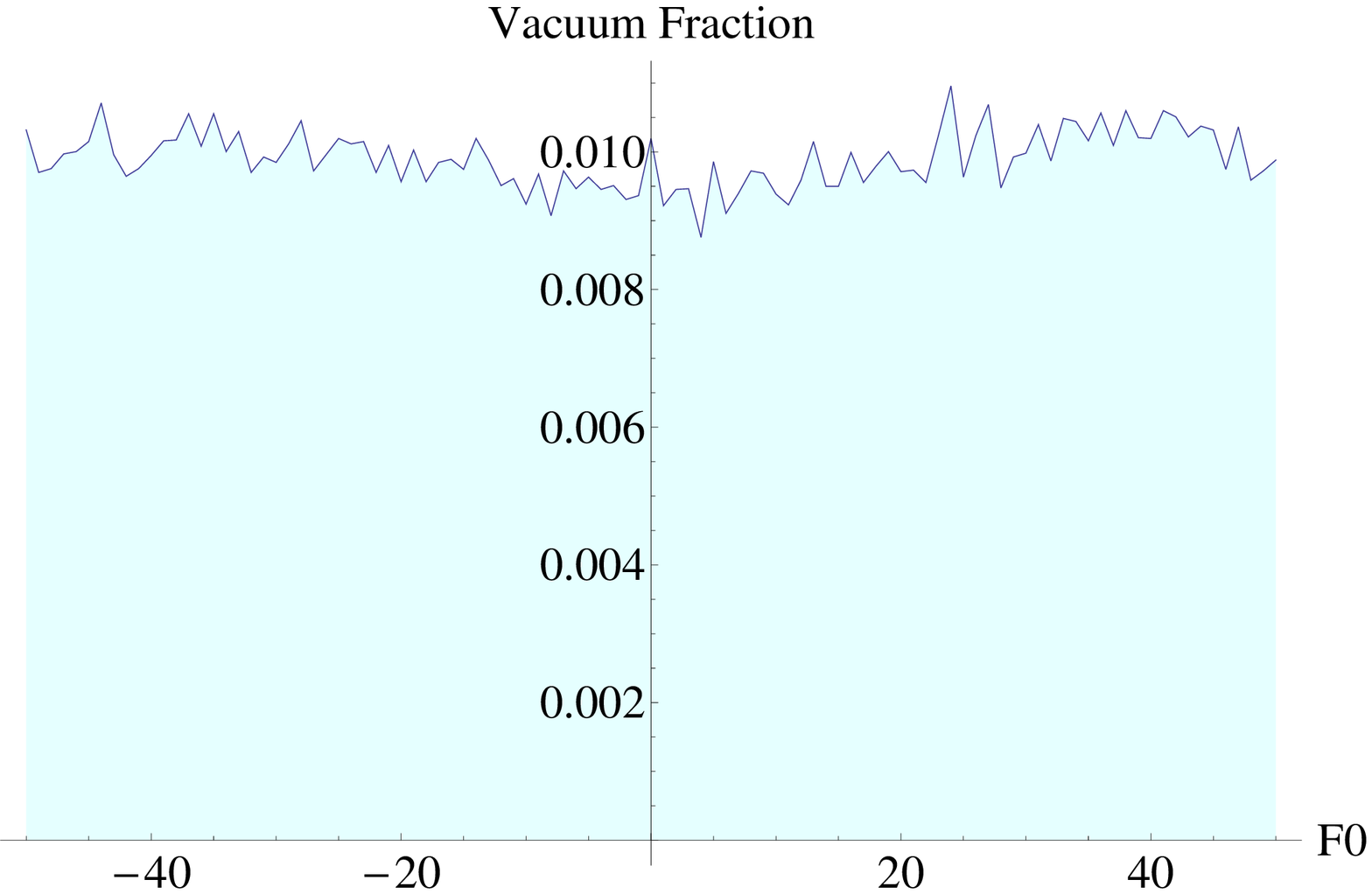}}
\subfigure[]
{\includegraphics[width=.5\textwidth]{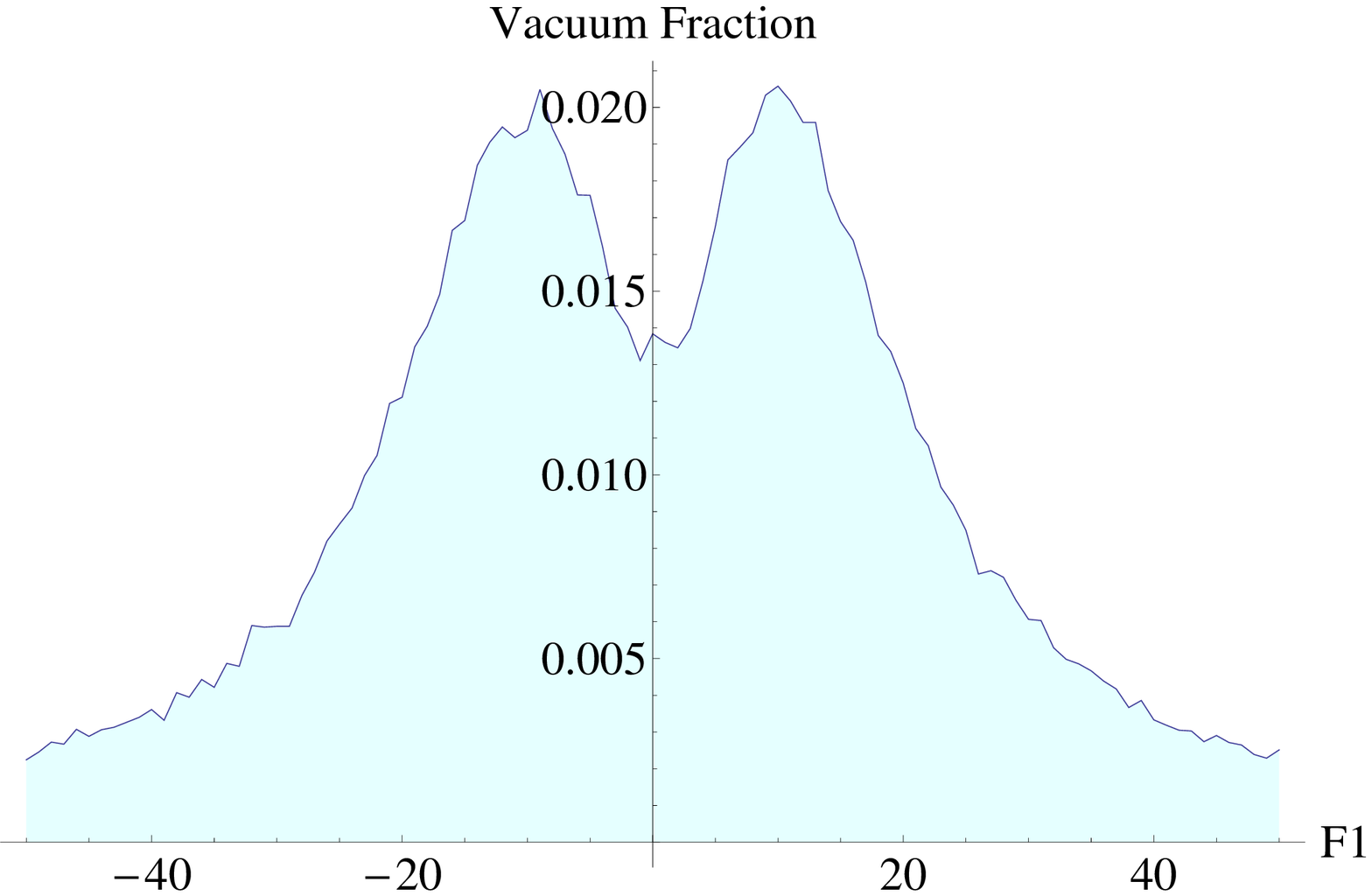}}
\subfigure[]
{\includegraphics[width=.5\textwidth]{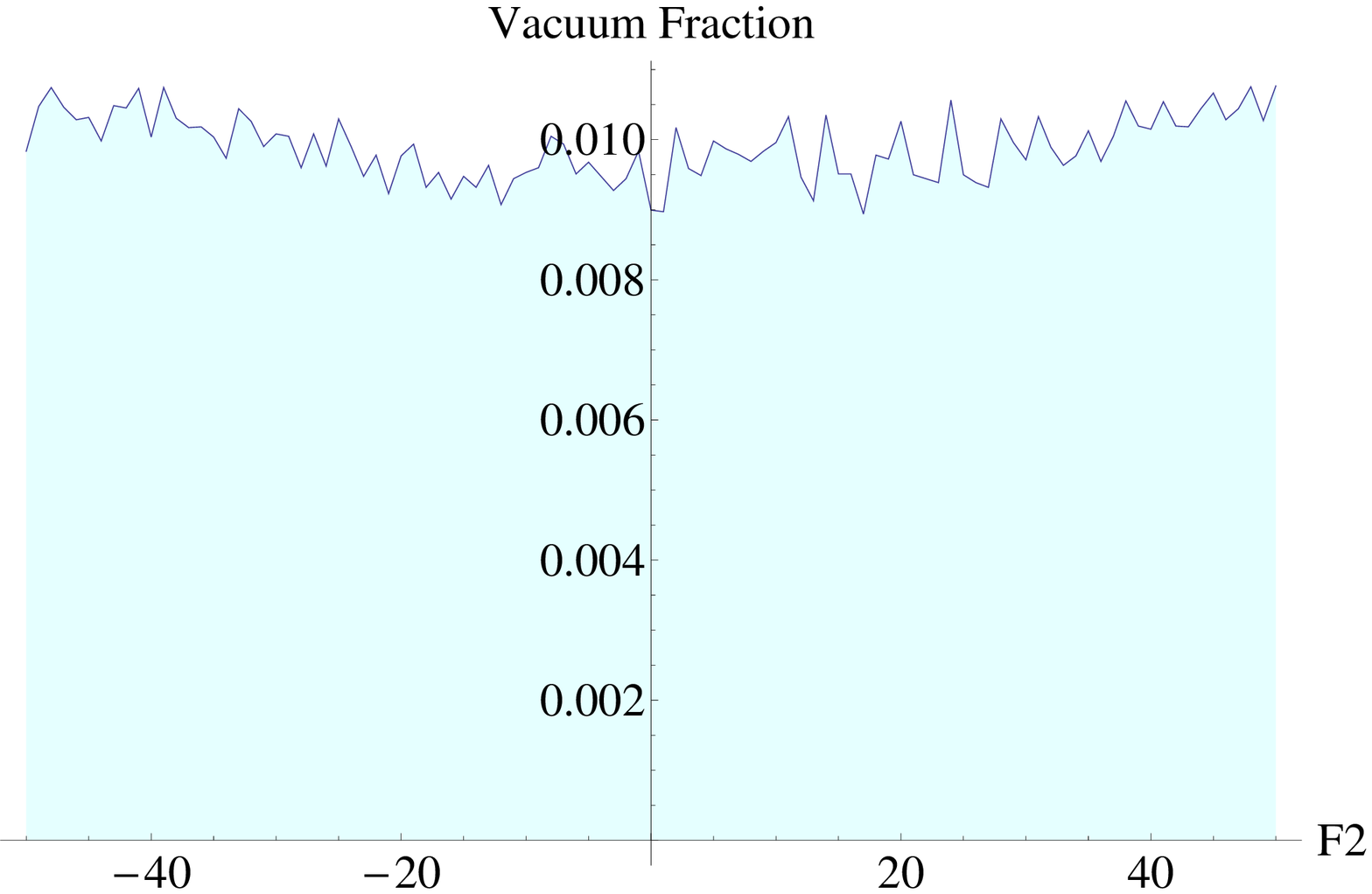}}
\subfigure[]
{\includegraphics[width=.5\textwidth]{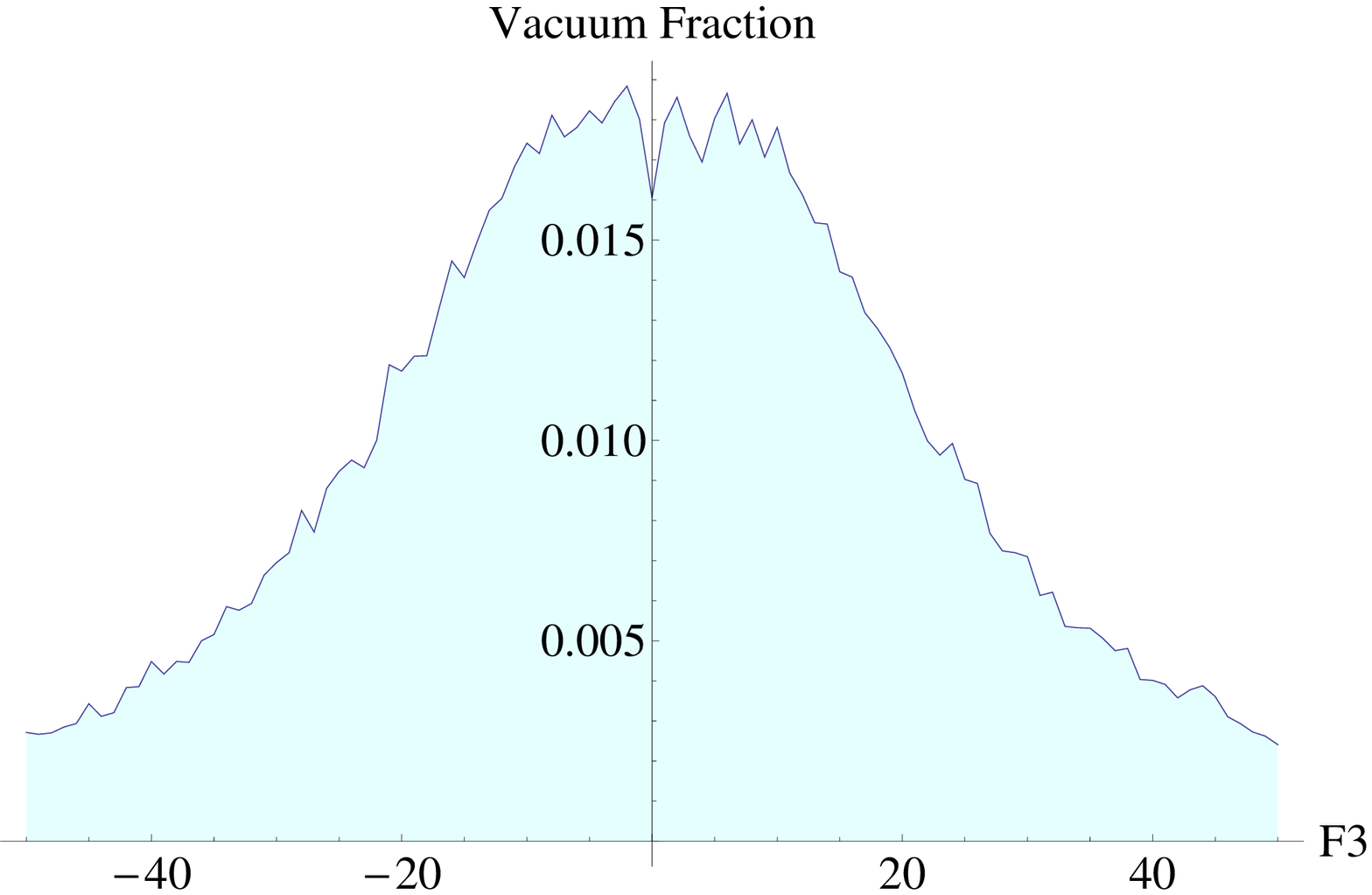}}
\caption{Distributions of flux vacua generated in $10^8$ runs with all fluxes in the interval $[-50,50]$. (a) The distribution sorted by $F_0$ appears uniform. (b) The distribution sorted by $F_1$ is bimodal and decays. (c) Sorting by $F_2$ appears also to yield a uniform distribution. (d) $F_3$ peaks around zero and decays.}
\label{fig-FfluxDistributions}
\end{figure}

\begin{figure}[t]
\subfigure[]
{\includegraphics[width=.5\textwidth]{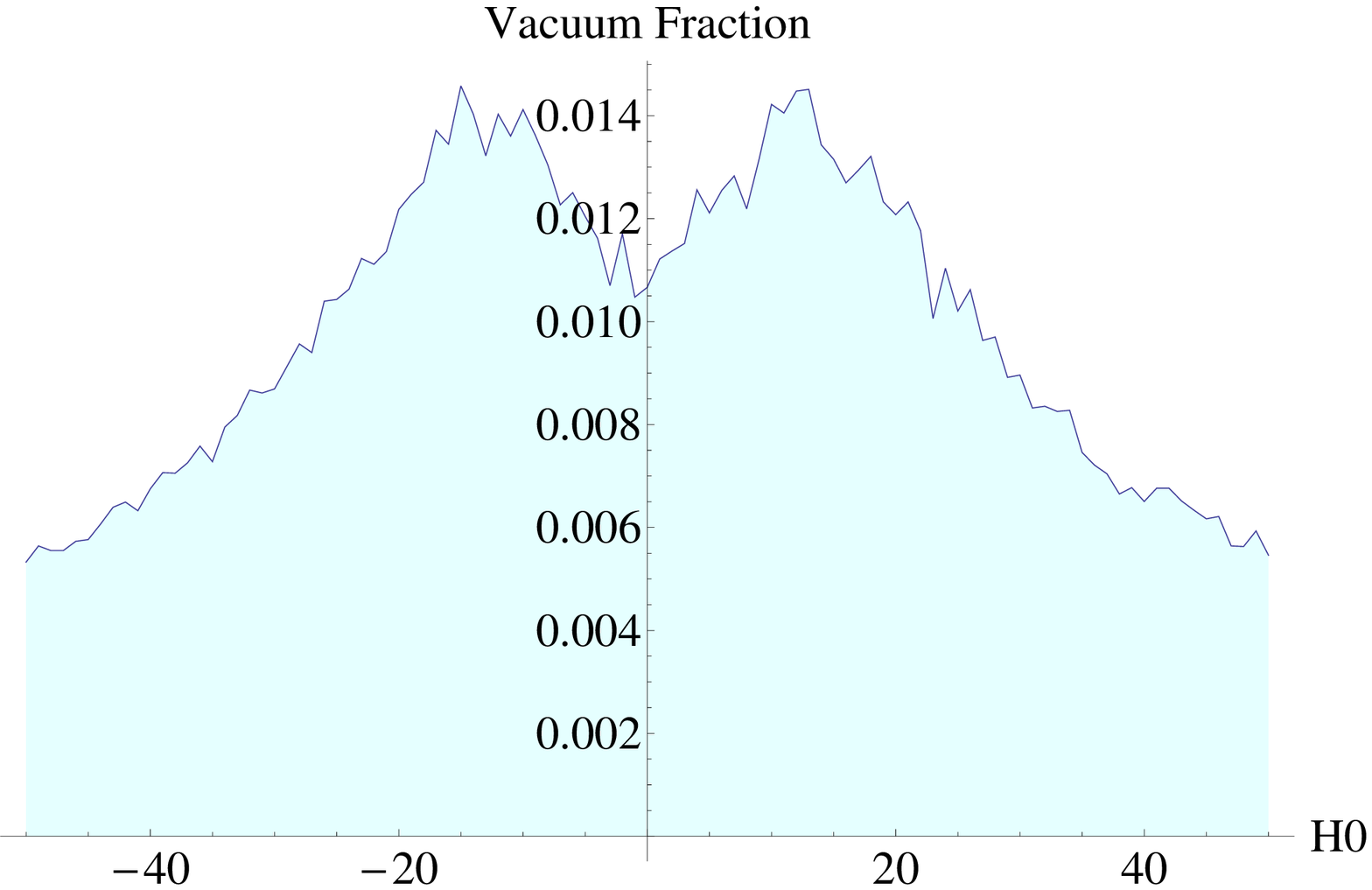}}
\subfigure[]
{\includegraphics[width=.5\textwidth]{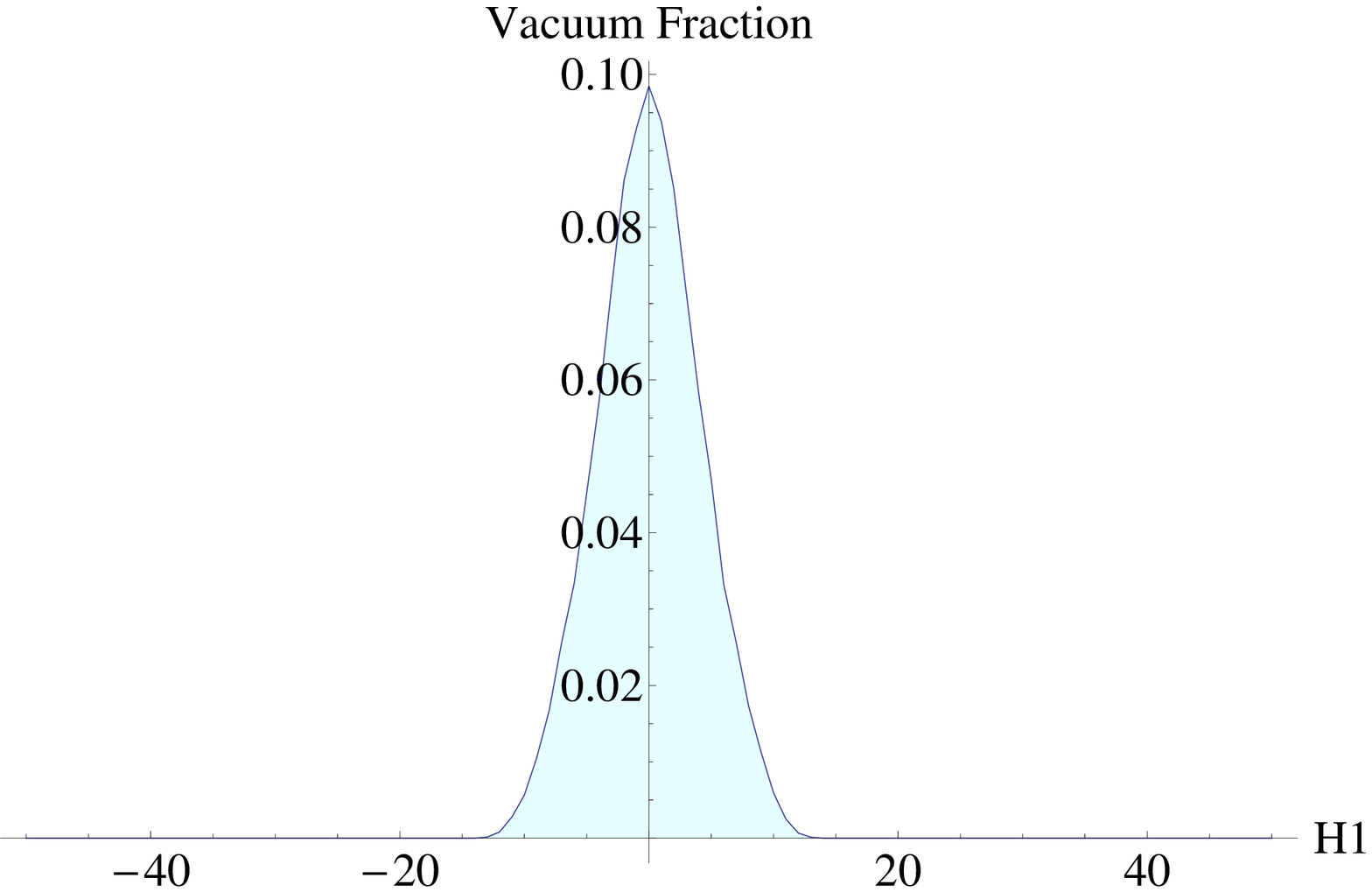}}
\subfigure[]
{\includegraphics[width=.5\textwidth]{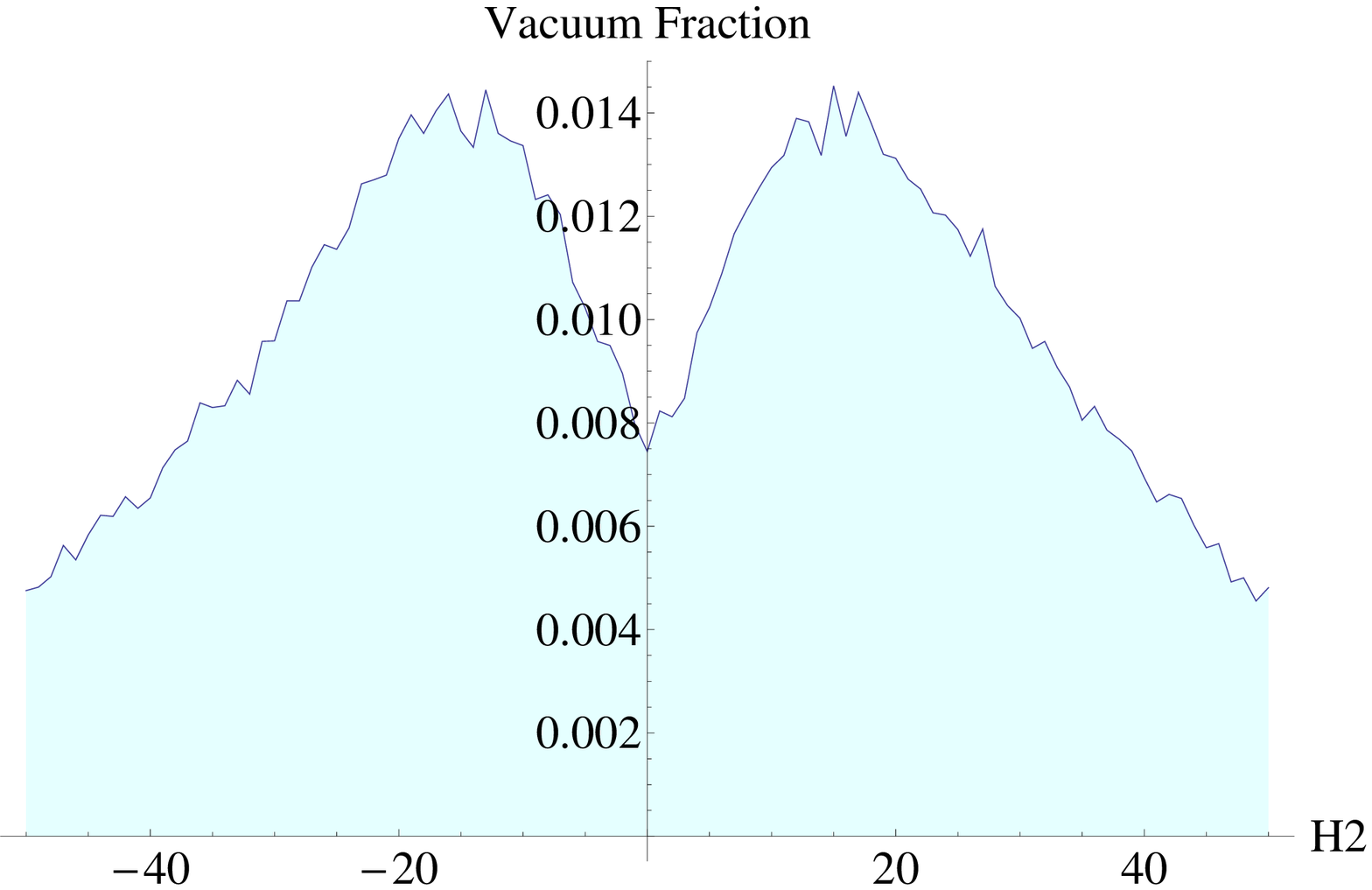}}
\subfigure[]
{\includegraphics[width=.5\textwidth]{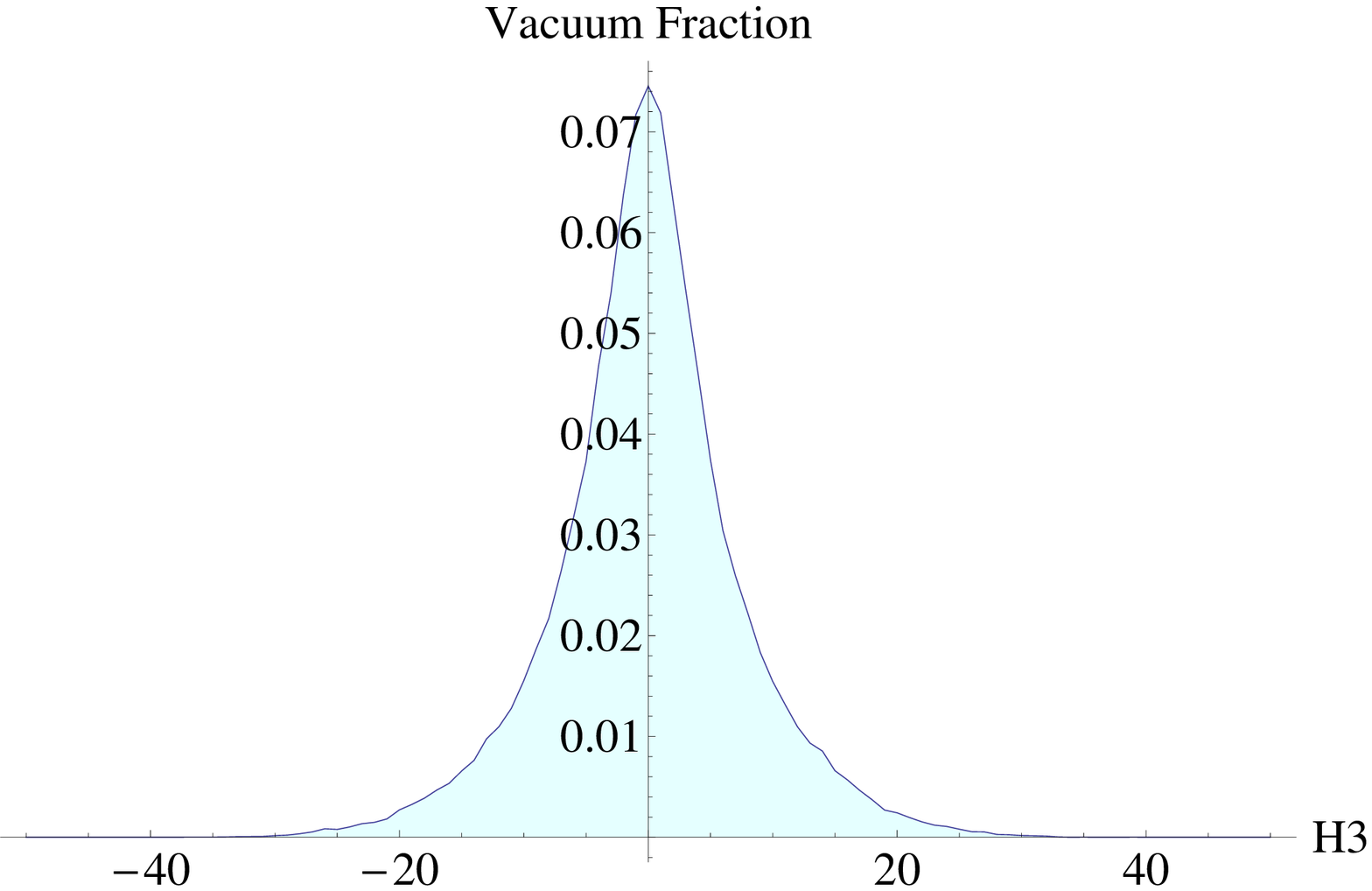}}
\caption{Distributions of flux vacua generated in $10^8$ runs with all fluxes in the interval $[-50,50]$. The distributions in (a) and (c) are sorted by $H_0$ and $H_2$, respectively. They are both bimodal and decay. Distributions (b) and (d) exhibit sharp peaks around $H_1 = 0$ and $H_3 = 0$, respectively with rapid decays.}
\label{fig-HfluxDistributions}
\end{figure}

In addition to recreating these older results, we find that the distribution of flux vacua around the conifold exhibits interesting structure when considered as a function of the value of the various fluxes. We see in figures \ref{fig-FfluxDistributions} and \ref{fig-HfluxDistributions} that the distribution of vacua is insensitive to the details of some of the fluxes ($F_0$ and $F_2$), while taken as functions of other fluxes we find non-trivial behavior. For example, the distribution of vacua drops off as $|F_3|$ grows, and does so even more severely for $H_1$ and $H_3$. The distributions exhibit bimodal behavior that then decays for $F_1$, $H_0$, and $H_2$. Understanding why these relationships exist is certainly worthy of further investigation.

\subsection{Vacuum statistics techniques and vacuum hunting}

For topographical explorations we can restrict to flux vectors of a certain form to investigate the connectivity of specific kinds of flux vacua. After generating a list of candidate vacua whose fluxes and locations satisfy all of the conditions of interest, we plug the flux vectors back into the numerical routines that generate the flux potential from the numerical period functions. So long as the vacua generated via the Monte Carlo method above are not too close to the conifold (where the numerics are inaccurate), we will find corresponding flux vacua in the numerically generated flux potential. If we fail to find the vacuum we are looking for there are three possibilities: 
\begin{itemize}
\item Since the vacuum positions receive corrections it is possible that when we plug the flux vectors back into the numerics, the vacuum position is corrected to be too close to the conifold.
\item It's possible that the vacuum, after receiving corrections is too far from the conifold point thus doesn't actually exist. This is possible since for some distances, order $\xi$ terms can compete with order $\log \xi$ terms. 
\item Oftentimes, the candidate vacua generated above will exist in the numerical potential but are found at flux vectors that differ by conifold monodromies. This is due to the fact that the log function is not single-valued.
\end{itemize}

An observation related to the last point is that in many cases, flux vacua near the conifold are connected via conifold monodromies to other flux vacua that need not be near the conifold. Thus, the above method often allows one to identify chains of SUSY flux vacua with several members far away from the conifold point.

In addition to the Monte Carlo method above, we have located some vacua via trial-and-error choices of flux vectors. Once again, given a set of fluxes that produce a potential with a minimum, monodromies can be used to search for connected vacua. We have found that this tends to produce chains of vacua which we exhibit below.

By comparing flux vacua for the different models, we find that given a set of values for $F$ and $H$ fluxes that produce a potential with a minimum within $|z|<1$ in one model, tend to produce potentials with similar minima in the other one-parameter models. We exhibit series of such analog vacua in each of the models. It's not hard to understand why this is so: the potentials are all dependent on the form of the Meijer functions used to compute the periods, and these functions do not differ drastically from model to model. In particular, the asymptotics near the LCS point are quasi-universal, taking the following general form
\begin{eqnarray}
\Pi_3 &\sim & \alpha_3 t^3 + \gamma_3 t + i \delta_3, \nn \\
\Pi_2 &\sim & \beta_2 t^2 + \gamma_2 t + \delta_2, \nn \\
\Pi_1 &\sim & t, \nn \\
\Pi_0 &\sim & 1,
\end{eqnarray}
where $t \sim \log z$ and the coefficients are all rational except $\delta_3 \sim \zeta(3)$. All of the coefficients and the precise behavior of $t$ conspire to ensure that the monodromy around $z=0$ in the Meijer basis is the same for all models (see appendix \ref{UniversalMonodromies}). On the other hand, the minima that fall outside the unit circle $|z| > 1$ do not appear to have analogs from model to model---or if they do, the analogs are less predictable and involve altering the fluxes. This is likely due to the fact that away from the LCS point, the models' periods exhibit more unique behaviors, which is most easily seen by observing that the monodromies around the Landau-Ginzburg point are unique to any given model.

To simplify our search for vacua, both using trial-and-error and Monte Carlo methods, we use the $SL(2,\mathbb{Z})$ symmetry of type IIB string theory to generally work with $H$ fluxes such that $H_3 = 0$. This choice ensures that the action of a conifold monodromy only alters the $F$ fluxes. Another useful trick involves exploiting the properties of the potential and string coupling under scaling of the flux vectors. When $F \rightarrow \lambda F$ and $H \rightarrow \rho H$, the potential goes like $V \rightarrow \lambda \rho V$, while the string coupling goes as $g_s \rightarrow (\rho/\lambda) g_s$. Thus, if while searching for flux vacua one encounters a candidate, but finds that it lies outside the bounds of validity due to $g_s$ being too large, one can generate a potential of equivalent shape, but with a suitably small $g_s$ by rescaling the fluxes. Note that this potential will not generically be connected to the original potential through monodromies; for example, one couldn't ever connect two such potentials through monodromies around the conifold point. It is unclear what this means: however, it is a useful shortcut for generating many examples of flux vacua. 

\subsection{Non-SUSY vacua}

We have tabulated several series of non-SUSY vacua for the various models. Vacua within the unit disk $|z|<1$ tend to have pretty clear analogs across the different models. Outside this region, but not too far from it vacua have more sporadic analogs. These involve tweaking the flux vectors in ways other than simple conifold monodromies.

\begin{center}
\begin{longtable}{|c|c|l|c|c|c|c|c|}
\hline
Model & Series & Vacuum & Flux Vectors & $z$ & $V_{\rm min}$ & $g_s$ & Monodromy\\
\hline \hline \endhead \hline \endfoot \hline
1   & 1 & i & $F = (2,9,-4,1)$; & $-0.286-0.485 i$ & 8.664 & 0.144 &  \\
	&    &  & $H = (-1,0,-7,0)$ &  &  &  &  \\ \cline{3-8}
    &    & ii & $F = (1,9,-4,1)$; & $-0.358-0.066 i$ & 8.452 & 0.159 & $T[1]^{-1}$  \\
    &    &  & $H = (-1,0,-7,0)$ &  &  &  &  \\ \cline{3-8}
    &    & iii & $F = (0,9,-4,1)$; & $-0.020+0.225 i$ & 8.236 & 0.177 & $T[1]^{-1}$  \\
    &    &  & $H = (-1,0,-7,0)$ &  &  &  &  \\ \cline{3-8}
\cline{2-8} \cline{2-8} 
 & 2 & i & $F = (4,-4,-1,2)$;  & $0.751-2.322 i$ & 3.042 & 0.215 & \\
 &  &  & $H = (1,1,6,0)$ &  &   & &  \\ \cline{3-8}
 &  & ii & $F = (2,-4,-1,2)$;  & $-2.442+1.865 i$ & 2.849 & 0.218 & $T[1]^{-1}$ \\
  & &  & $H = (1,1,6,0)$ &  &    & &  \\ \cline{3-8}
  & & iii & $F = (0,-4,-1,2)$;  & $-0.163+0.598 i$ & 2.778 & 0.207 & $T[1]^{-1}$ \\
  & &  & $H = (1,1,6,0)$ &  &    & &  \\
\cline{1-8} \\ \\ \\ 
\cline{1-8}
2 & 1 & i & $F = (2,9,-4,1)$; & $-0.184-0.418 i$ & 8.738 & 0.199 &  \\
 &	&  & $H = (-1,0,-7,0)$ &  &    & &  \\ \cline{3-8}
  & & ii & $F = (1,9,-4,1)$; & $-0.196-0.063 i$ & 8.499 & 0.218 & $T[1]^{-1}$  \\
  & &  & $H = (-1,0,-7,0)$ &  &  &  &  \\ \cline{3-8}
  & & iii & $F = (0,9,23,32)$; & $0.049+0.053 i$ & 8.252 & 0.248 & $T[0]^{-1} T[1]^{-1}$ \\
  & &  & $H = (-1,1,-1,3)$ &  &  &  &   \\ 
\cline{2-8} \\ \cline{2-8}
& 2 & i & $F = (4,-4,-1,2)$; & $-3.199-3.622 i$ & 2.510 & 0.198 &  \\
& 	&  & $H = (1,1,5,0)$ &  &  &  &   \\ \cline{3-8}
&	& ii & $F = (2,-4,-1,2)$; & $-2.148+2.032 i$ & 2.377 & 0.203 & $T[1]^{-1}$\\
&	&  & $H = (1,1,5,0)$ &  &  &  &  \\ \cline{3-8}
&	& iii & $F = (0,-4,-1,2)$; & $ -0.490+0.484 i $ & 2.349 & 0.201 & $T[1]^{-1}$\\
&	&  & $H = (1,1,5,0)$ &  &  & &  \\ 
\cline{1-8} \\ \\ \\
\cline{1-8}
8 & 1 & i & $F = (2,9,-4,1)$; & $-0.139-0.772 i$ & 8.689 & 0.106 &  \\
&	&  & $H = (-1,0,-7,0)$ &  &  &  &   \\ \cline{3-8}
 & & ii & $F = (1,9,-4,1)$; & $-0.617-0.239 i$ & 8.502 & 0.117 & $T[1]^{-1}$  \\
 & &  & $H = (-1,0,-7,0)$ &  &  &  &  \\ \cline{3-8}
 & & iii & $F = (0,9,-4,1)$; & $-0.444+0.329 i$ & 8.308  & 0.131 & $T[1]^{-1}$ \\
 & &  & $H = (-1,0,-7,0)$ &  &  &  &  \\ \cline{3-8}
 & & iv & $F = (-1,9,-4,1)$; & $0.078+0.461 i$ & 8.126 & 0.148 & $T[1]^{-1}$ \\
 & &  & $H = (-1,0,-7,0)$ &  &  &  & \\ 
\cline{2-8} \\ \cline{2-8}
& 2 & i & $F = (4,-4,-1,2)$; & $0.925-1.892 i$ & 3.606 & 0.237 &  \\
&	&  & $H = (1,1,7,0)$ &  &    & & \\ \cline{3-8}
&	& ii & $F = (2,-4,-1,2)$; & $-2.9+1.25 i$ & 3.354 & 0.237 & $T[1]^{-1}$\\
&	&  & $H = (1,1,7,0)$ &  &  & & \\ \cline{3-8}
&	& iii & $F = (0,-4,-1,2)$; & $ -0.047+0.673 i $ & 3.412  & 0.209 & $T[1]^{-1}$\\
&	&  & $H = (1,1,7,0)$ &  &  & & \\ 
\cline{1-8} \\
\cline{1-8}
12 & 1 & i & $F = (3,9,-4,1)$; & $-0.284-0.443 i$ & 8.838 & 0.174 &  \\
&	&  & $H = (-1,0,-7,0)$ &  &  &  &  \\ \cline{3-8}
 & & ii & $F = (2,9,-4,1)$; & $-0.350-0.018 i$ & 8.606 & 0.190 & $T[1]^{-1}$  \\
&	&  & $H = (-1,0,-7,0)$ &  &  &  & \\ \cline{3-8}
&  & iii & $F = (1,9,-4,1)$; & $-0.117+0.169 i$ & 8.373 & 0.209 & $T[1]^{-1}$  \\
&  &  & $H = (-1,0,-7,0)$ &  &  &  & \\
\hline
\end{longtable}
\vspace{-8pt}

\hspace{-250pt} {\small {\bf Table 1:} Non-supersymmetric flux vacua.}\label{NonSUSYVacuaTable}
\end{center}

\subsection{SUSY vacua and SUSY chains}

\begin{figure}[htp]
\subfigure[]
{\includegraphics[width=.5\textwidth]{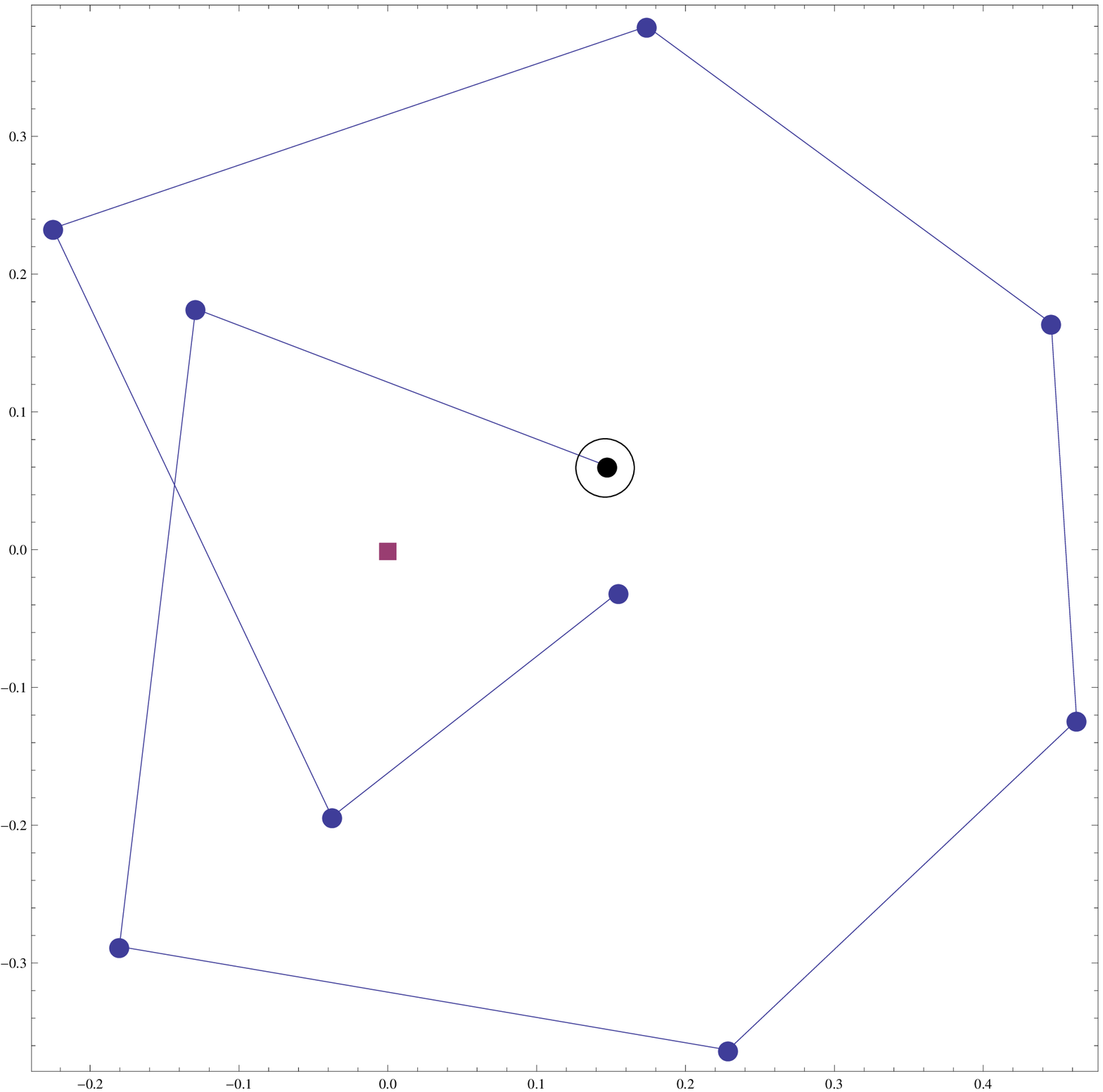}}
\subfigure[]
{\includegraphics[width=.5\textwidth]{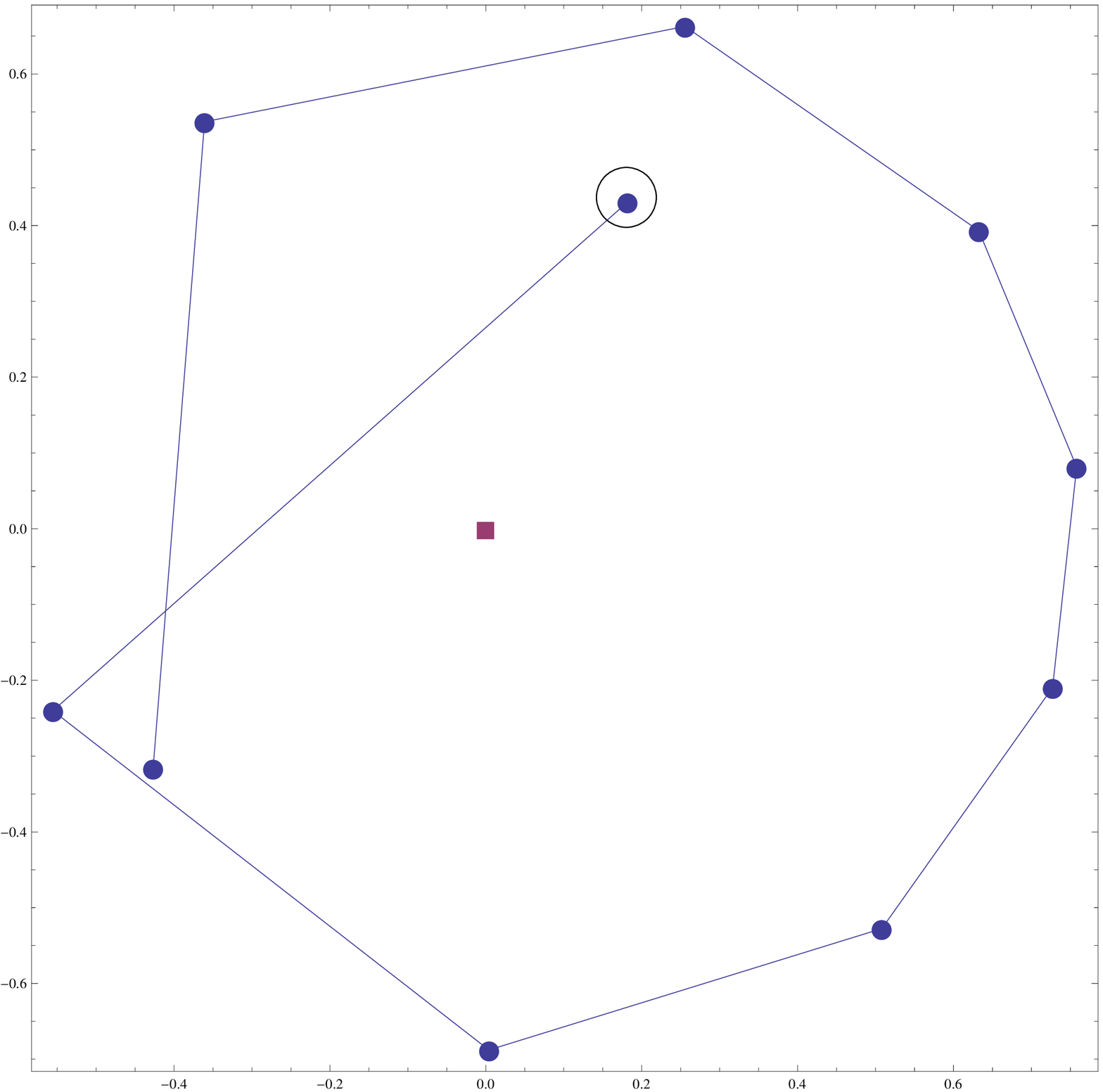}}
\subfigure[]
{\includegraphics[width=.5\textwidth]{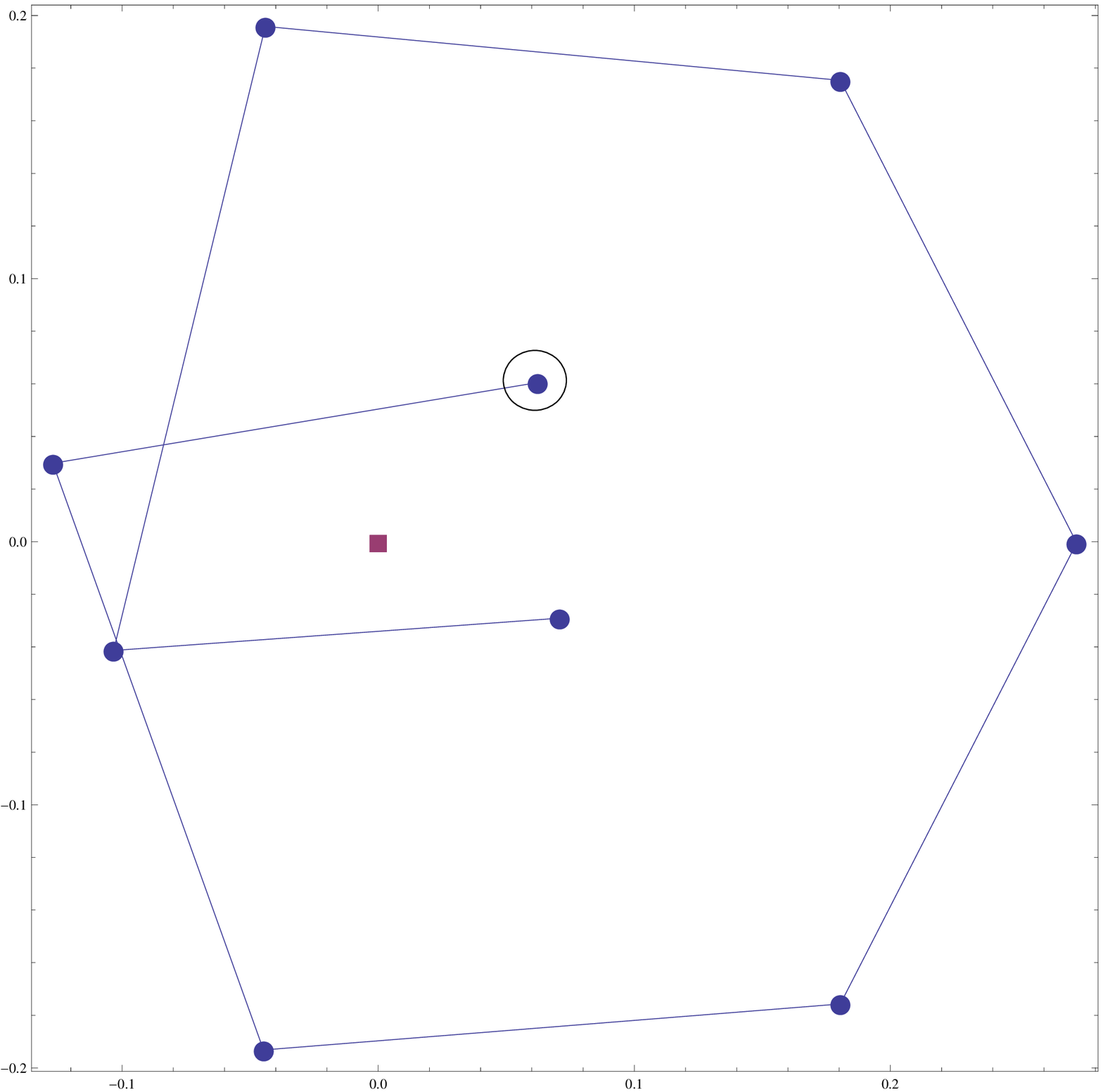}}
\subfigure[]
{\includegraphics[width=.5\textwidth]{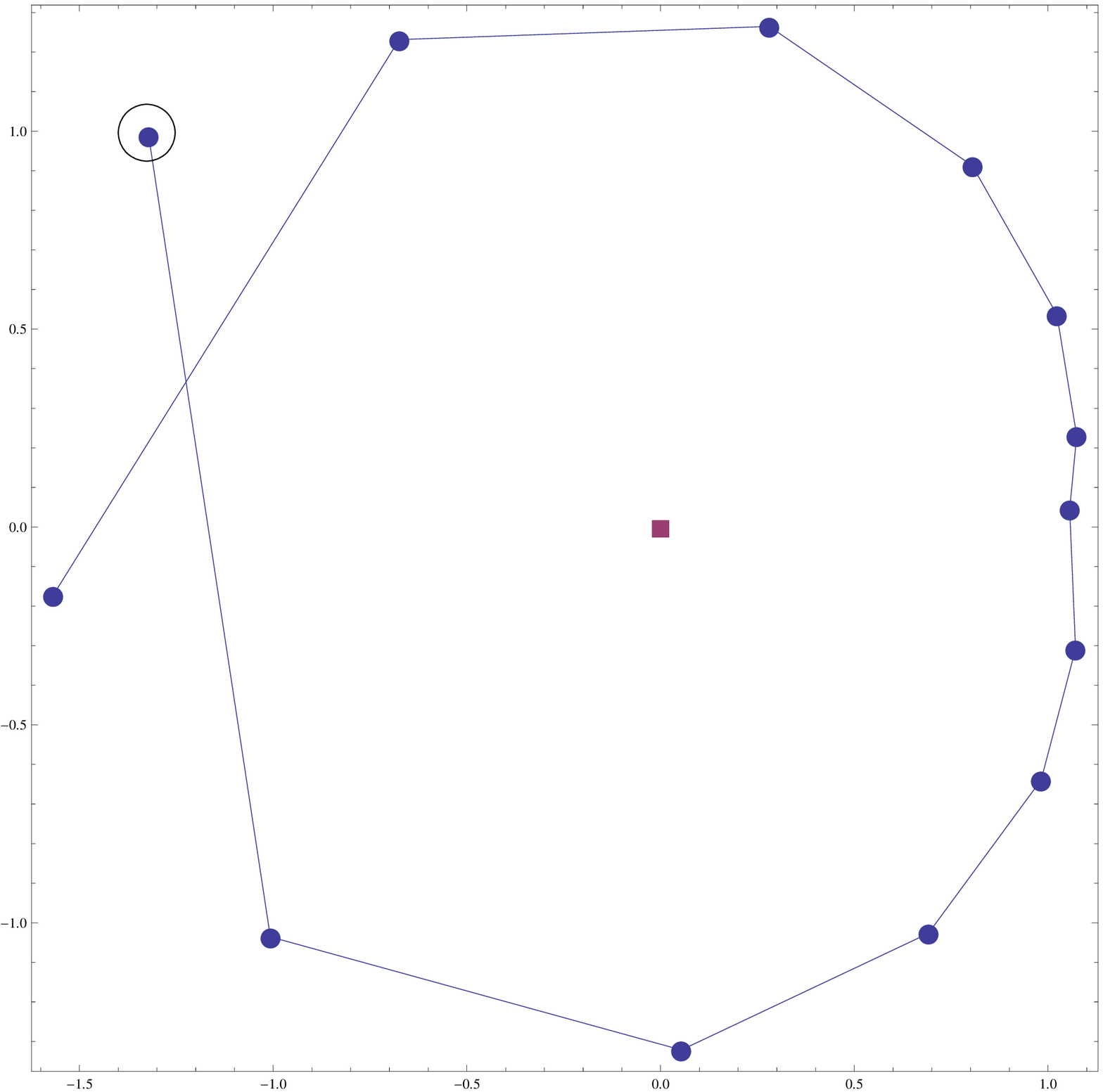}}
\caption{The supersymmetric chains with $H = (-1,0,-7,0)$ for four models. The circular dots are the locations of the vacua in the $z$-plane. The square dot indicates the LCS point $z=0$. In each panel the initial dot is circled. Panel (a) is the chain for the mirror quintic starting with $F = (-1,-6,-9,-1)$. Panel (b) is the chain for model 8 with initial $F = (0,-6,-9,-1)$. Panel (c) is the chain for model 12 with initial $F = (-1,-6,-9,-1)$. Panel (d) is the chain for model 14 with initial $F = (2,-6,-9,-1)$.}
\label{fig-SUSYvacs}
\end{figure}

Some of the SUSY vacua for the mirror quintic found in \cite{Danielsson} exhibited an interesting chained structure: given a SUSY vacuum, winding around the conifold point would usually take you to another one. Eventually, the vacua cease to be supersymmetric and have $V \neq 0$. These vacua, uncovered in \cite{Danielsson}, are arrayed in quasi-circular chains around the LCS point. In table \ref{SUSYTable} below we exhibit analog SUSY vacua in models 1, 8, 12, and 14. These vacua are plotted on the $z$-plane in figure \ref{fig-SUSYvacs}.

On examining these chains of vacua, one can see an approximate conjugation symmetry; if there is a vacuum at some complex $z$ there is usually a partner near to $\bar{z}$. This is also apparent in the example found in \cite{Danielsson}.

\begin{table}[hb]
\begin{tabular}{|l|c|c|c|c|}
\hline
NS-NS Flux & Model 1 & Model 8 & Model 12 & Model 14 \\
\hline 
(-1,-6,-9,-1) & 0.147+0.061\,i & --		               & 0.0622 + 0.060\,i & -- \\ 
(0,-6,-9,-1) & -0.129+0.175\,i & 0.182 + 0.431\,i      & -0.127 + 0.030\,i & -- \\
(1,-6,-9,-1) & -0.180-0.288\,i & -0.555 - 0.240\,i     & -0.045 - 0.193\,i & -- \\
(2,-6,-9,-1) &  0.229-0.363\,i & 0.004 - 0.688\,i      & 0.180 - 0.176\,i  &-1.321 + 0.988\,i\\
(3,-6,-9,-1) &  0.463-0.124\,i & 0.508 - 0.528\,i      & 0.273 - 0.000\,i  & -1.007 - 1.036\,i \\
(4,-6,-9,-1) &  0.446+0.165\,i & 0.727 - 0.210\,i      & 0.180 + 0.175\,i  & 0.054 - 1.321\,i \\
(5,-6,-9,-1) &  0.174+0.380\,i & 0.758 + 0.081\,i      & -0.044 + 0.196\,i & 0.691 - 1.027\,i \\
(6,-6,-9,-1) & -0.225+0.233\,i & 0.633 + 0.393\,i      & -0.103 - 0.041\,i & 0.982 - 0.639\,i \\
(7,-6,-9,-1) & -0.038-0.194\,i & 0.256 + 0.663\,i      & 0.071 - 0.029\,i  & 1.071 - 0.308\,i \\
(8,-6,-9,-1) &  0.155-0.031\,i & -0.361 + 0.537\,i 	   & --                & 1.056 + 0.045\,i \\
(9,-6,-9,-1) &  -- 			   & -0.427 - 0.316\,i     & -- 			   & 1.074 + 0.231\,i \\
(10,-6,-9,-1) & -- 			   & -- 				   & -- 			   & 1.023 + 0.536\,i \\
(11,-6,-9,-1) & -- 			   & -- 				   & -- 			   & 0.805 + 0.913\,i \\
(12,-6,-9,-1) & -- 			   & -- 				   & -- 			   & 0.280 + 1.265\,i \\
(13,-6,-9,-1) & -- 			   & -- 				   & -- 			   & -0.675 + 1.231\,i \\
(14,-6,-9,-1) & -- 			   & -- 				   & -- 			   & 1.569 - 0.174\,i \\
\hline
\end{tabular}
\caption{Chains of supersymmetric vacua connected via conifold monodromies. The complex numbers in the table are the locations of the vacua in the $z$-plane. Note that all vacua have $H = (-1,0,-7,0)$.}
\label{SUSYTable}
\end{table}

Looking at the table, one sees that model 14 appears to be a bit of an outlier. It exhibits the same patterns as the other models, but all of its vacua lie outside the unit circle in the $z$-plane. Model 14 turns out to be a rather special case. On comparing the asymptotic expansions around the LCS and Landau-Ginzburg points in this model, one finds that the periods around either point are related by a rescaling of $z^{-1/2}$ and a basis change \cite{Lazaroiu}. This suggests that the model may have an analog of T-duality which exchanges $z$ and $1/z$. One can actually think of two equivalent models in which either $z$ or $1/z$ is taken as the natural coordinate around the LCS point. Just as with T-duality, one imagines these to be glued along the unit circle, so that when $|z|$ becomes too large, the proper description shifts to the model in terms of $1/z$. Such a construction would eliminate the vacua for model 14 listed in table \ref{SUSYTable} and depicted in figure \ref{fig-SUSYvacs}. However, this is mostly speculative and we will not delve into this issue further in this paper.

\begin{table}[tbp]
\begin{tabular}{|l|c|c|c|}
\hline
NS-NS Flux & Model 1 & Model 8 & Model 12 \\
\hline 
(3, -18, 9, -1)  & $-0.001 - 0.071\,i$ & --    				   & -- \\
(4, -18, 9, -1)  & $ 0.011 - 0.083\,i$ & --     			   & -- \\
(5, -18, 9, -1)  & $ 0.024 - 0.093\,i$ & --     			   & --  \\
(6, -18, 9, -1)  & $ 0.067 - 0.134\,i$ & --     			   & --  \\
(7, -18, 9, -1)  & $ 0.111 - 0.156\,i$ & $ 0.137 - 0.315\,i$   & $ 0.0673958 - 0.0806928\,i$  \\
(8, -18, 9, -1)  & $ 0.165 - 0.179\,i$ & $ 0.192 - 0.361\,i$   & $ 0.10864 - 0.090557\,i$ \\
(9, -18, 9, -1)  & $ 0.227 - 0.195\,i$ & $ 0.254 - 0.406\,i$   & $ 0.150319 - 0.0941855\,i$ \\
(10, -18, 9, -1) & $ 0.298 - 0.204\,i$ & $ 0.323 - 0.450\,i$   & $ 0.194668 - 0.091672\,i$	\\
(11, -18, 9, -1) & $ 0.375 - 0.206\,i$ & $ 0.398 - 0.493\,i$   & $ 0.23935 - 0.0830528\,i$ \\
(12, -18, 9, -1) & $ 0.459 - 0.200\,i$ & $ 0.477 - 0.534\,i$   & $ 0.293778 - 0.0684983\,i$ 	\\
(13, -18, 9, -1) & $ 0.546 - 0.185\,i$ & $ 0.560 - 0.574\,i$   & $ 0.347221 - 0.0483759\,i$	 \\
(14, -18, 9, -1) & $ 0.635 - 0.162\,i$ & $ 0.646 - 0.616\,i$   & $ 0.401897 - 0.0209002\,i$ 	 \\
(15, -18, 9, -1) & $ 0.724 - 0.131\,i$ & $ 0.734 - 0.660\,i$   & $ 0.4568 + 0.0119374\,i$ 	\\
(16, -18, 9, -1) & $ 0.810 - 0.093\,i$ & $ 0.827 - 0.712\,i$   & $ 0.510894 + 0.0506745\,i$ 	 \\
(17, -18, 9, -1) & $ 0.889 - 0.050\,i$ & $ 0.925 - 0.774\,i$   & $ 0.563092 + 0.0949499\,i$ 	 \\
(18, -18, 9, -1) & $ 0.957 - 0.005\,i$ & $ 1.033 - 0.851\,i$   & $ 0.612196 + 0.144716\,i$ 	 \\
(19, -18, 9, -1) & $ 1.041 + 0.044\,i$ & $ 1.154 - 0.950\,i$   & $ 0.657017 + 0.200672\,i$	 \\
(20, -18, 9, -1) & $ 1.129 + 0.092\,i$ & $ 1.296 - 1.077\,i$   & $ 0.69653 + 0.264578\,i$ 	 \\
(21, -18, 9, -1) & $ 1.243 + 0.165\,i$ & $ 1.466 - 1.244\,i$   & $ 0.729777 + 0.33926\,i$	 \\
(22, -18, 9, -1) & $ 1.379 + 0.268\,i$ & $ 1.676 - 1.464\,i$   & $ 0.756487 + 0.428029\,i$ 	 \\
(23, -18, 9, -1) & $ 1.544 + 0.414\,i$ & $ 1.940 - 1.760\,i$   & $ 0.774739 + 0.534383\,i$ 	 \\
(24, -18, 9, -1) & $ 1.740 + 0.621\,i$ & $ 2.281 - 2.167\,i$   & $ 0.78168 + 0.661604\,i$ 	 \\
(25, -18, 9, -1) & $ 1.968 + 0.916\,i$ & $ 2.728 - 2.741\,i$   & $ 0.772258 + 0.812786\,i$ 	 \\
(26, -18, 9, -1) & $ 2.225 + 1.342\,i$ & $ 3.332 - 3.585\,i$   & $ 0.738964 + 0.990706\,i$ 	 \\
(27, -18, 9, -1) & $ 2.498 + 1.963\,i$ & $ 4.168 - 4.880\,i$   & $ 0.671106 + 1.19711\,i$ 	 \\
(28, -18, 9, -1) & $ 2.743 + 2.883\,i$ & -- 				   & $ 0.554217 + 1.43178\,i$ 	 \\
(29, -18, 9, -1) & $ 2.853 + 4.259\,i$ & -- 				   & $ 0.369054 + 1.69091\,i$ 	  \\
(30, -18, 9, -1) & -- 			       & -- 				   & $ 0.0910431 + 1.9635\,i$  	  \\
(31, -18, 9, -1) & -- 			       & -- 				   & $-0.309701 + 2.22908\,i$ 	 \\
(32, -18, 9, -1) & -- 			       & -- 				   & $-0.863008 + 2.44355\,i$	 \\
(33, -18, 9, -1) & -- 			       & -- 				   & $-1.60096 + 2.54456\,i$  	 \\
(34, -18, 9, -1) & -- 			       & -- 				   & $-2.53126 + 2.42602\,i$ 	 \\
(35, -18, 9, -1) & -- 			       & -- 				   & $-3.61356 + 1.94421\,i$ 	 \\
(36, -18, 9, -1) & -- 			       & -- 				   & $-4.72581 + 0.913417\,i$ 	  \\
\hline
\end{tabular}
\caption{New type of chains of SUSY vacua connected via conifold monodromies. The complex numbers in the table are the locations of the vacua in the $z$-plane. Note that all vacua have $H = (-2, -4, -33, 0)$. The entries marked -- indicate the end of the series according to the numerics, but it is possible that these series extend indefinitely.}
\label{SUSYTable2}
\end{table}

\begin{figure}
\subfigure[]
{\includegraphics[width=.5\textwidth]{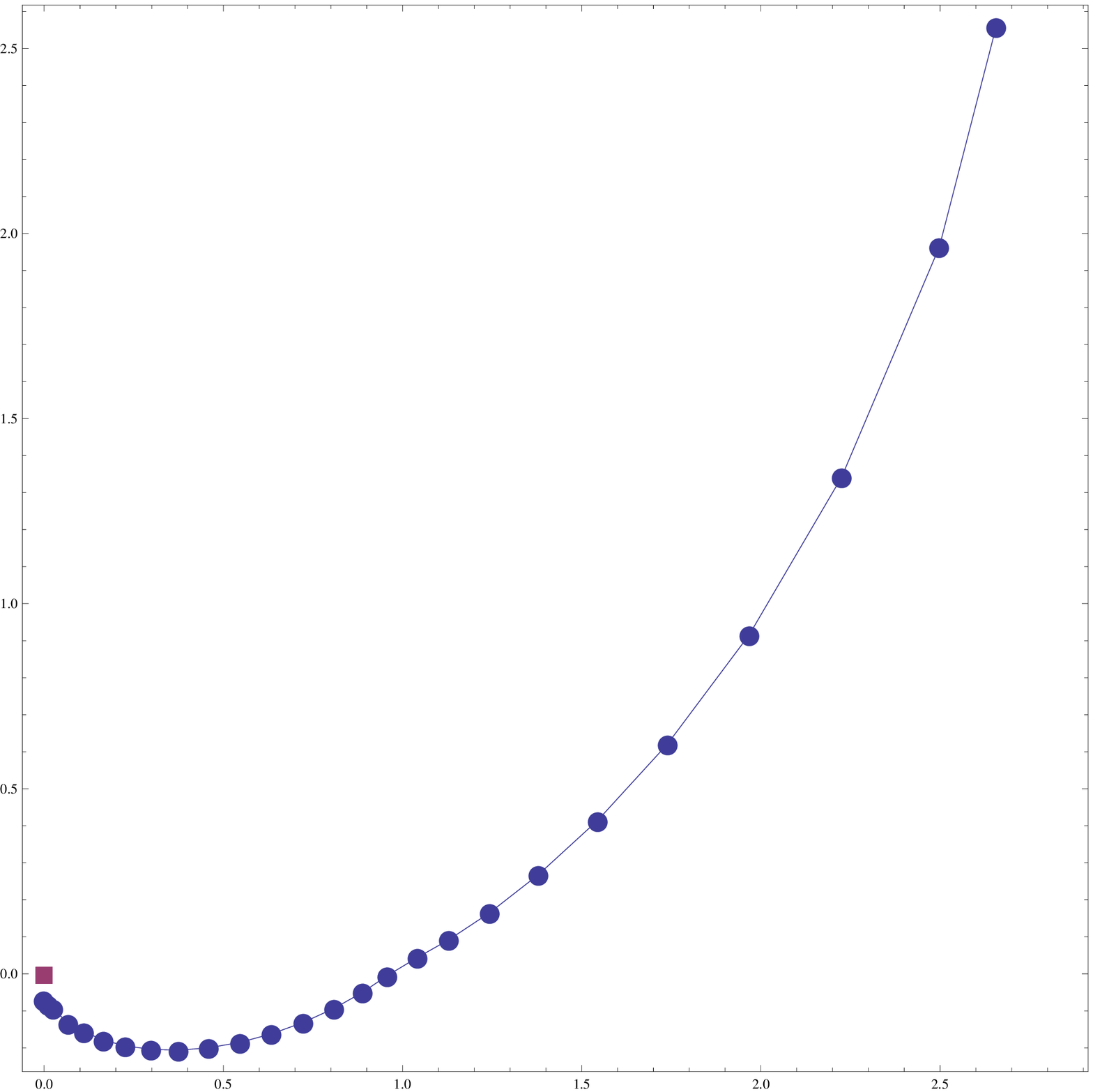}}
\subfigure[]
{\includegraphics[width=.5\textwidth]{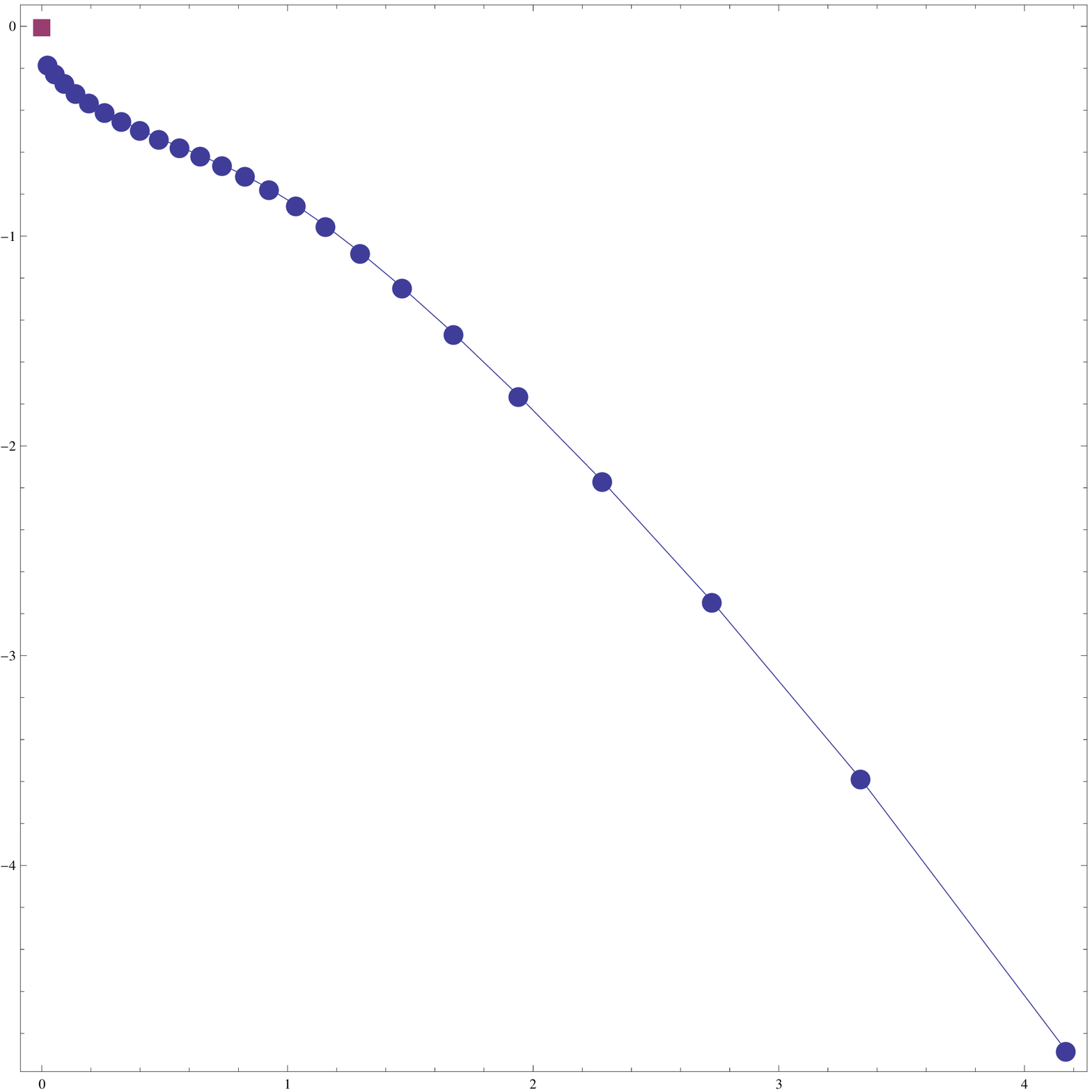}}
\subfigure[]
{\includegraphics[width=0.9\textwidth]{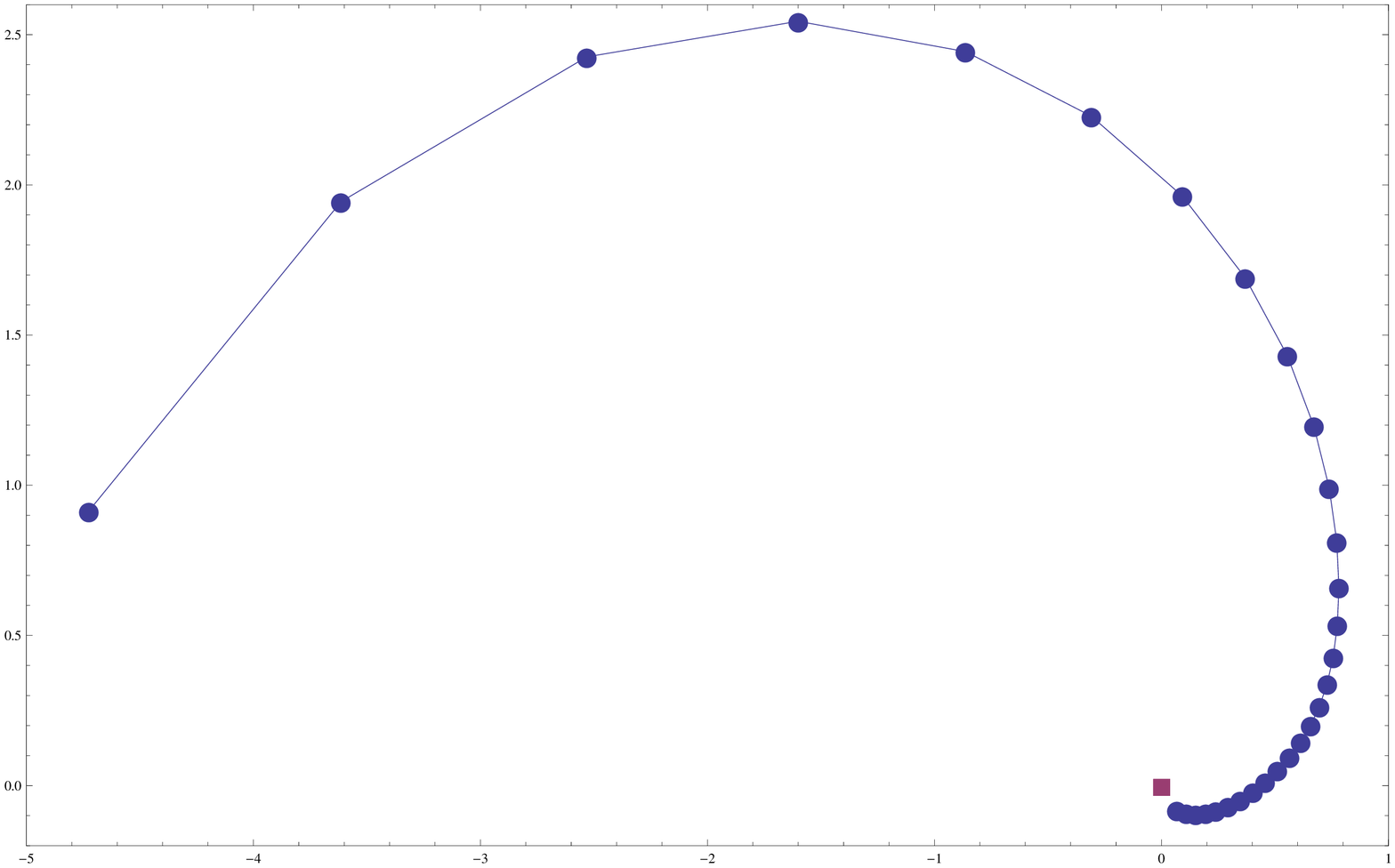}}
\caption{The supersymmetric chains with $H = (-2,-4,-33,0)$ for three models. The circular dots are the locations of the vacua in the $z$-plane. The square dot indicates the LCS point $z=0$. Panel (a) is the chain for the mirror quintic starting with $F = (3,-18,9,-1)$. Panel (b) is the chain for model 8 with initial $F = (7,-18,9,-1)$. Panel (c) is the chain for model 12 with initial $F = (7,-18,9,-1)$.}
\label{fig-SUSYvacs2}
\end{figure}

Our Monte Carlo search for vacua also turned up a new type of SUSY vacuum chain, exhibited in table \ref{SUSYTable2} and figure \ref{fig-SUSYvacs2}. These appear to arc away from $z=0$ in ever larger jumps from the conifold point as one performs inverse conifold monodromies. There appear to be accumulations of vacua approaching $z=0$, but since the numerical methods only give us access to a finite grid with finite resolution on the moduli space, we cannot tell if the vacua that lie outside the unit circle extend forever, and we cannot resolve whether vacua continue to accumulate near $z=0$ indefinitely.

\section{Vacuum Transitions}
\label{sec-CDL}

Having found several examples of supersymmetric and non-supersymmetric vacua, we now turn our attention to vacuum transitions.  Although different vacua have different flux configurations, monodromies can connect certain vacua in a single continuous potential with multiple sheets as exemplified in tables \ref{NonSUSYVacuaTable} and \ref{SUSYTable} above. As pointed out in \cite{Yan09,AJL,BroDah10}, tunneling with multiple fields can be quite subtle---we should not expect to easily guess a path connecting two vacua in the potential and then construct the Coleman-deLuccia (CDL) instanton \cite{Coleman:1980aw}.  Thus, before we look for the actual string theory instantons, here we give a general overview concerning multifield tunneling.

 \subsection{A nontrivial generalization}
 \label{sec-onefield}
 
Consider a potential $V(\phi)$ with false vacuum $\phi^{(1)}$ and true vacuum $\phi^{(2)}$.  In field theory, the tunneling transition is given by an instanton---a field configuration whose Euclidean action
\begin{equation}
S_E = \int dx^4 \left(\frac{(\partial\phi)^2}{2}+V\right)
\end{equation}
is a saddle point with exactly one negative mode.  It was proved to be an $O(4)$ symmetric, Euclidean solution of the following equation of motion with given boundary conditions:
\begin{equation}
\frac{d^2\phi}{dr^2}+\frac{3}{r}\frac{d\phi}{dr}=\frac{dV}{d\phi}~, \ \ \ \ 
\dot{\phi}(0)=0, \ \ \ \ \phi(\infty)\rightarrow\phi^{(1)}~,
\label{eq-coleman}
\end{equation}
where $\phi(0)$ is somewhat close to $\phi^{(2)}$.  Including gravity, the solution is similarly a topological 4-sphere with $O(4)$ symmetry:
\begin{eqnarray}
\frac{d^2\phi}{dr^2}+\frac{3}{a(r)}\frac{da}{dr}\frac{d\phi}{dr}&=&\frac{dV}{d\phi}~, 
\nonumber \\
\frac{1}{2}\left(\frac{d\phi}{dr}\right)^2+V&=&\frac{3}{a^2}\left(\frac{da}{dr}\right)^2~.
\label{eq-CDL}
\end{eqnarray}
Here the boundary conditions are $d\phi/dr=0$ at the two poles (where $a=0$) of this topological 4-sphere.

Such an instanton solution is numerically quite tractable through the overshoot/undershoot strategy given by Coleman.  Basically, one starts from a field value near the true vacuum, integrates the equations and see what happens at the other boundary.  The result may either fail to reach the false vacuum (undershoot), or go past the false vacuum (overshoot).  Repeating the process can get you arbitrarily close to a true solution which just reaches the false vacuum.

The same strategy fails for multiple fields.  When you miss the false vacuum in a multidimensional field space, it is not simply an undershoot or an overshoot.  You may try to keep shooting and see if you get lucky, but we know even if the numerics are very precise, a small change in the initial condition can be magnified (in a practically unpredictable way) while integrating the equations of motion. In addition to the unwieldy errors that propagate from initial conditions, other numerical errors also accumulate.  Although one can envision Monte Carlo processes that would aid one in applying the shooting strategy, such approaches will not be necessary for our purposes and we do not pursue them here.

Instead of the undershoot/overshoot approach, we consider a special class of simple and widely studied instanton solutions---the thin wall instantons. In these situations the action can be written as
\begin{eqnarray}
S_E &=& \int_0^{\tilde{r}} \frac{4\pi}{3} r^3 dr V_2 
    + \int_{\rm thin\ wall} \frac{4\pi}{3} r^3 dr \left(\frac{\phi'^2}{2}+V\right)
    + \int_{\tilde{r}}^\infty 4\pi r^3 dr V_1 \nonumber \\
    &=& \frac{\pi}{3}\tilde{r}^4 (V_2-V_1) + \frac{4\pi}{3}\tilde{r}^3\sigma
    + S_{\phi^{(1)}}~.
\end{eqnarray}
The dependence of $S_E$ nicely factorizes into the size of the bubble $\tilde{r}$, and the dependence of the domain wall tension $\sigma$ on the field configuration $\phi$. With fixed tension, we find that $S_E$ is maximized by choosing 
\begin{equation}
\tilde{r}=r_c=\frac{3\sigma}{(V_1-V_2)}~.
\end{equation}
This is the only negative mode in the action.  Namely, it reduces our problem to minimizing the tension with respect to different field configurations.  Generalizing to multiple fields, we look for a path in the multidimensional field space which minimizes
\begin{equation}
\sigma = \int dz \left(\frac{G_{ij}}{2}\frac{d\phi_i}{dz}\frac{d\phi_j}{dz}+V-V_1\right)~.
\end{equation}
Since we are dealing with a minimization problem (rather than looking for a saddle point), we can apply the ``relaxation method''\footnote{It was first introduced to this problem in\cite{AJL} and applied in\cite{GLHY}.} which will be described in more detail in section \ref{sec-conifunneling}.  Note that we introduced a nontrivial metric $G_{ij}$ that depends on the fields $\phi_i$, since with multiple fields we cannot in general absorb it via field redefinition.  In fact, the non-trivial metric on field-space plays a crucial role in the dynamics. Despite this, basic aspects of the problem do not change; for example, once we find the minimizing path, it solves the equations of motion and gives us a familiar formula
\begin{equation}
\sigma = \int_{\rm path} \sqrt{2(V-V_1)}\sqrt{G_{ij}d\phi_id\phi_j} ~.
\end{equation}

We will always assume that we can have a thin wall bubble. Therefore, the search for an instanton reduces to the search for a tunneling path through the multidimensional field space.

\subsection{Intuitions for multifield tunneling}
\label{sec-path}
 
Our primary method for finding the correct path is the relaxation technique mentioned above. Usually, applying relaxation involves cleverly designing a quantity to be relaxed which may have little connection to any physical quantity we truly care about. In our case, relaxation is much more natural since we wish to address the physical problem of finding dominant contributions to a path integral.  The quantity being relaxed, $\sigma$, enters the action directly and indeed a path contributes more as $\sigma$ varies more and more slowly. That being said, we cannot totally depend on numerical techniques, however naturally suited to our problem they might be.  Here we will provide some intuitions for analytical approaches. The combination of numerical and analytical reasoning will eventually lead to a satisfying answer.

One way to simplify the problem of finding the correct path is to reduce the effective number of relevant fields, even down to a single field if possible.  For example, given a potential $V(\phi_i)$ of $n$ fields with two vacua $\phi_i^{(1)}$ and $\phi_i^{(2)}$, we can look for a subspace containing those vacua, parametrized by $\psi_j$
\begin{equation}
\phi_i=f_i(\psi_j)~, \ \ \ \  j=1\sim m~,\ \ \ \  m<n~,
\end{equation}
such that this subspace is a local minimum along all orthogonal directions.
\begin{eqnarray}
& & \forall~l=1\sim(n-m)~, \ k=1\sim m~, \nonumber \\
& & G_{ij}\,\theta_i^l \frac{\partial f_j}{\partial\psi_k}=0~, \ \ 
G_{ij}\,\theta_i^l  \theta_j^l=1~,
\nonumber \\
& & \frac{\partial V}{\partial\theta} = 0~, \ \ \ \  
\frac{\partial^2 V}{\partial\theta^2}>0~.\nonumber 
\end{eqnarray}
Then we can solve the problem with fewer fields,
\begin{eqnarray}
\bar{G}_{kl}&=&G_{ij}\frac{\partial f_i}{\partial\psi_k}\frac{\partial f_j}{\partial\psi_l}~,
\nonumber \\
\bar{V}(\psi_j)&=&V(f_i(\psi_j))~.
\end{eqnarray}

This formalism looks especially promising when the $(n-m)$ degrees of freedom removed are heavy, namely when
\begin{equation}
\frac{\partial^2 V}{\partial\theta^2}\gg 
\frac{1}{\bar{G}_{kl}}\frac{\partial^2V}{\partial\psi_k\partial\psi_l}~.
\end{equation}
This is essentially the same as integrating out the UV spectrum to study the low energy effective theory.  In the simplest case where $n=2, m=1$, this can be visualized as a valley that connects two vacua. 

Unfortunately, this does not work in general.  First of all, there is no guarantee that the potential has a valley.  Even if it does, recall that a tunneling path is a path of classical motion in the inverse potential, $-V$.  A valley in the potential corresponds to a mountain ridge in the inverse potential.   In the case of heavy transverse fields, this mountain ridge is narrow and steep.  Obviously it is very easy to roll down the slope and there is no guarantee that a classical path follows the ridge (this is only possible when it is extremely straight).  Examples where no paths follow a valley can be found in appendix~\ref{sec-toy}.

Even though na\`ive integrating out of fields is dangerous, low energy effective theories should be safe as long as we treat them carefully.  In the case where a na\`ive tunneling path within the low energy theory is fake, like the ones in appendix~\ref{sec-toy}, either it is illegal to integrate out along the path in the first place, or you will find that the tension of the fake path is comparable to the mass of the heavy fields, namely the UV cutoff scale of your low energy theory\footnote{We thank Alberto Nicolis and Eduardo Ponton for pointing out this issue.  We also thank Erick Weinberg for making it clear.}.

On the other hand, the analogy to classical motion implies that we can ignore light enough degrees of freedom.  Imagine adding a flat direction to a standard $(\phi^2-1)^2$ potential. Obviously it is still a single field problem.  Now imagine that the extra direction is not exactly flat, but instead, it varies in energy scale much less than the $(\phi^2-1)^2$ potential.  The induced dynamics in the orthogonal direction will be so weak that we can still ignore them.  Basically, the path wants to stay straight, because the change in the potential is too small to justify the gradient of a curvy path.  So, in searching for a tunneling path, we can freeze light degrees of freedom.  In the string theory context, we imagine that the Kahler moduli will be stabilized by something akin to the KKLT mechanism \cite{KKLT} at a much lower scale, so we can just freeze them and study tunneling paths in the 4D field space of the complex structure modulus $z$ and the axio-dilaton $\tau$.

Typically we cannot reduce a problem all the way down to a single field, but even so, we do not need to resort to a purely numerical investigation.  Recently, several authors realized that the global properties of a potential play important roles in tunneling paths\cite{Yan09,AJL,BroDah10}.  In potentials with a run-away direction---for example with decompactifying extra dimensions or a dilaton field---it was shown that tunneling paths tend to take an excursion through those directions.  As demonstrated in appendix~\ref{sec-toy} and with the mirror quintic Calabi-Yau in section~\ref{sec-conifunneling}, numerical methods help to find these global paths.
Once they are found, we can study them analytically and gain deeper insights as in section~\ref{sec-analysis}.

\section{Numerical Conifunneling}
\label{sec-conifunneling}

%{\it
%\begin{enumerate}
%\item Briefly introduce the numerical potential and Kahler metrics(calculations of fluxes and Meiyer functions etc), and provide the exact number of the model we are using, also some Figures.
%\item Relaxation on the numerical data grid.  Mention that it does not stabilize at the Johnson-Larfors path.  ( I suppose Pontus can write this part so he won't feel so left out. )
%\item Parameters of the near conifold behavior of the potential and Kahler metrics.
%(David's note should contain all relevent materials for this.)
%\item Relaxation on the near conifold potential.
%\end{enumerate}
%}

In this section we apply the numerical relaxation method to find domain wall solutions in degenerate vacua---these solutions are excellent approximations to instantons with weakly non-generate vacua. We will focus on the technicalities of the method, and demonstrate that the solutions we find are robust. We postpone discussing the physical interpretation of our solutions to the next section \ref{sec-analysis}, and hence this section can be skipped by those who do not have a strong interest in the numerical analysis.

Nevertheless, we cannot resist providing a brief description of the results here. Our goal is to look for instantons between vacua that reside on separate sheets of the flux potential. In our construction, such vacua are associated with the monodromy transformations around the conifold point. If we take the perspective of the monodromies acting on the fluxes, the instantons describe tunneling between different flux compactifications.

It will turn out that the bounce solution connecting these flux vacua \emph{generically} passes very close to the conifold point. The instanton solution is driven there by the presence of non-trivial kinetic terms despite the seeming lack of an obvious path in the potential. This behavior is natural: flux transitions are associated with the nucleation of branes and the moduli dynamics appear to ``know'' that it is energetically easier to nucleate a brane near the conifold point.

Seeing this behavior numerically requires knowledge of both the Kahler metric and the potential in both the bulk (i.e. far away from the conifold and near the vacua), and near the conifold point in the field space. In sections \ref{sec-MQ} and \ref{sec-VacuumHunt} we have calculated the potential numerically, but the technique becomes computationally prohibitive near the conifold point. Fortunately, the near conifold behavior can be treated analytically as detailed in appendix \ref{sec-NearConifoldPotential}. The bulk and near-conifold constructions could then be glued to obtain a full flux potential that describes both the bulk and the near conifold region. However, relaxation is rather easier to implement when we split the problem into two parts: relaxation in the bulk and relaxation in near the conifold point.

\subsection{Equations of motion and set-up}

Our goal is to find domain wall solutions using the relaxation method for this system given an action in the Einstein frame
\begin{equation}
{\cal L} = -\left(K_{z\bar{z}}\partial_{\mu}z\partial^{\mu}\bar{z}+ K_{\tau\bar{\tau}}\partial_{\mu}\tau\partial^{\mu}\bar{\tau} \right)+V(z,\tau)  \label{eqn:relaxaction}
\end{equation}
where $z$ is the complex structure modulus and $\tau$ is the axio-dilaton field. We assume that the Kahler moduli fields $\rho$ are frozen by some mechanism (perturbative or non-perturbative) and that they do not contribute to the dynamics, so we will simply treat them as constants in this section. Hence, we have a system with four real fields. It is convenient to choose the following parameterization
\begin{equation}
re^{i\theta}\equiv z-1~,~\tau \equiv u+iv
\end{equation}
where $\phi\equiv\{r,\theta,u,v\}$ are all real dynamical fields which we have collected into a vector $\phi$ for notational simplicity. In the same vein, we also define
\begin{equation}
K_{z\bar{z}} \equiv \frac{1}{2}f(r,\theta)
\end{equation}
and remind the reader that
\begin{equation}
K_{\tau\bar{\tau}}=\frac{1}{2}\frac{1}{4v^2}.
\end{equation}

We are looking for domain wall solutions which are effectively 1+1 dimensional, hence without any loss of generalities we can choose $(x,t)$ as coordinates. Ignoring gravity, the equations of motion are 
\begin{eqnarray} 
f(\ddot{r}-r'')+\frac{1}{2}(\dot{r}^2-r'{}^2)-\frac{1}{2}\partial_r(r^2 f)(\dot{\theta}^2-\theta'{}^2)+\partial_{\theta}f(\dot{\theta}\dot{r}-\theta'r')+\partial_r V&=&0 \nn \\
fr^2(\ddot{\theta}-\theta'')+\frac{1}{2}r^2\partial_{\theta}f(\dot{\theta}^2-\theta'{}^2)-\frac{1}{2}\partial_{\theta}f(\dot{r}^2-r'{}^2)+\partial_r(fr^2)(\dot{r}\dot{\theta}-r'\theta')+\partial_{\theta}V&=&0 \nn \\
-\frac{1}{2v^3}(\dot{v}\dot{u}-v'u')+\frac{1}{4v^2}(\ddot{u}-u'')+\partial_u V&=&0 \nn \\
-\frac{1}{4v^3}(\dot{v}^2-v'{}^2)+\frac{1}{4v^3}(\dot{u}^2-u'{}^2)+\frac{1}{4v^2}(\ddot{v}-v'')+\partial_vV&=&0 \label{eqn:EOMrelax}
\end{eqnarray}
with dots and primes denoting space derivatives.

Domain wall solutions, $\phi_{*}(x)$, are static solutions to the set of differential equations (\ref{eqn:EOMrelax}) with boundary conditions
\begin{equation}
\phi_*(x\rightarrow-\infty)=\phi_1~,~\phi_*(x\rightarrow\infty)=\phi_2
\end{equation}
where $\phi_1$ and $\phi_2$ are the locations of the minima. 

We solve (\ref{eqn:EOMrelax}) on a finite 1-dimensional grid, with a domain $\{x_{min},x_{max}\}$ where the domain's size is much larger than $1/m$, $m$ being the characteristic mass of the domain wall\footnote{This is not known in advance of course, but one can make a good guess at a value just after a few iterations of our prescription.}. In practice, we choose the size of the domain to be a balance between accuracy and computational efficiency. Once a solution is found, we vary the size of the domain to ensure that the results are robust.

We insert a test solution at some initial time $t_0$, $\phi_0(x,t_0) = \phi_*(x) + \Delta \phi(x,t_0)$, where the difference $\Delta \phi$ is preferably, but not necessarily, small compared to $\phi_*(x)$. In addition to possessing the correct boundary conditions, we fix the first derivatives at the boundaries to be identically zero at all time
\begin{equation}
\dot{\phi}(x_{max},t)=\dot{\phi}(x_{min},t)=0.
\end{equation}
Given these boundary conditions, we then guess several initial profiles for $r_0(x,t_0)$ and $\theta_0(x,t_0)$ that interpolate between the two vacuum positions. Using these test profiles we find the corresponding \emph{minimum} points of a given $r$ and $\theta$ for $u_0(x,t_0)$ and $v_0(x,t_0)$, which we can find by solving for\footnote{This choice for $u_0$ and $v_0$ is motivated by the fact that we expect that in the actual domain wall solution $u$ and $v$ do not deviate radically from this global minimum solution.  However, they do deviate in general, which we can easily see by their equations of motion (\ref{eqn:EOMrelax}): the spatial derivatives must be supported by a non-zero derivative of the potential.}
\begin{equation}
\frac{\partial V}{\partial u} (u_0,v_0)=0~,~\frac{\partial V}{\partial v} (u_0,v_0)  =0. \label{eqn:tauglobalmin}
\end{equation}

The total energy functional of the system is the integral of the Hamiltonian over the domain\footnote{We have suppressed two spatial dimensions -- the energy functional is formally infinite if integrated over these suppressed dimensions.}
\begin{equation}
E[\phi(x)] = \int_{x_{\min}}^{x_{\max}} dx \left[\frac{1}{2}f(\dot{r}^2-r'{}^2+r^2\dot{\theta}^2-r^2\theta'{}^2)+\frac{1}{8v^2}(\dot{u}^2-u'{}^2+\dot{v}^2-v'{}^2)+V(r,\theta,u,v)\right].
\end{equation}
Bogomolny's bound tells us that the true domain wall solution, if it exists, minimizes the total energy of the system
\begin{equation}
E[\phi(x)] \geq E[\phi_*(x)].
\end{equation}

Hence any deviation from the true solution means that there is additional energy in the system, which manifests itself as scalar radiation as the fields seek to relax to their true minimum energy configuration. In a perfect world, the radiation propagates to spatial infinity, never to be seen again. However, our fixed boundary conditions act as a rigid barrier at finite distance, and hence radiation will bounce back from this barrier and remain in the system. To remove this radiation, we introduce friction terms into the equations of motion 
\begin{equation}
\ddot{\phi}+\phi'' +...=0 \rightarrow \ddot{\phi}+\phi'' +...+\lambda(t) \dot{\phi}=0,
\end{equation}
allowing the fields to relax into the true minimum energy configuration (i.e. a domain wall). Note that we allow the friction term to be a function of time; we will say a bit more about how we engineer the friction term later. In principle, the friction term turns itself off once the static solution has been found. We check for the robustness of our solution by manually turning off the friction term.

The test solutions themselves are not very important. In practice, we find that a well chosen initial profile may speed up the computation marginally, but most guesses find identical static solutions in the end. More insidious however, is the possibility that there exist multiple static solutions which are \emph{not} the minimum energy solution $\phi_*$. To test for that, we choose several different initial profiles with different initial total energy and check that they all relax to the correct solution\footnote{More amusingly, we check for the robustness of our relaxation code by constructing potentials where there exist more than one static solution, i.e. one of them has a higher total energy so it is the subdominant path. We find that indeed different initial profiles will relax to different static solution.}.

In addition, there may be \emph{no} solution. The simpler case of this possibility is that the total energy becomes negative after some time. Since our potential is bounded from below and positive, $V>0$, this never happens. More difficult to detect is the possibility is that the field approaches, but never quite converges to, a static configuration. In this case, the system never completely relaxes and long code run times may be mistaken for a true solution. We can check for this by taking the time derivative of the total energy, but in practice we never encounter such a situation.

In the following sections, we separate the field space into two regimes: far away from the conifold point $r>0.1$ which we call the \emph{bulk} and the near conifold regime where $r<0.1$. The cut-off at $r=0.1$ is arbitrary, motivated by the fact that we lack accurate numerical data for the potential below this point. Near the conifold point, the calculation of the potentials is tractable analytically as demonstrated in appendix \ref{sec-NearConifoldPotential}. Note that the numerical bulk potential does not include the effects of warping, but since the data is only really accurate up to $r>0.1$ and strong warping is not expected to be important until $r\ll 1$, this is not a problem.

In summary, we find that in the bulk relaxation phase, the field profile for $r$ relaxes towards the conifold point rapidly, reaching $r<0.1$ where we do not possess numerical data for the potential. To investigate the near conifold behavior, we use our analytic potentials and find that the field profiles do indeed continue to be driven to near $r=0$, but then making a turn-around back into the bulk. While deep inside the near conifold regime, we find that $\theta$ makes a rapid transition across the sheets, hence tunneling across a monodromy transition. We dub this behavior, where the fields are driven towards the conifold point in order to transition into a new flux configuration \emph{conifunneling}.

\subsection{Relaxation in the bulk} \label{sect:relaxbulk}

We first look for domain wall solutions between two SUSY vacua related by a conifold monodromy in the bulk. For concreteness, we choose our vacua in the mirror quintic compactification (model 1) from table \ref{SUSYTable}:
\begin{eqnarray}
F_1=(3,-6,-9,-1)&\rightarrow &F_2=(2,-6,-9,-1) \\
H_1=(-1,0,-7,0)&\rightarrow &H_2=(-1,0,-7,0).
\end{eqnarray}
The potentials at each corresponding sheet in $z$-space are generated using numerically computed Meijer functions as described in section \ref{sec-MQ}. The vacuum positions are essentially those given in table \ref{SUSYTable}, but to greater precision, they are
\begin{eqnarray}
z_1& =& -0.4628+0.1237i \\
z_2&=& 0.2286+0.3631i.
\end{eqnarray}
We use coordinates $(r,\theta)$ around the conifold point to unwrap the potential and stitch the data together across the sheets. From this, we generate the effective super- and Kahler potentials. The vacua positions in these coordinates are then, with $\theta=0$ being the branch cut,
\begin{eqnarray} \label{eqn:monodromyflux} \label{eqn:vacuapos}
\phi_1&=&(r_1=0.55,\theta_1 =-2.92,u_1=-3.41,v_1=4.22)\\
\phi_2&=&(r_2=0.85,\theta_2=3.58,u_2=-3.37,v_2=4.17).
\end{eqnarray} 

\begin{figure*}[htp]
\centering
\subfigure[] 
{\includegraphics[width=0.4\textwidth]{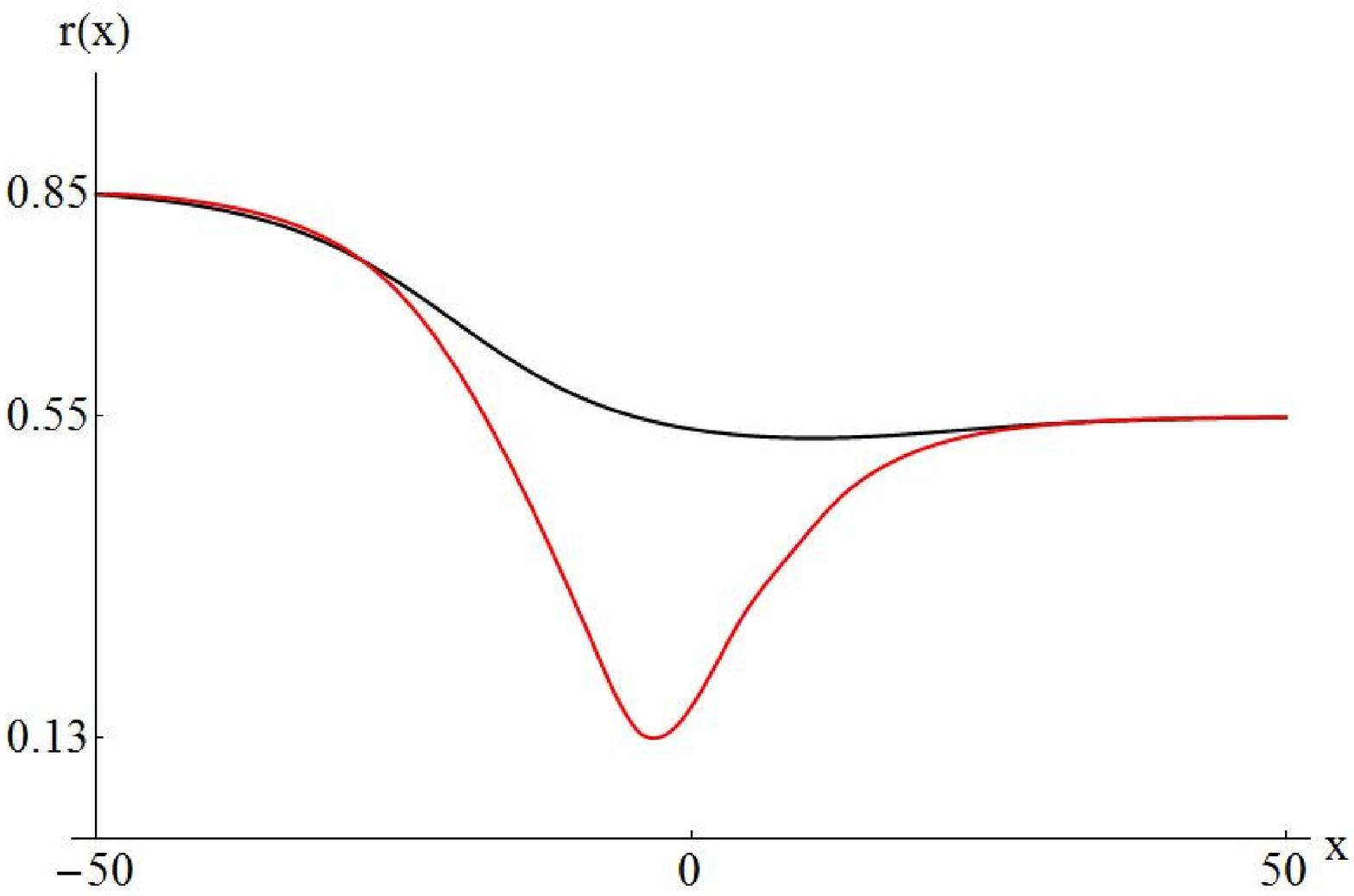}} \label{fig:rbulk}
\subfigure[] 
{\includegraphics[width=0.4\textwidth]{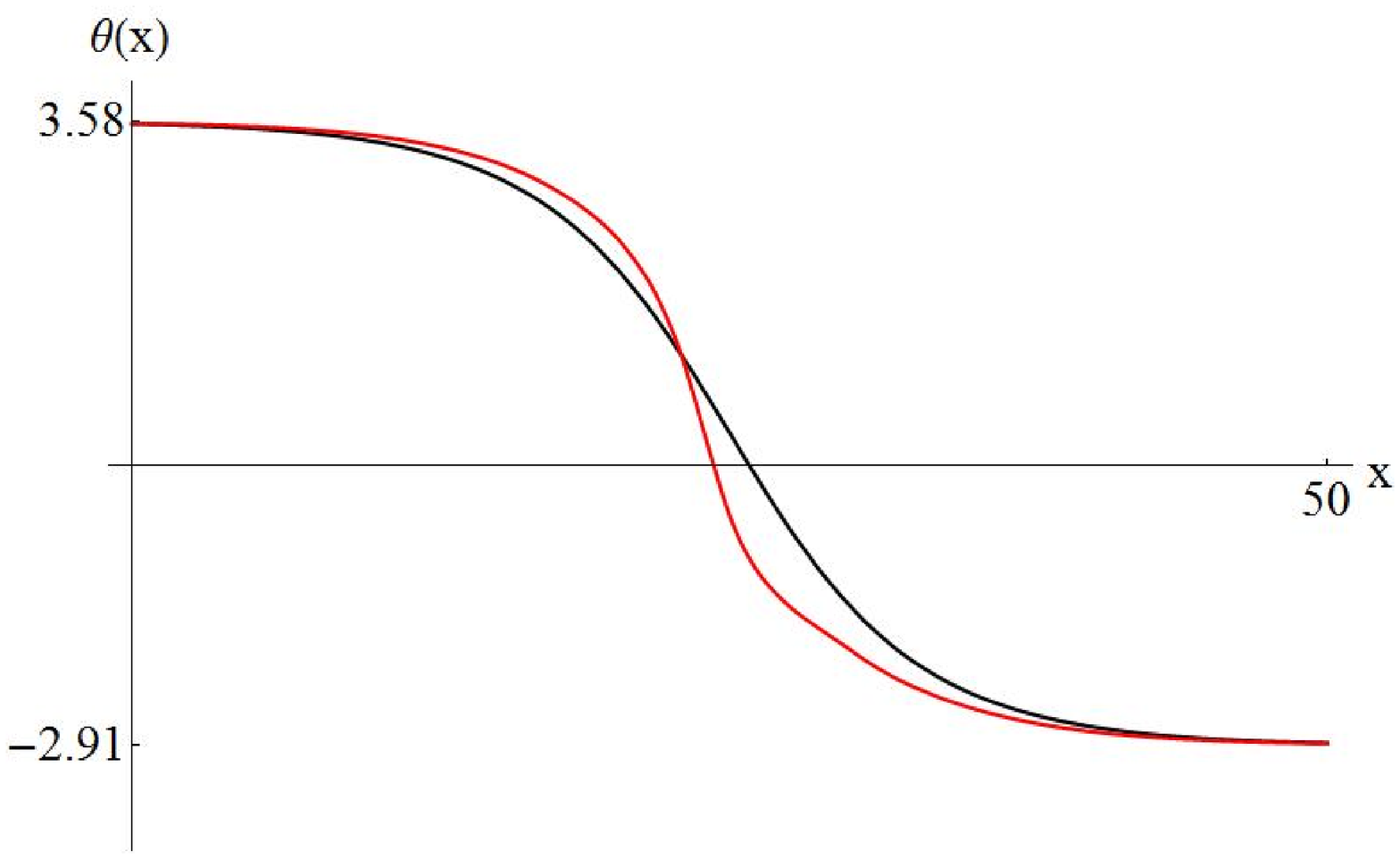}} \label{fig:thetabulk}
\caption{Initial (red) and final (blue) profiles for the complex structure field $r$ (left) and $\theta$ (right) in the bulk relaxation phase. We stopped the simulation at the final configuration, $r<0.1$ even though the fields are still not static (indeed they are highly dynamical) since  we do not have numerical data for the potentials. Nevertheless, it is clear that the path between the two vacua is rapidly relaxing to the conifold point. We will replace the numerical potential with a near-conifold analytical potential in the next section.  }
\label{fig-bulkprofile}
\end{figure*}

The vacuum positions for $\tau=u+iv$ are found using conditions (\ref{eqn:tauglobalmin}).  We use the following test profile 
\begin{eqnarray} 
\theta(x,t_0)&=&\frac{2(\theta_1-\theta_2)}{\pi}\tan^{-1}\left(e^{x/\delta}\right)+\theta_2 \label{eqn:test2} \\
r(x,t_0)& =& \frac{2(r_{min}-r_1)}{\pi}\tan^{-1}\left(e^{(x-x_2)/\Delta_1}\right)+\frac{2(r_1-r_{min})}{\pi}\tan^{-1}\left(e^{(x-x_2)/\Delta_2}\right)+r_1 \nn \\ \label{eqn:test1}
\end{eqnarray}
where $\Delta,\delta$ are parameters which control the initial test thickness of the walls, while $r_{min}$ sets the datum for the turn around point (see figure \ref{fig-bulkprofile}).

We then run relaxation simulations, using uniform and constant friction for all 4 dynamical fields, varying both the initial test profiles and the magnitude of the friction (ranging from  $\lambda=0.1$ to $\lambda=10$) to ensure that our general conclusions are robust.

\begin{figure*}[htp]
\centering
\subfigure[] 
{\includegraphics[width=6 cm]{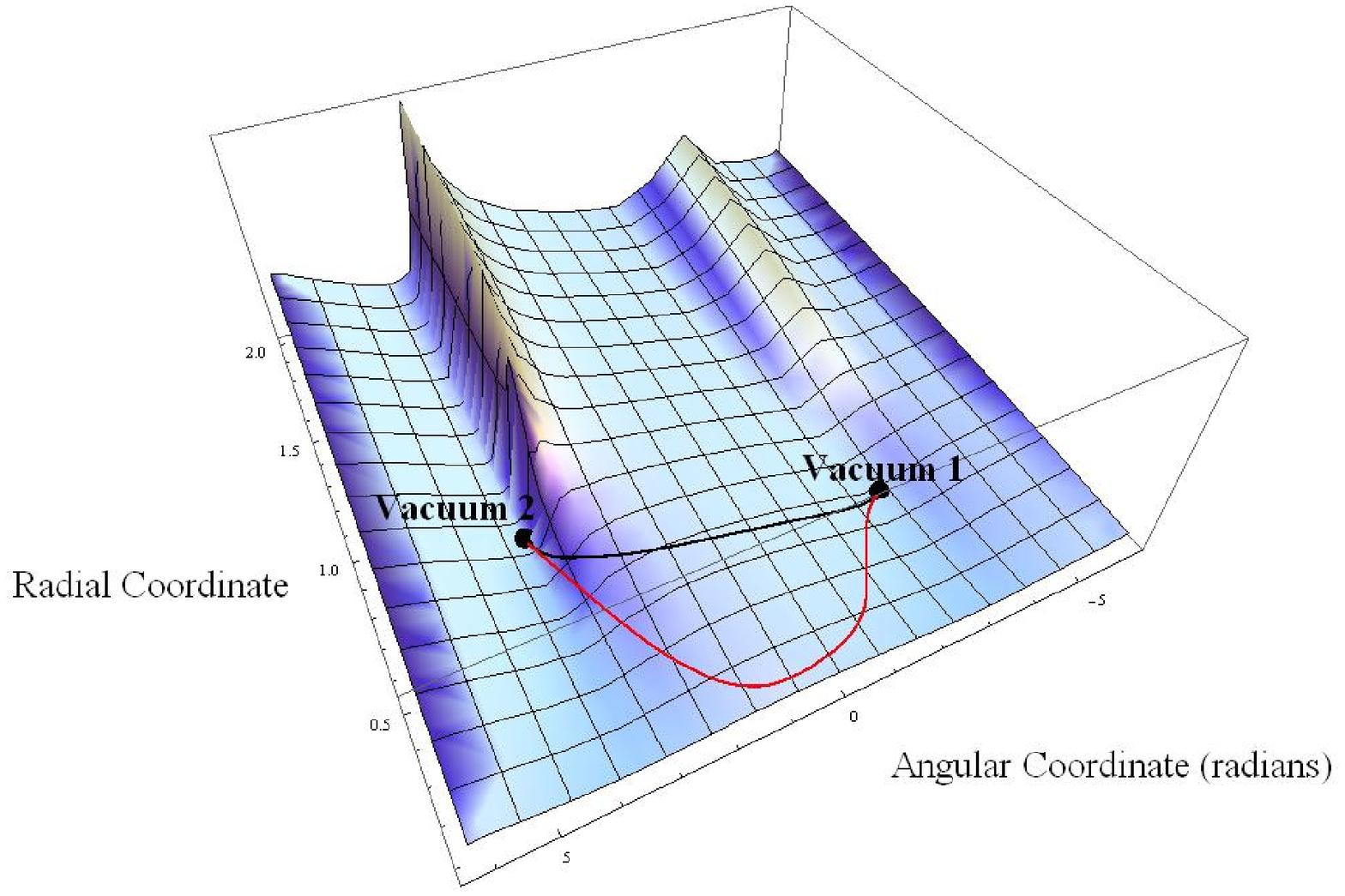}} \label{fig:Vbulk}
\subfigure[] 
{\includegraphics[width=6 cm]{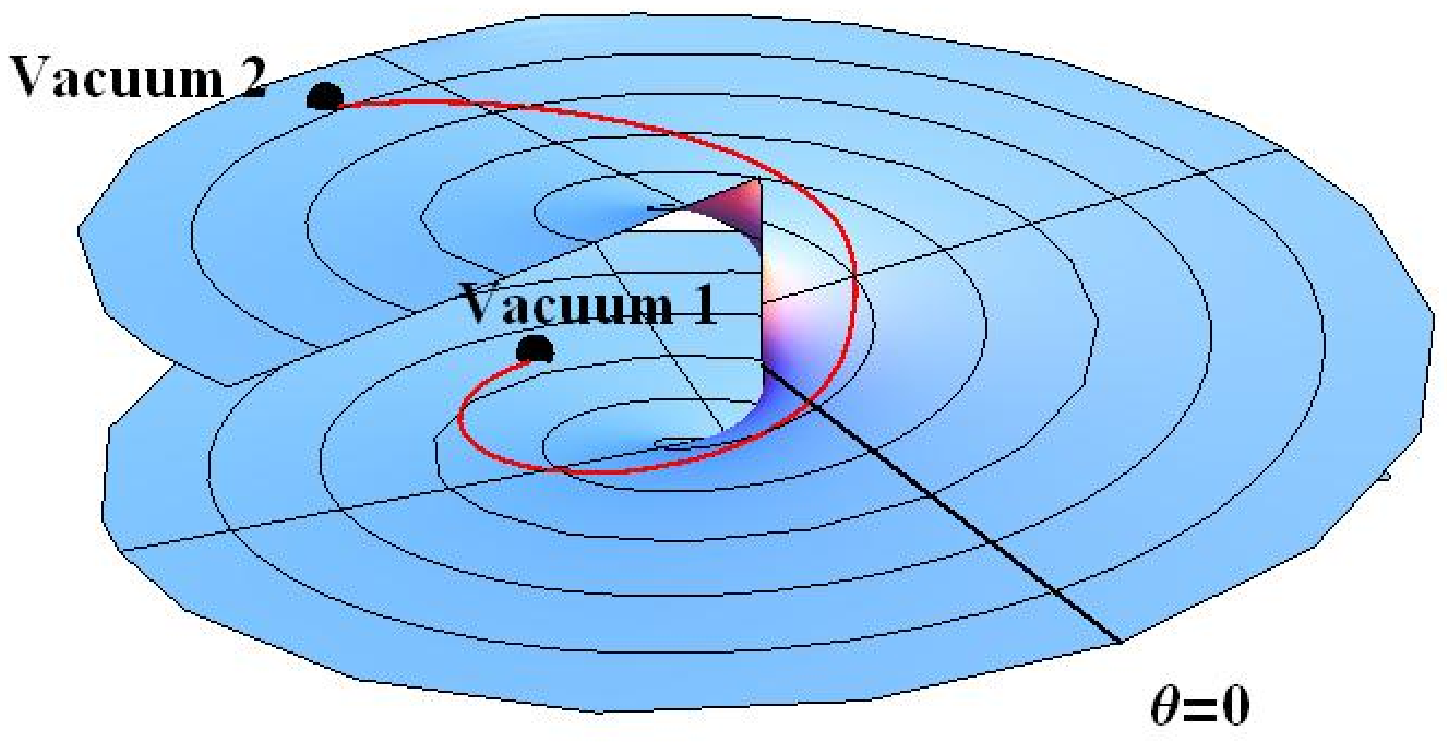}} \label{fig:funnelbulk}
\caption{Figures showing the path of the complex structure $(r,\theta)$. On the left, we superimposed the initial (black) and final (red) profiles of the bulk relaxation phase over the reduced potential $V(r,\theta, u_{\min}(r,\theta),v_{\min}(r,\theta))$, where $u_{\min}$ and $v_{\min}$ are global minima for $\tau$ found using (\protect\ref{eqn:tauglobalmin}). On the right, we suppress the structure of the potential, but instead plot the final path in polar coordinates. The two sheets are joined at $\theta=0$ with $r=0$ being the conifold point. It is clear from this picture that the path traverses close to the conifold point as it wanders down the ``funnel''.}
\label{fig-bulkfunnel}
\end{figure*}

Generically, the field profile for $r$ rapidly relaxes to near the conifold point $r<0.1$ where we do not possess good numerical data for the bulk flux potential (see figure \ref{fig-bulkfunnel}), hence the simulation breaks down at this point. We emphasize that this behavior is driven by the presence of the non-trivial Kahler factor in the kinetic terms of the equations of motion, in particular $K_{z\bar{z}}$. We will discuss this further in the section \ref{sec-shortest}. 
%This behavior is driven by the presence of the Kahler factor in the kinetic terms of the equations of motion. Energetically one can understand it by noting that the Kahler term is a roughly a decaying exponential in the direction $r\rightarrow 0$, and hence it is more profitable for the field to ``lose energy'' in this particular direction [CHECK].  In fact, this behavior dominates over the shape of the potential $V$ -- we will discuss this point in the next section.
	
\subsection{Results from relaxation in the vicinity of the conifold point}

In order to investigate the behavior of the solutions near the conifold point, we use the analytical approximation described in appendix \ref{sec-NearConifoldPotential}:
\begin{equation}\label{FluxPotentialNearConifoldtex}
V_{nc} = {1 \over 16 \tau_I \rho_I^3} \left( \left({1\over 2\pi}\log {\Lambda_0^6 \over |0.35r|^2} + K_1 + {C_1 \over |0.35r|^{4/3}} \right)^{-1} 
\left|{F_\Asc \over 2\pi i} \log0.35r + A_1 - \tau B_1 \right|^2 
 + |A_2+\bar{\tau}B_2|^2 \right).
\end{equation}
Note that this is simply equation (\ref{FluxPotentialNearConifold}), with the rescaling $|\xi|=0.35r$ for consistency with the notation we are using in this section. The parameters of this near conifold potential are derived assuming that the two vacua are associated by the monodromy described by (\ref{eqn:monodromyflux}). See appendix \ref{sec-NumericalParameterData} for the values of the parameters derived from our numerical data for the mirror quintic with the fluxes chosen as above. 

\begin{figure*}[htp]
\centering
\subfigure[] 
{\includegraphics[width=6 cm]{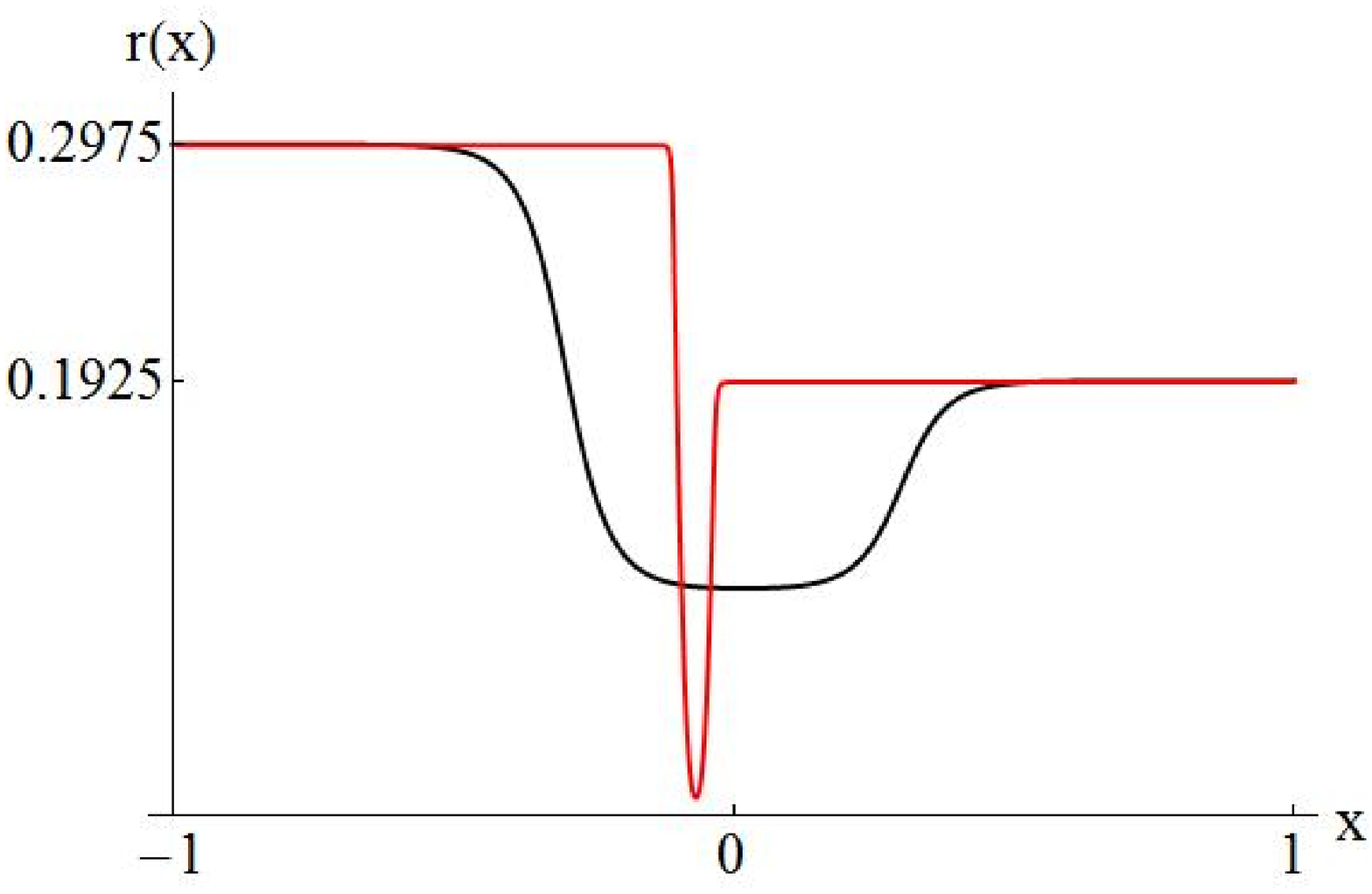}}\label{fig:rnc}
\subfigure[] 
{\includegraphics[width=6 cm]{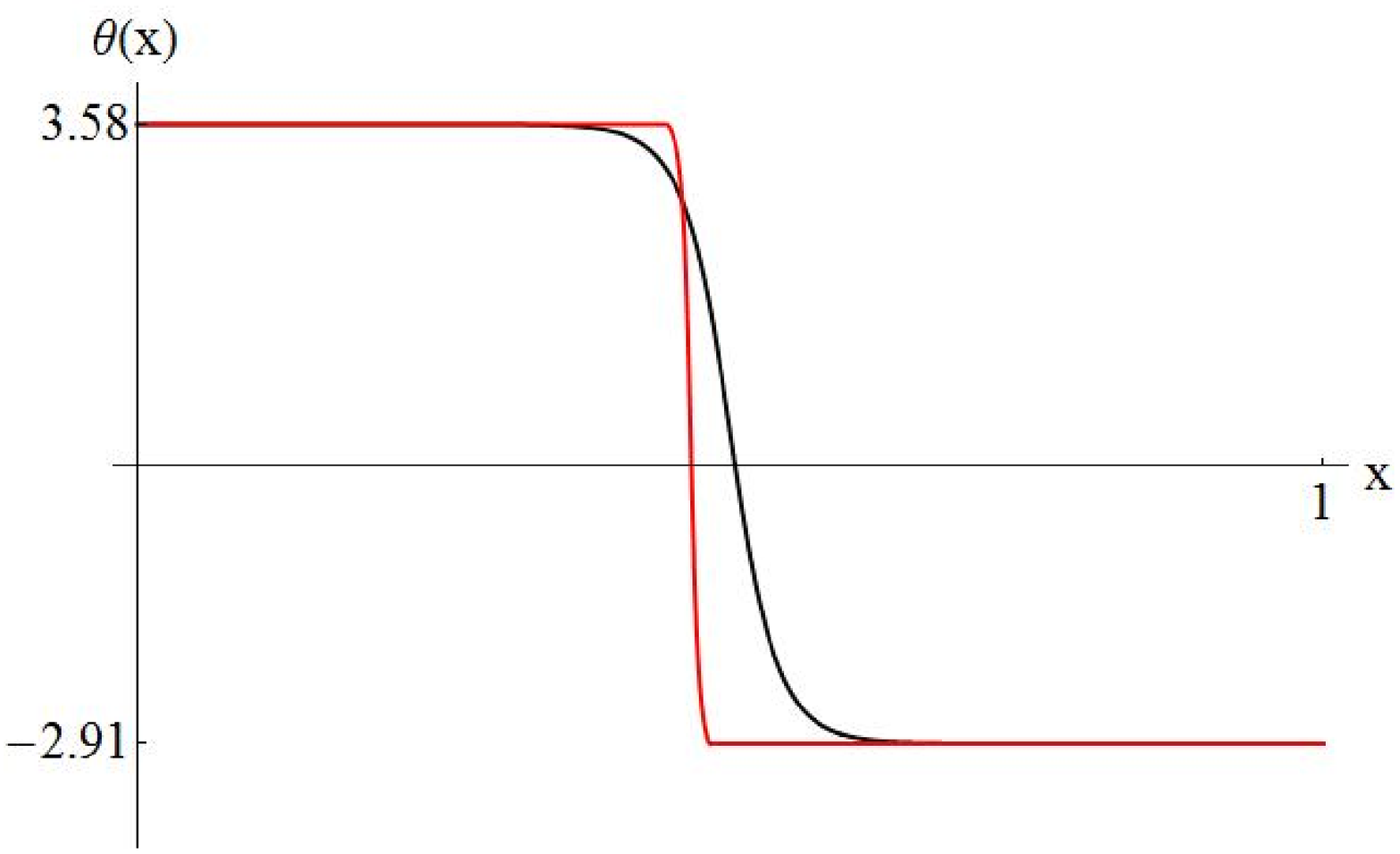}}\label{fig:thetanc}
\caption{Initial (blue) and final (red) profiles for the complex structure field $r$ (left) and $\theta$ (right) in the near-conifold relaxation phase. In the final configuration, a static solution is achieved and hence is a true domain wall solution. The complex structure modulus funnels very close to the conifold point, the proximity depending on how strong the warping is. In terms of the $r$ and $\theta$ fields, it is clear that a very sharp $\theta$ transition occurs when the field is near the conifold point $r\ll 1$. This indicates that there are three clear phases in the entire process---a shrinking of the 3-cycle associated with the formation of the conifold, a monodromy transition as $\theta$ tunnels into the next sheet, and then a return of the 3-cycle to near its original size. We will discuss this further in section \protect\ref{sec-analysis}. } 
\label{fig-ncrthetaprofiles}
\end{figure*}

\begin{figure*}[htp]
\centering
\subfigure[] 
{\includegraphics[width=.4\textwidth]{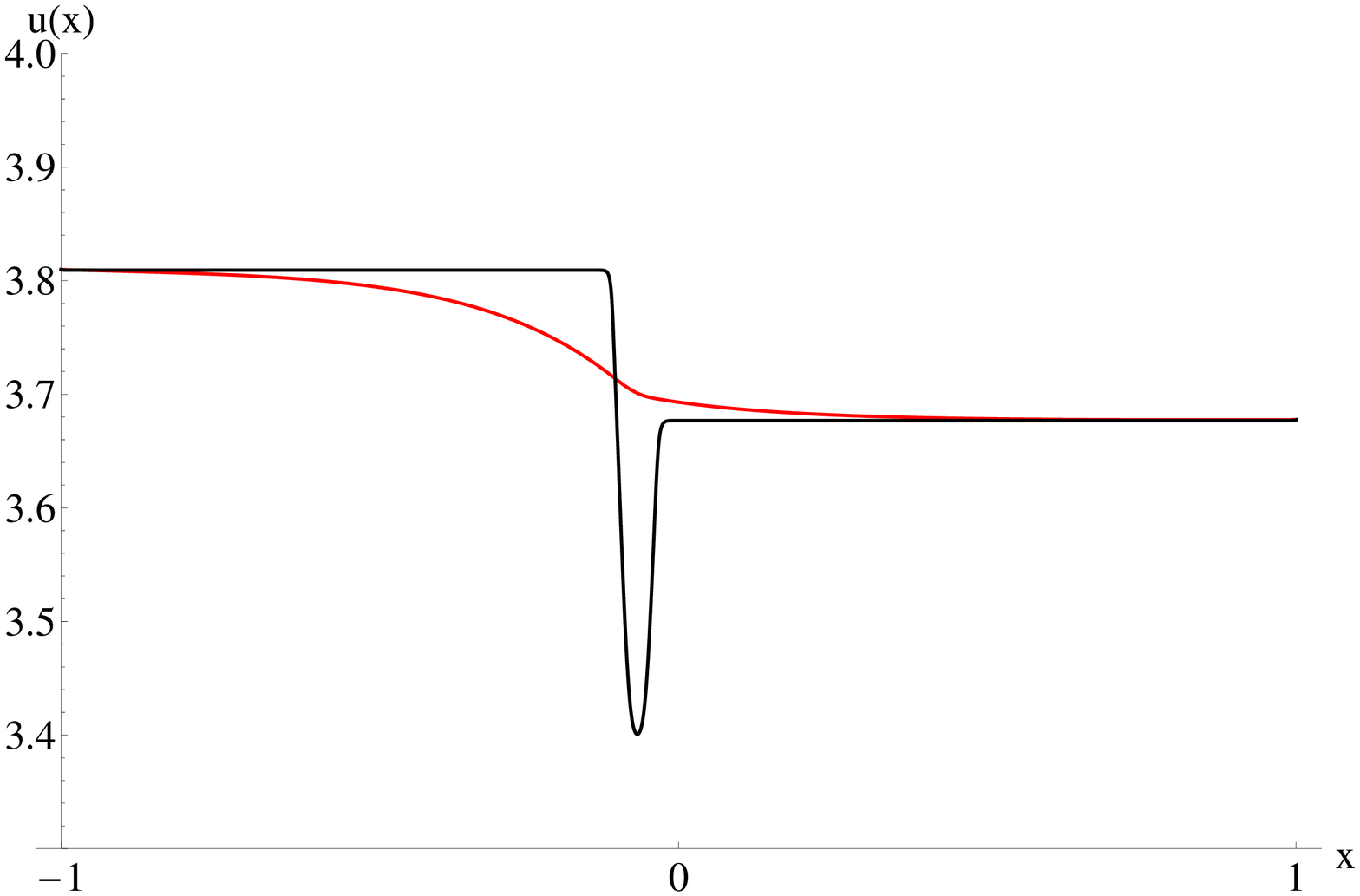}} %\label{fig:unc}
\subfigure[] 
{\includegraphics[width=.4\textwidth]{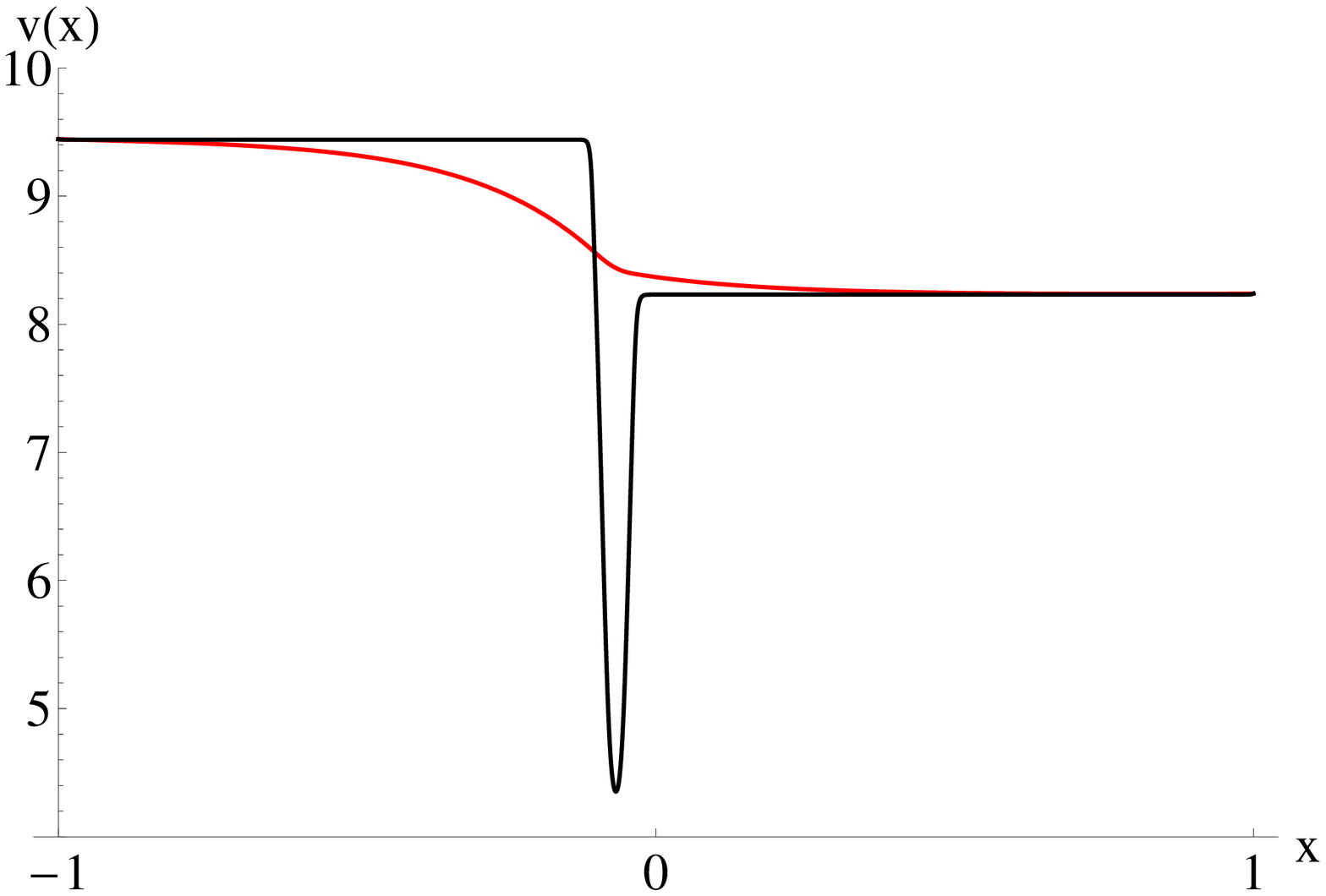}} %\label{fig:vvnc}
\caption{The \emph{final} (red) profiles for the axio-dilaton field $u$ (left) and $v$ (right), and the global minima (black) $u_{min}$ (left)  and $v_{min}$ (right) in the near-conifold relaxation phase. In the final configuration, a static solution is achieved and hence is a true domain wall solution. It is clear from the equations of motion (\protect\ref{eqn:EOMrelax}) that $u_{min}$ and $v_{min}$ are not static solutions. The actual final static domain wall solutions are mildly localized. The global minimum solution exhibits a sharp feature as expected from the highly localized nature of the complex structure $z$ domain wall solution.}
%\caption{The \emph{final} (black) profiles for the axio-dilaton field $u$ (left) and $v$ (right), and the global minima (red) $u_{min}$ (left)  and $v_{min}$ (right) in the near-conifold relaxation phase. In the final configuration, a static solution is achieved and hence is a true domain wall solution. It is clear from the equations of motion (\protect\ref{eqn:EOMrelax}) that $u_{min}$ and $v_{min}$ are not static solutions. The actual final static domain wall solutions exhibit a sharp feature. In particular, since the string coupling $g_s = 1/v$, we see that the path generically heads towards \emph{stronger} couplings, though we are still in regimes where things are still under perturbative control.}
\label{fig-ncuvprofiles}
\end{figure*}

This approximation becomes almost exact near the conifold $r<0.1$, but breaks down in the bulk. The key feature that is lost is the existence of the original vacua. To stabilize the vacuum positions, we drill Gaussian SUSY vacua into the potential 
\begin{equation}
\tilde{V} = V_{nc}+V_1+V_2
\end{equation}
with
\begin{equation}
V_i = -V_{nc}(r_i,\theta_i,u_i,v_i)e^{[-(\theta-\theta_i)^2-(r-r_i)^2]/\sigma^2}
\end{equation}
where $\sigma$ is the width of the Gaussian holes. At these vacuum positions, $V_i=0$. The positions of the holes are matched to the actual vacuum positions for their respective flux configurations, as given in (\ref{eqn:vacuapos}). Note that we do not drill holes in the $\tau$ directions; we simply solve for the minima of $\tau$ via equation (\ref{eqn:tauglobalmin}) as in section \ref{sect:relaxbulk}. Since the behavior of the domain wall will be dominated by the near conifold regime, we do not try to reproduce the shape of the potentials beyond this modification. As long as the solution conifunnels towards $r\rightarrow 0$ when it is at $r>0.1$ we are satisfied with the overall bulk behavior. 

Nevertheless, there remain two subtleties involved in choosing the exact coefficient for the strong warping factor $C_1$. First, in principle it may depend on $\tau$, although such a dependence will not greatly effect the behavior of $\tau$. Second, the exact numerical value of this coefficient is treated as a free parameter related to the overall volume of the Calabi-Yau manifold. Consistency requires that the parameter be chosen small enough so as not to have any effects on the bulk of the moduli space. For the purpose of our numerical simulation, we choose $C_1$ such that the warping term is subdominant when $r\approx r_{1,2}$ i.e.
\begin{equation}
C_1 \ll \left({1\over 2\pi}\log {\Lambda_0^6 \over |0.35r|^2} + K_1\right) |0.35r|^{4/3}~\mathrm{at}~r=r_1,r_2.
\end{equation}

Again, we use the test solutions (\ref{eqn:test2}) and (\ref{eqn:test1}), varying the test parameters to ensure robustness of our conclusions (figures \ref{fig-ncrthetaprofiles} and \ref{fig-ncuvprofiles}). However, due to the strong warping term $r^{-4/3}$ in the flux potential, instead of inserting constant friction terms for all our field equations, we use instead an exponentially damping friction
\begin{equation}
\lambda_{\phi}(t) = \lambda_{\phi}(t_0)e^{-\alpha(t-t_0)}
\end{equation}
where $\alpha$ is some parameter which governs how rapidly friction is turned off\footnote{We also imposed a hard cut-off of the friction when we check for stability after a solution is obtained.}. 

\begin{figure*}[htp]
\centering
\subfigure[] 
{\includegraphics[width=6 cm]{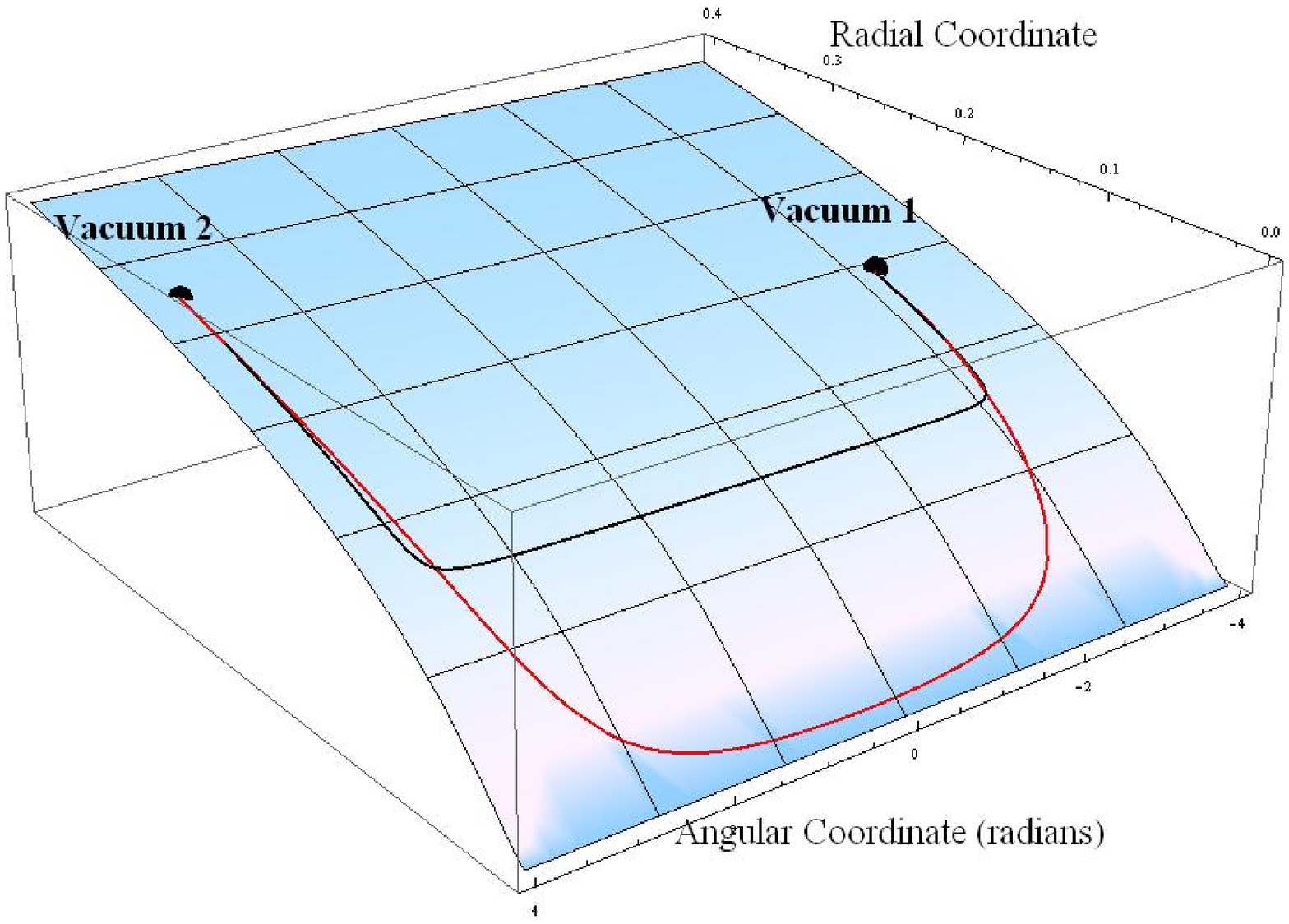}}\label{fig:Vnc}
\subfigure[]
{\includegraphics[width=6 cm]{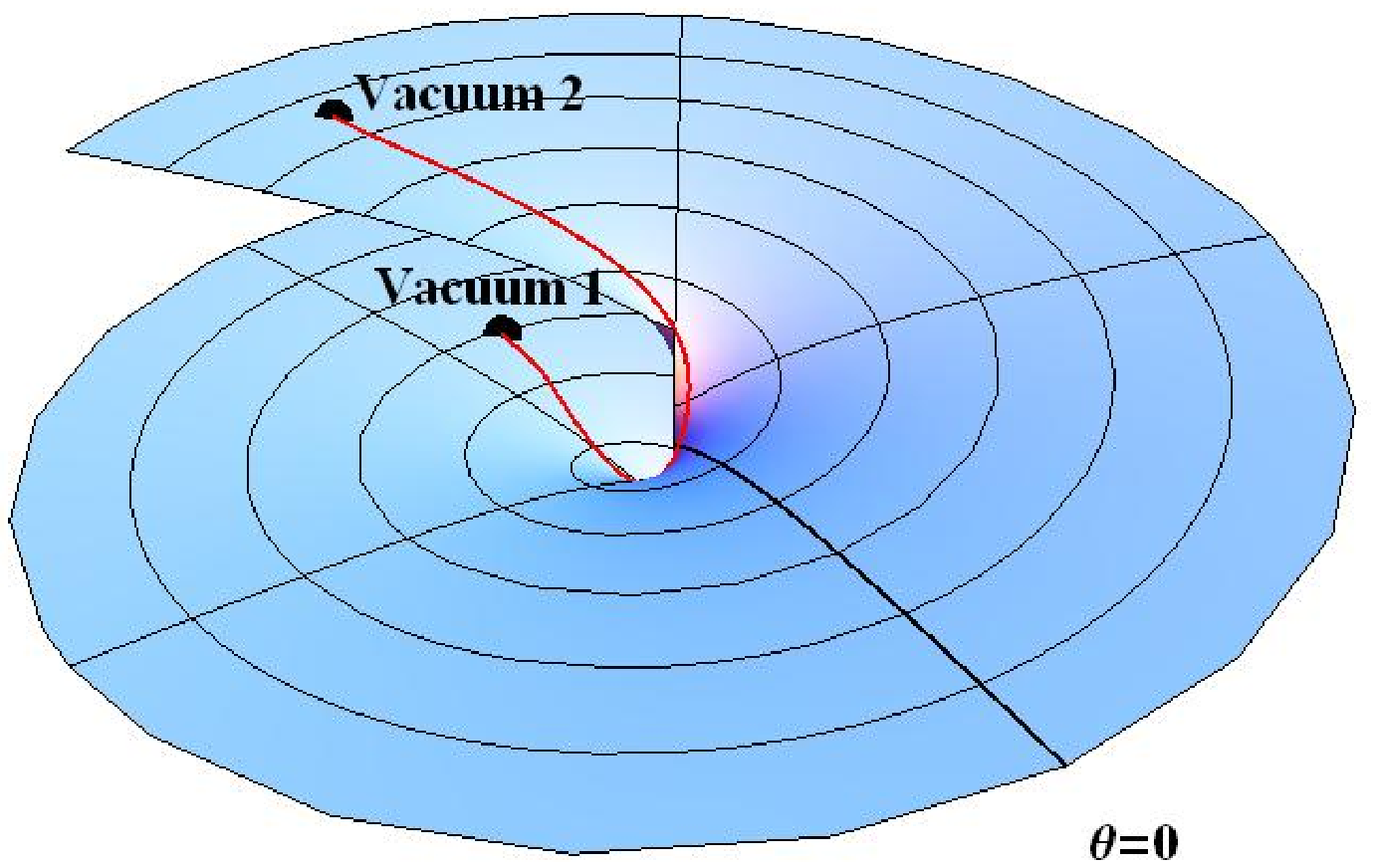}} \label{fig:funnelnc}
\caption{Figures showing the path of the complex structure $(r,\theta)$. On the left, we superimposed the initial (black) and final (red) profiles of the bulk relaxation phase over the ``reduced'' analytic near-conifold potential $V_{nc}(r,\theta, u_{\min}(r,\theta),v_{\min}(r,\theta))$, where $u_{\min}$ and $v_{\min}$ are global minima for $\tau$ found using (\protect\ref{eqn:tauglobalmin}). On the right, we plot the final path in polar coordinates, suppressing the gaussian vacuum holes but keeping the structure of the potential near the conifold visible -- the strong warping term $r^{-4/3}$ suppresses the potential deep inside the conifold point, resulting in a potential that looks like a true ``funnel''. The two sheets are joined at $\theta=0$ with $r=0$ being the conifold point. The final static path falls deep into the funnel but reemerges on the other side of the monodromy---the conifold funnels the path across the monodromy, hence our moniker \emph{conifunneling}.}
\label{fig-ncfunnel}
\end{figure*}

The static solutions are shown in figure \ref{fig:funnelnc}. The solution relaxes towards the conifold point as we have seen in the previous section using the bulk potential. However, instead of falling into an abyss, the domain wall solution passes very close to the conifold point, and then turns back up into the bulk. In other words, a stable static domain wall solution exists between two vacua related by a monodromy transformation. Moreover, the domain wall passes very close to the conifold point---the exact proximity being determined by the coefficient in front of  the strong warping term $r^{-4/3}$ in the flux potential. The smaller the coefficient, the later the turn-around occurs\footnote{We note that although both $\log r$ and $r^{-4/3}$ blow up as $r\rightarrow 0$, the rate at which this blow up occurs is crucial in determining whether the domain wall will turn around sufficiently quickly (see section \ref{sec-shortest}).}.

%Using the potential above with and without the corrections from strong warping, we investigated the existence of tunneling paths passing near the vicinity of the conifold point in the Calabi-Yau complex moduli space. We modeled the vacua arising from the flux compactifications described above by smoothly deforming our potential a reasonably far distance away from the conifold point. In the polar-coordinates centered on the conifold point, the angular separation of the vacua typical of flux compactifications are a little less than $2\pi$.

%The simulations matched our expectations from the shooting argument, namely, for vacua separated by an angular distance greater than $\pi$, the uncorrected near-conifold potential was incapable of exhibiting a real path for tunneling. However, the corrected potential does indeed allow relaxation to converge on a path. These scenarios are depicted in figures XXXYYYZZZ

\section{The Physics of Conifunneling}
\label{sec-analysis}

Despite trying to find a tunneling path through non-singular parts of the Calabi-Yau moduli space, numerical relaxation drove our solutions into the vicinity of the conifold. These instantons represent conifunneling. Relaxation was only able to succeed due to the crucial effects of strong warping, analyzed in \cite{DouglasSUSYBreaking, DouglasWarped, DouglasKinetic}, and described both in general and for the conifold in appendices \ref{Reduction} and \ref{sec-ConifoldWarping}, respectively. The usual simplification of assuming away such effects is a problematic strategy when one is interested in studying the dynamics of fields in the string theory landscape. In this section we draw some lessons for dealing with more general landscape tunneling problems.

\subsection{Geometric interpretation}

\begin{figure}
\begin{center}
\includegraphics[width=.9\textwidth]{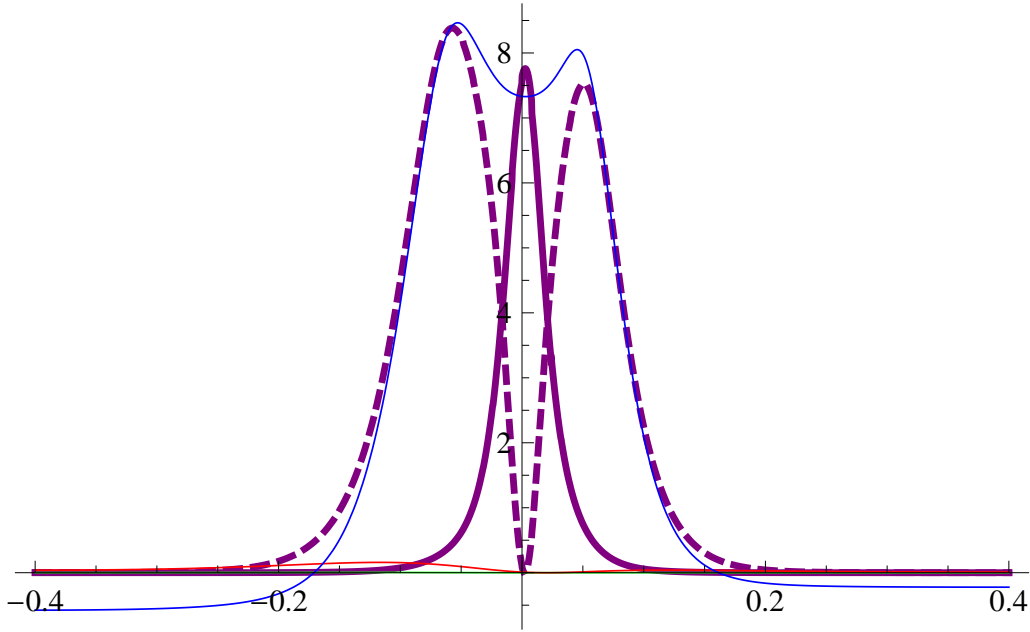}
\end{center}
\caption{The action integrand (Lagrangian) broken up into contributions from five terms.  The thin-blue line is the potential term, the dashed-purple line is the kinetic term in $r$, the solid-purple line is the kinetic term in $\theta$.  Kinetic terms in the two components of $\tau$ are colored red and green, and barely contribute. Note that the potential dips below zero for the two vacua, this is an artifact of our procedure for drilling these vacua in the near-conifold potential.}
\label{fig-AB}
\end{figure}

Figure \ref{fig-AB} shows the value of the tension integrand in terms of five separate terms: kinetic terms in each field and the flux potential term.  We can see that the kinetic terms in $\tau$ are much smaller, which means the dynamics are mostly in the complex structure moduli, $z$\footnote{This means that we would have got similar results if we had reduced the problem from 4D to 2D and only focused on $z$.  But it is not obvious that $\tau$ would essentially act as a spectator field, so we included it in the analysis for completeness.}. In particular, the dynamics separates into three distinct parts: radial changes toward and away from the conifold point occur in the beginning and the end of the transition, while angular changes around the conifold occur in the middle.

This situation is very similar to \cite{Yan09,AJL,BroDah10}.  The two vacua are connected by a monodromy transformation, namely, a change of $\Delta\theta\sim2\pi$.  We just need to determine the most economical way to perform this transformation.  Namely, minimizing the tension with 3 terms,
\begin{equation}
\sigma=\sigma_1 + \sigma_m + \sigma_2~,
\label{eq-3act}
\end{equation}
where $\sigma_m$ is the tension for a monodromy transformation in the vicinity of a point in the moduli space---i.e. keeping close to some geometrical configuration ${\cal M}_*$ for the Calabi-Yau. The first term $\sigma_1$ comes from deforming the Calabi-Yau from vacuum 1 to the geometry ${\cal M}_*$, and $\sigma_2$ for deforming ${\cal M}_*$ back to vacuum 2.  Near the vacua, $\sigma_1$ and $\sigma_2$ depend on the deformation only to second order.  However $\sigma_m$ has nothing to do with the initial and final vacua, so the leading order change will be linear.  It is always more economical if $\sigma_m$ can be reduced by deforming away from the vacua.

For our case, $\sigma_m$ is essentially the action integral in the $\theta$ direction,
\begin{equation}
\sigma_m = \int \sqrt{2V-2V_0}\sqrt{K_{z\bar{z}}}\,r\,d\theta~,
\label{eq-Sm}
\end{equation}
and $\sigma_1$, $\sigma_2$ are like integration in the $r$ direction,
\begin{equation}
\sigma_i = \int \sqrt{2V-2V_0}\sqrt{K_{z\bar{z}}}\,dr~.
\end{equation}
We can see that since large deformations in the $\tau$ direction always increase tension, they are highly suppressed.  On the other hand reducing $r$ decreases $\sigma_m$, so the path wants to go near the conifold point.

\begin{figure}
\begin{center}
\includegraphics[width=.5\textwidth]{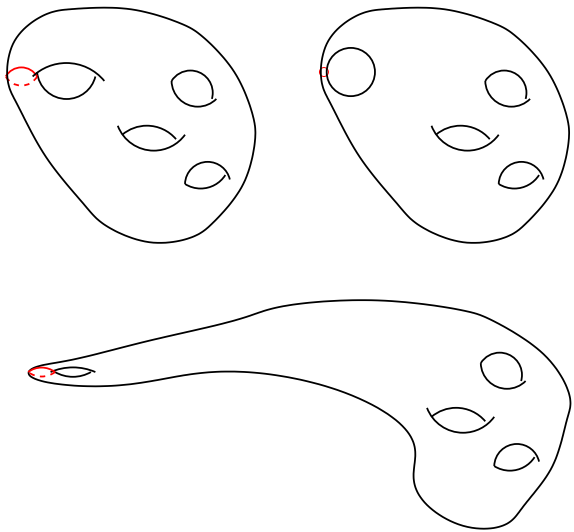}
\end{center}
\caption{Top left is the vacuum configuration of the Calabi-Yau manifold.  A monodromy transformation (on the red cycle) contributes less to the action if the 3-cycle is small (top right), and even less if it happens on the tip of the strongly warped conifold (bottom).}
\label{fig-warped}
\end{figure}

The geometric picture is quite straight forward.  The shrinking 3-cycle near the conifold point is exactly the cycle which we cut and twist in the monodromy transformation.  Physically the shrinking 3-cycle (with flux) cannot go to zero size, so eventually it becomes strongly warped and the monodromy happens at the tip of the strongly warped conifold, as shown in figure \ref{fig-warped}.  Both shrinking and warping help to reduce $\sigma_m$.

\begin{figure}
\begin{center}
\includegraphics[width=.5\textwidth]{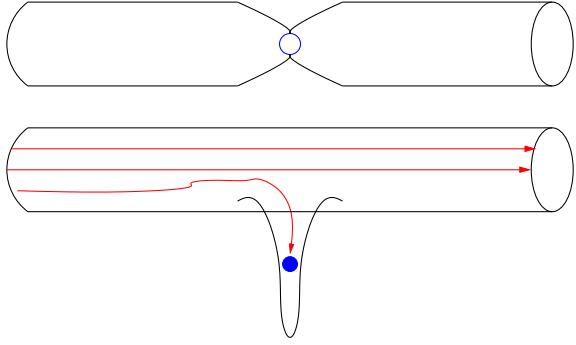}
\end{center}
\caption{The extended horizontal direction represents 4D spacetime.  The top tube comes with the 3-cycle wrapped by the D5 brane, which wants to shrink.  The bottom tube represent the other 3-cycle where the flux is changing, in which the brane is a point like object where the flux line can end.  Placing the charge on a locally warped region also reduces 4D tension.}
\label{fig-brane}
\end{figure}

We can also understand this process in the dual picture, where the flux is changed by nucleating a charged brane instead of monodromy.  The dual 7-form flux which is orthogonal to the shrinking 3-cycle changes by nucleating a 5-brane. Three legs of this 5-brane will wrap the shrinking cycle leaving two spatial directions for the (2+1)D domain wall in the 4D spacetime.  As depicted in figure \ref{fig-brane}, the monodromy contribution is replaced by a brane,
\begin{equation}
\sigma^{\rm 2-brane}_{\rm 4D}=\sigma^{\rm 5-brane}_{\rm 10D}
(V_{\rm shrinking\ 3-cycle})({\rm volume\ factor})({\rm warp\ factor})~.
\label{eq-efften}
\end{equation}
The volume factor corresponds to the dimensional reduction from the 10D theory string frame to the 4D theory Einstein frame.  It is a constant in our case since we have frozen the Kahler moduli.  It is also easy to see why the shrinking 3-cycle volume and the warp factor help to reduce the effective 4D tension\footnote{One may expect us to match Eq.~(\ref{eq-efften}) and Eq.~(\ref{eq-Sm}) to determine $\sigma^{\rm 5\ brane}_{\rm 10D}$.  We cannot do that because in the monodromy picture the effective brane is smeared and cannot be assigned an exact location in the warped throat.}.

Of course, it is very surprising to see that the balance between reducing $\sigma_m$ and increasing $\sigma_1+\sigma_2$ happens at such an extreme geometry---a strongly warped Calabi-Yau.  In the next section we will provide a more quantitave analysis in this particular case.  Here we want to suggest a good intuition for general multi-field tunneling.  The roughly equal separate contributions shown in figure \ref{fig-AB} suggest an equipartition among the three terms $\sigma_1, \sigma_2,$ and $\sigma_m$ that make up the action, (\ref{eq-3act}).  This is quite natural assuming that the three terms depend on a parameter in the same way (say polynomially or exponentially). What we have is essentially a generalized virial theorem telling us that the three terms should have similar orders of magnitude. Knowing this in advance, we could use this to estimate how big the deformation of the vacuum geometry is. 

 \subsection{The shortest path}
 \label{sec-shortest}
Our numerical results suggest a simple analytical argument for conifunneling.  As noted previously, although the axio-dilaton $\tau$ changes during tunneling, it contributes very little to the action integral.  Therefore the dynamics is similar to a 2D problem in just the complex structure moduli $z$.

Starting from the simplest case with 2D canonical kinetic term in the polar coordinate.
\begin{equation}
L = \frac{1}{2}\left(\dot{r}^2+r^2\dot{\theta}^2\right)-V_{\rm inverse}(r,\theta)~.
\end{equation}
Let us first assume that the inverse potential $V_{\rm inverse}$ doesn't have any special properties near the conifold point (taken to be at the origin, $r=0$). In this case, minimizing the action is like finding the shortest path, which is of course a straight line. If there are multiple sheets through branch cuts emanating from the conifold point along $\theta=2\pi n$ there is an additional constraint.  When the angular separation between two points is larger than $\pi$, a straight line will be obstructed by the branch cut.  Therefore the maximum angular separation is $\pi$ if two vacua are to be connected by a tunneling path.

The strongly warped behavior near the conifold point in our mirror quintic case tells us that we must modify the above with non-canonical kinetic terms
\begin{equation}
L = \frac{K(r)}{2}\left(\dot{r}^2+r^2\dot{\theta}^2\right)-V_{\rm inverse}(r,\theta)~.
\end{equation}
Assuming that the dominant behavior of the Kahler metric is of the form
\begin{equation}
K(r)=r^{2\beta}~,
\end{equation}
we may change to a more natural set of variables, defining
\begin{equation}
\tilde{r}=\frac{r^{\beta+1}}{\beta+1}~.
\end{equation}
This yields
\begin{equation}
L = \frac{1}{2}\left(\dot{\tilde{r}}^2+\tilde{r}^2(\beta+1)^2\dot{\theta}^2\right)
-\tilde{V}_{\rm inverse}(\tilde{r},\theta)~,
\end{equation}
where
\begin{equation}
\tilde{V}_{\rm inverse}(\tilde{r},\theta)=
V_{\rm inverse}\left([(\beta+1)\tilde{r}]^{1/(\beta+1)},\theta\right)~.
\end{equation}
Ignoring the inverse potential, we can see that
\begin{equation}
\Delta\theta_{\rm max}=\frac{\pi}{\beta+1}~.
\end{equation}
In our case $\beta=-2/3$, so $\Delta\theta_{\rm max}=3\pi$. In addition, if the inverse potential has a peculiar behavior near $r=0$, it will modify this result.  For example, a standard $-1/r$ attractive core will double the maximum angle.  A repulsive core, which means $V_{\rm inverse}(0)$ is larger than the conserved energy of the path, in general reduces the maximum angle. In our case,
\begin{equation}
\lim_{r\rightarrow0}V=e^K K^{\tau\overline{\tau}}|D_\tau W|,
\end{equation}
has a positive global minimum in the $\tau$ space.  This means that the inverse potential is at most a finite attractive core, which has neglible effects.  So $\Delta\theta_{\rm max}=3\pi$, namely $3/2$ of monodromy transformation, is the best we can get. We have confirmed this with numerical simulations.

Also, note that if we did not include the strong warping correction, we would have had 
\begin{equation}
K(r)\sim \log r~, \ \ \ \ \ \ \ \ V(r)\sim \log r~,
\end{equation}
near $r=0$.  Since $\log r$ diverges slower than any $r^\beta$ with $\beta<0$, it should give us roughly $\Delta\theta_{\rm max}=\pi$.  Also the uncorrected $V$ has a logarithmic divergence, which corresponds to an attractive core in the inverse potential.  It is also weaker than, for example, $V_{\rm inverse}=-1/r$.  As we saw in the simulation, there is no reason that a path can make $\Delta\theta\sim2\pi$.

From this point of view, conifunneling happens because the path needs the strong warping correction to the Kahler metric in order to make a monodromy angle change of $2\pi$. For our particular choice of fluxes $F_3=-1$, this is the minimum amount which $F_0$ can change.  With $|F_3|>1$, one might expect to see several vacua on one sheet, which would correspond to changing $F_0$ by 1 several times. We have not seen such things in any of the examples we have investigated\footnote{Multiple vacua in a given sheet have been observed in other analyses \cite{TorrobaMultiVacs}, but $\tau$ is treated as a fixed parameter. Our $\tau$ is dynamical and we know of no physical reason that requires multiple vacua on a single sheet.}. However, let us for the moment simply assume that there are cases with multiple vacua for $|F_3|>1$.  From our result, it is quite natural to make the following 3 conjectures:
\begin{itemize}
\item For angular changes less than $\pi$, which means $|\Delta F_0|<|F_3|/2$, tunneling is possible regardless of warping and the path does not get close to the conifold point.
\item For angular changes larger than $\pi$, which means  $|\Delta F_0|>|F_3|/2$, we will see conifunneling.
\item For angular changes larger than $3\pi$, which means $|\Delta F_0|>3|F_3|/2$, there will be no tunneling path.
\end{itemize}

\subsection{The BPS Path}

Earlier work on similar string models focused on (near) BPS paths \cite{CerDal06}. Here we will demonstrate that in this model, there is no clear notion of being near-BPS.  Conifunneling is the general behavior and the BPS path is a special case in good agreement with conifunneling between two SUSY vacua.

The path we found numerically is a spatial planar interpolation between two vacua that minimizes the action integral,
\begin{equation}
S = -\int d^4x (K_{z\bar{z}}\partial_\mu z \partial^\mu \bar{z}
+ K_{\tau\bar{\tau}}\partial_\mu \tau \partial^\mu \bar{\tau}-V)~,
\end{equation}
Dropping the (2+1) planar integral, we get the tension of this transition domain wall,
\begin{equation}
\sigma = \int dx~(K_{z\bar{z}}\dot{z} \dot{\bar{z}}
+ K_{\tau\bar{\tau}}\dot{\tau}\dot{\bar{\tau}}-V)~.
\end{equation}
Comparing to the similar problem in supergravity\cite{CerDal06}, we dropped the Kahler sector and 4D gravity.  It is effectively the field theory limit of the same problem.  In particular, we can similarly define $Z=e^{K/2}W$ and rewrite the tension integral in the BPS form
\begin{equation}
\sigma = \mp 2\Delta|Z| + \int dx~\left(K_{z\bar{z}}\left|{\dot{z}} \mp 2K^{z\bar{z}}\partial_{\bar{z}}|Z|\right|^2
+K_{\tau\bar{\tau}}\left|{\dot{\tau}} \mp 2K^{\tau\bar{\tau}}\partial_{\bar{\tau}}|Z|\right|^2\right)~.
\end{equation}
It is straight forward to see that the tension is minimized by BPS equations,
\begin{eqnarray}
\dot{z}&=& \pm 2K^{z\bar{z}}\partial_{\bar{z}}|Z|~, \\
\dot{\tau}&=& \pm 2K^{\tau\bar{\tau}}\partial_{\bar{\tau}}|Z|~.
\end{eqnarray}

Note that these BPS equations are first order, so they only solve the equations of motion (which are second order) with certain boundary conditions.  The paths of BPS equations are field lines of $\partial|Z|$ and general vacua, local minima in $V$, are not necessarily distinguished in $|Z|$, so there is no reason that two vacua sit on the same field line.  In general we cannot see criteria that ensure a tunneling path is necessarily near BPS.

If the vacua are supersymmetric, $V=0$, they are critical points of $|Z|$.  Now the BPS equations agree with equation of motions with zero initial velocity, so potentially a BPS path can be a tunneling path.  But if this critical point is just a saddle point, then it does not have BPS paths in all directions.  We will need a local maximum or minimum of $|Z|$, which is a focal point for BPS paths, so it covers all directions and matches one to one with the possible tunneling paths.

We have checked and found that all of the mirror quintic SUSY vacua listed in table \ref{SUSYTable} are local minima of $|Z|$\footnote{There might be deeper physical reason for this. We defer the study of that question to future work.}.  Another property of a BPS path is that $|Z|$ must be monotonic along the path, so it cannot smoothly connect two local minima in $|Z|$. Instead, it must go through at least one other special point, for example another critical point in $|Z|$. However, that means that another supersymmetric vacuum would exist, and we do not observe this.  Another possibility is that the path goes through a region where $|Z|=0$ and $\partial|Z|$ is not well-defined.  However, $|Z|=0$ implies that $\tau=-A/B$, but the condition $D_\tau W = 0$ requires $\tau=-\overline{A}/\overline{B}$. Thus, if we go from a vacuum to a point where $|Z|=0$, it must go through Im($\tau)=0$, where the string coupling diverges and is dynamically forbidden.

That leaves us with one choice, going through the conifold point where the Kahler metric $K_{z\bar{z}}$ diverges.  Note that with this analysis alone, a BPS path would not be convincing since going through such a singular point often implies complications.  It is because we found that there are paths near the conifold in general that it is reasonable to view the BPS paths from either vacua joined at the conifold as a special limit for the tunneling path. 

Note that the lack of an obvious BPS path dovetails with studies of supersymmetric branes probing the conifold \cite{RamalloBraneProbe}. In this work, the authors construct explicit supersymmetric D3, D5, and D7 brane solutions wrapping various cycles in the singular conifold geometry. In particular, they find supersymmetric solutions that represent D3 branes wrapping the collapsing 3-cycle and a D5 wrapping the dual $S^2$. However, they can only construct stable, non-supersymmetric D5 branes that wrap the collapsing 3-cycle. This provides further support for our interpretation of conifunneling as involving the nucleation of a 5-brane wrapped on the collapsing 3-cycle.

\section{Discussion}

In this paper, we've undertaken the most detailed and refined study to date of particular loci in the Type IIB string landscape.
We began by studying flux potentials arising from various one-parameter Calabi-Yau compactifications. Using a topographical approach similar to that of \cite{Danielsson, Larfors}, we established a general similarity between models that have a single complex structure modulus within the unit circle in the complex moduli space with coordinate $z$. Outside of the unit circle, the models show more unique behavior. Vacua are more sporadic and involve more drastic changes to the fluxes across models. At least for the first family of models whose $\alpha$-parameters are all distinct, studies of flux vacuum statistics in \cite{KachruTaxonomy,DeWolfeKachru,GiryavetsKachru} demonstrate that the behavior of these models strongly depends on their details near the Landau-Ginzburg point. This is likely to hold true for all of the models we have examined.

After finding connected chains of SUSY vacua in various models, we turned our attention to flux transitions between such vacua. The results of our numerical investigation indicated that any initial guess profile fed into the relaxation algorithm is funneled into the conifold region of the moduli space---what we have called ``conifunneling". The stability of these conifunneling solutions hinges critically on incorporating strong warping corrections to the Kahler metric. Physically, these corrections become important near the conifold point because a shrinking 3-cycle pierced through by flux leads to ever denser field strength, which strongly distorts the surrounding Calabi-Yau geometry. Such corrections would be completely washed out by taking the infinite volume limit of our geometry; we work instead with a large but finite volume.

The usual intuition might have suggested that since these flux vacua are near each other in the complex moduli space, it should be simple to construct an appropriately charged brane whose nucleation brings one from one vacuum to another. Our findings suggest that competing effects between the energy required to nucleate a brane wrapping the appropriate cycles and the contraction or growth of these cycles are an essential aspect of the dynamics. Thus, it seems clear that given an initial flux vacuum, finding the most probable tunneling path is more subtle than simply looking at the separation in $z$-space and concluding that a brane can connect two neighboring vacua (including vacua on the $z$-plane that are not on a flux potential connected through LCS, conifold, or LG monodromies). Rather, such a brane would still need to wrap appropriate cycles in order to absorb the appropriate charges, and thus, we expect that the dynamics of these cycles  will play a crucial role in determining when such a brane can be nucleated. Because of this, we expect conifunneling to be important in determining how tunneling occurs through a ``discretuum" \cite{Susskind} of flux vacua such as that envisioned by Bousso and Polchinski \cite{Bousso:2000xa}, and hence in the complexion of the multiverse such processes generate.

\acknowledgments

We thank Simon Judes for calculations contributed at an early state of this work. We also thank Erick Weinberg, Lam Hui, Alberto Nicholis, Eduardo Ponton, Adam Brown, Ulf Danielsson, Puneet Batra, Hooman Davoudiasl, Matt Johnson, Richard Easther, Henry Tye, Liam McCallister, Gary Shiu, Heng-Yu Chen, Thomas van Riet, Charles Doran, Shamit Kachru, Mina Aganagic, and Gonzalo Torroba for numerous helpful discussions. Saswat Sarangi was supported by grant RFP2-08-26 from The Foundational Questions Institute (fqxi.org). Pontus Ahlqvist is partly supported by a graduate fellowship from the Sweden America Foundation. This work is supported in part by DOE grant DE-FG-02-92ER40699.

\begin{appendix}

\section{Properties of One-Parameter Models} \label{UniversalMonodromies}

Below we derive the general form for the monodromies about $z=0,1,$ and
$\infty$, and describe the transformation to the symplectic basis (in
which
the intersection form is canonical) which allows us to determine the
intersection form in the basis of Meijer periods. First recall that the
four
Calabi-Yau periods are solutions to the Picard-Fuchs equation. This can be
written as a differential equation of hypergeometric type:
\begin{equation}
\left[\delta^4 - z \prod_{r=1}^4 (\delta + \al_r)\right] u = 0,
\end{equation}
where $\delta = z d/dz$, The fourteen models classified in
\cite{Doran:2005gu} have $\al$-parameters that satisfy the relations
$\al_r = 1 - \al_q$ with $r,q \in \{1,2,3,4\}$ and $r \neq q$, giving
two related pairs of the $\al$-parameters. We will use these relationships
to simplify some of the results arrived at below.

\subsection{Finding monodromies and the symplectic basis}

We can organize a complete set of solutions $\{U_j\}_{j=0}^3$ into a
period vector $U$. The monodromy around $z=0$ is defined by the relation
\begin{equation}
U(e^{2\pi i} z) = T[0] U(z).
\end{equation}
To facilitate the calculation of the monodromy matrix, the Picard-Fuchs
equation can be recast as a set of first-order differential equations
which
are solved by the fundamental matrix $\Phi_{ij} = \delta^i U_j$. This
matrix
can be written in the form
\begin{equation}
\Phi(z) = S(z) z^{R[0]}.
\end{equation}
The monodromy around $z=0$ is then simply
\begin{equation}
T[0] = e^{2\pi i R[0]^\top}.
\end{equation}
Under a general change of basis $U = M \widetilde{U}$ will transform the
other matrices as follows:
\begin{equation}
\Phi = \widetilde{\Phi} M^\top, \ \ \ S = \widetilde{S} M^\top,
\ \ \ R = M^{-\!\top} \widetilde{R} M^\top, \ \ \ T = M \widetilde{T}
M^{-1}.
\end{equation}
It's useful to transform to a canonical basis in which $S_{\textrm{can}} =
S(0) = \mathbb{I}$. The matrix $R_{\textrm{can}}$ turns out to have a
particularly simple Jordan form in this basis,
\begin{equation}
R_\textrm{can}[0] = \begin{pmatrix}
0 & 1 & 0 & 0 \\
0 & 0 & 1 & 0 \\
0 & 0 & 0 & 1 \\
0 & 0 & 0 & 0
\end{pmatrix}
\end{equation}

Given the period vector $U$ in some basis, the fundamental matrix near
$z=0$
can be written as
\begin{equation}
\Phi(z) = Z(z) q(z),
\end{equation}
with the row-vector $Z(z) = (1, \log z, (\log z)^2, (\log z)^3)$ and
$q(z)$
a matrix related to the solutions in a way that will be described more
explicitly below in the case of the monodromy around $z=\infty$. In the
canonical basis, $U_\textrm{can} = Z(z) q_\textrm{can}(z)$. Given the
transformations above, it is not hard to show that $M = q(0)^\top
q_\textrm{can}^{-\!\top}$ with $U = M U_\textrm{can}$.
It is obvious that the monodromy in the canonical basis is independent of
the $\al$ parameters since $R_\textrm{can}[0]$ is. A somewhat more
surprising result is that $T[0]$ remains independent of the $\al$
parameters
in the basis given by the Meijer functions
\begin{equation}
U_j(z) = {1 \over (2 \pi i)^j} \oint_\gamma \phi_j(s,z) ds.
\end{equation}
The integrand is
\begin{equation}
\phi_j(s,z) = \left((-1)^{j+1} z \right)^s{\Gamma(-s)^{j+1} \over
\Gamma(s+1)^{3-j}}
\prod_{\al_r} {\Gamma(s + \al_r)^{D_r} \over \Gamma(\al_r)^{D_r}},
\end{equation}
where $D_r$ is the multiplicity parameter of $\al_r$ and the product is
over {\em unique} instances of the $\al$ parameters. Since our interest
is the monodromy around $z=0$, we may close the contour $\gamma$ to the
right, picking up residues at each non-negative integer $s = n$. For the
details of this particular calculation see \cite{Greene} (note that
our basis here differs by factors of $(2 \pi i)^{-j}$). One finds that for
all the models
\begin{equation}
T[0] = \begin{pmatrix}
1 & 0 & 0 & 0 \\
-1 & 1 & 0 & 0 \\
1 & -1 & 1 & 0 \\
0 & 0 & -1 & 1 \\
\end{pmatrix}.
\end{equation}

Similar remarks apply when we consider the monodromy around $z=\infty$.
One major difference is that $R_\textrm{can}[\infty]$ is not the same as
$R_J[\infty]$ where $R_J[\infty]$ is the Jordan form of the matrix
$R[\infty]$.
To go to the Jordan basis, we transform $U_\textrm{can} = P^\top U_J$.
The transformations above imply that $S_J(\infty) = P$ and
$R_\textrm{can}[\infty] = P^{-1} R_J[\infty] P$. The transformation $P$
and $R_J[\infty]$ are given explicitly below.

Since we are now interested in the monodromy around $z = \infty$ we close
the contour to the left, picking up residues at $s = -\al_r - n$ for all
non-negative integers $n$. It is useful to rewrite $\phi_j(s,z)$ in a way
that reflects the pole structure:
\begin{equation}
\phi_j(s \sim -\al_r-n,z) =
\left(   \Gamma(s + \al_r + n + 1)\over (s+\al_r+n) (s+\al_r)_{(n)}
\right)^{D_r}
{\Gamma(-s)^{j+1} \prod_{\al_q \neq \al_r} \Gamma(s + \al_q) \over
\Gamma(s+1)^{3-j}
\prod_{\al_p} \Gamma(\al_p)^{D_p}} \left((-1)^{j+1} z \right)^s,
\end{equation}

Let $((-1)^{j+1})^s = \delta_{j,\textrm{odd}} + \delta_{j,\textrm{even}}
e^{\pi i s}$ with $\delta_{j,\textrm{odd}} = 1$ when $j$ is odd and
vanishing otherwise (and similarly for $\delta_{j,\textrm{even}}$).
It is useful to split off the $z$-dependence from the rest of the factors
in the integrand as it is expressed above:
\begin{equation}
\phi_j(s \sim -\al_r - n, z) = {\phi_{n,r,j}(s) \over
(s+\al_r+n)^{D_r}}\,z^s
\end{equation}
The periods for $|z| > 1$ can be written as a sum over the residues of
$\phi_j(s,z)$:
\begin{equation}
U_j(z)
= {1 \over (2\pi i)^{j-1}} \sum_{n = 0}^\infty \sum_{r = 1}^4
\textrm{Res}_{s\,= -\al_r - n}\left( \phi_j(s,z) \right)
\end{equation}
Expanding out the residue:
\begin{equation}
U_j(z) = {1 \over (2\pi i)^{j-1}}\sum_{n=0}^\infty z^{-n} \sum_{r=1}^4
\sum_{i=0}^{D_r - 1}
{{D_r - 1}\choose i} {\phi_{n,r,j}^{(D_r - 1 - i)}(-n-\al_r) \over (D_r -
1)!}
z^{-\al_r} \log(z)^{i}.
\end{equation}
The above can be re-expressed as
\begin{equation}
U_j(z) = \sum_{r=1}^4 Z_r(z) q_{rj}(z)
\end{equation}
or $U^\top = Z q$, where $Z(z)$ is a row vector that is specific to the
situation. There are four cases of interest:
\begin{equation}
Z(z) = \left\{
\begin{array}{ll}
(z^{-\al_1}, z^{-\al_2}, z^{-\al_3}, z^{-\al_4}),
& \mbox{if all $\al$'s differ} \\
(z^{-\al_1}, z^{-\al_2}, z^{-\beta}, z^{-\beta} \log z),
& \mbox{if $\al_3 = \al_4 = \beta$} \\
(z^{-\al}, z^{-\al}\log z, z^{-\beta}, z^{-\beta}\log z),
& \mbox{if $\al_1=\al_2=\al$ and $\al_3 = \al_4 = \beta$} \\
z^{-\al} (1,\,\log z,\,(\log z)^2,\,(\log z)^3),
& \mbox{if $\al_1=\al_2=\al_3=\al_4=\al$}. \\
\end{array}
\right.
\end{equation}
Define the log-index vector
\begin{equation}
\rho = \left\{
\begin{array}{ll}
(0,0,0,0), & \mbox{if all $\al$'s differ} \\
(0,0,0,1), & \mbox{if $\al_3 = \al_4 = \beta$} \\
(0,1,0,1), & \mbox{if $\al_1=\al_2=\al$ and $\al_3 = \al_4 = \beta$} \\
(0,1,2,3), & \mbox{if $\al_1=\al_2=\al_3=\al_4=\al$}.
\end{array}
\right.
\end{equation}
and the multiplicity vector
\begin{equation}
D = \left\{
\begin{array}{ll}
(1,1,1,1), & \mbox{if all $\al$'s differ} \\
(1,1,2,2), & \mbox{if $\al_3 = \al_4 = \beta$} \\
(2,2,2,2), & \mbox{if $\al_1=\al_2=\al$ and $\al_3 = \al_4 = \beta$} \\
(4,4,4,4), & \mbox{if $\al_1=\al_2=\al_3=\al_4=\al$}.
\end{array}
\right.
\end{equation}
Then
\begin{equation}
q_{rj}(z) = {1 \over (2\pi i)^{j-1}}\sum_{n=0}^\infty z^{-n}
{{D_r - 1}\choose \rho_r} {\phi_{n,r,j}^{(D_r - 1 - \rho_r)}(-n-\al_{r})
\over (D_r - 1)!} ,
\end{equation}
note that $r = 1,\ldots,4$ while $j = 0,\ldots,3$. We are interested in
the value around $z = \infty$. The sum over integers reduces to
\begin{equation}
q_{rj} = {1 \over (2\pi i)^{j-1}}
{{D_r - 1}\choose \rho_r} {\phi_{0,r,j}^{(D_r - 1 - \rho_r)}(-\al_{r})
\over (D_r - 1)!}.
\end{equation}

It is easy to compute $z^{R_J[\infty]}$ given the $R_J$ and the change
of basis $P$ discussed above:
\begin{equation}
R_J[\infty] =
\begin{pmatrix}
-\al_1 & {\rho_2 \over \rho_1 + 1} & 0 & 0 \\
0 & -\al_2 & {\rho_3 \over \rho_2 + 1} & 0 \\
0 & 0 & -\al_3 & {\rho_4 \over \rho_3 + 1} \\
0 & 0 & 0 & -\al_4
\end{pmatrix},\ \ \
P =
\begin{pmatrix}
{1\over \al_1^3} & {2+\rho_1\rho_2 + 4\rho_2 \over 2\al_1^{\rho_2}
\al_2^3} & {2+\rho_2 \rho_3 + 4\rho_3 \over 2\al_2^{\rho_3} \al_3^3}
& {2+\rho_3 \rho_4 + 4\rho_4 \over 2\al_3^{\rho_4} \al_4^3} \\
{1 \over \al_1^2} & {1+\rho_2 \over \al_1^{\rho_2} \al_2^2}
& {1+\rho_3 \over \al_2^{\rho_3} \al_3^2} & {1+\rho_4 \over
\al_3^{\rho_4} \al_2^2} \\
{1\over \al_1} & {1 \over \al_1^{\rho_2} \al_2} & {1 \over
\al_2^{\rho_3} \al_3} & {1 \over \al_3^{\rho_4} \al_4} \\
1 & 1-{\rho_2 \over \rho_1 + 1}& 1-{\rho_3 \over \rho_2 +1 }&
1-{\rho_4 \over \rho_3 +1}
\end{pmatrix},
\end{equation}
recall that the $\al_r$ need not be distinct in this expression. We
have $\Phi_{\textrm{can}} = \Phi_J P$, which implies that $S_J = P$.
Using the fact that the period matrix can be written in the Jordan
basis as $U_J(z)^\top = Z(z) q_J(z)$ and that $(\Phi_J)_{ij} = \delta^i
(U_J)_j$ allows us to find an expression for $q_J$. In our original
Meijer-basis $U(z)^\top = Z(z) q(z)$ so the two bases are related by
the transformation $U = M U_J$, $M =  q^\top q_J^{-\top}$.

The period matrix is then $T[\infty] = M T_J[\infty] M^{-1}$. Remarkably,
this can be expressed in a universal way in terms of simple functions of
the $\al$ parameters:
\begin{equation}
T[\infty] = \begin{pmatrix}
m_1 & m_2 & m_3 & m_4 \\
-1 & 1 & 0 & 0 \\
1 & -1 & 1 & 0 \\
0 & 0 & -1 & 1
\end{pmatrix},
\end{equation}
where
\begin{eqnarray}
m_2 &=& 4\left(\sin(\pi \al_1)^2 + \sin(\pi \al_2)^2\right),
\ \ \ m_1 = 1 - m_2, \\
m_4 &=& 16 \sin(\pi \al_1)^2 \sin(\pi \al_2)^2, \ \ \ \ \ \ \ \ m_3 =
-m_4.
\end{eqnarray}
Showing this involves some arduous calculations best done using a computer
algebra system such as {\em Mathematica}.

The monodromy around the conifold point $z=1$ follows from the fact that
going around a loop that encompasses all three special points results in a
trivial transformation of the periods. Therefore,
\begin{equation}
T[1] = \left\{
\begin{array}{ll}
T[\infty] T[0]^{-1}, & \mbox{if Im$\,z < 0$} \\
T[0]^{-1} T[\infty], & \mbox{if Im$\,z > 0$}
\end{array}
\right.
\end{equation}

In order to find the general intersection form in the Meijer-basis we
pass to the symplectic basis via $\Pi(z) = L\,U(z)$. In this basis the
intersection form takes on a natural form:
\begin{equation}
Q = \begin{pmatrix}
0 & 0 & 0 & -1 \\
0 & 0 & 1 & 0 \\
0 & -1 & 0 & 0 \\
1 & 0 & 0 & 0
\end{pmatrix}.
\end{equation}
The requirement that $\Pi^\dag Q^{-1} \Pi \propto U^\dag Q_U^{-1} U$
implies
that $Q_U = L^{-1} Q L^{-\dag}$ up to an overall scaling factor.

Our goal now is to determine the change of basis $L$. First, we observe
that
in the symplectic basis, the monodromy $T_\Pi[1]$ has a simple form. The
argument is as in \cite{Candelas:1990rm}: choose a pair of cycles $A$ and
$B$ such that $A$ corresponds to the $S^3$ that degenerates at the
conifold
point $z=1$, and $B$ is the cycle that intersects it. By transporting the
cycles around $z=1$, $A$ remains unambiguous, but $B$ is only defined up
to possible contributions by $A$, that is $B \rightarrow B + n A$ for some
integer $n$. The other pair of cycles are left unchanged since they can be
chosen to lie ``far away" (i.e. outside the neighborhood) of $A$ and $B$.
This implies that in the symplectic basis,
\begin{equation}
T_\Pi[1] = \begin{pmatrix}
1 & 0 & 0 & 0 \\
0 & 1 & 0 & 0 \\
0 & 0 & 1 & 0 \\
n & 0 & 0 & 1
\end{pmatrix}.
\end{equation}
Examining $T[1]$ in the Meijer basis shows that it can always almost be
diagonalized, bringing it to the above form with $n=1$. The relation
between
the two forms of the monodromy is $T_\Pi[1] = L T[1] L^{-1}$, which
constrains
 the choice of $L$. A further constraint is that all of the monodromies
should act as symplectic transformations in the symplectic
basis. Therefore
for each monodromy $T$ in the Meijer basis, $T_\Pi = L T L^{-1}$ should
preserve the intersection form $Q$. We find that
\begin{equation}\label{SymplecticTransform}
L = \kappa \begin{pmatrix}
0 & m_2 & 0 & m_4 \\
0 & C & -m_4 & 0 \\
0 & -1 & 0 & 0 \\
1 & 0 & 0 & 0
\end{pmatrix}
\end{equation}
satisfies these conditions, with $\kappa$ and $C$ undetermined constants.
The general expression for the intersection form of all the models in the
Meijer basis is
\begin{equation}
Q_U = {1 \over \kappa^2 m_4^2} \begin{pmatrix}
0 & 0 & 0 & m_4 \\
0 & 0 & -m_4 & 0 \\
0 & m_4 & 0 & -m_2 \\
-m_4 & 0 & m_2 & 0
\end{pmatrix}.
\end{equation}

\subsection{Summary of models, their monodromies and intersection forms} \label{TheModels}

The models we are considering have been classified in
\cite{Doran:2005gu} and are summarized in the table below:

\begin{center}
\begin{tabular}{|l|l|llll|llll|}
\hline
\# & Model & $\al_1$ & $\al_2$ & $\al_3$ & $\al_4$ & $m_1$ & $m_2$ &
$m_3$ & $m_4$ \\
\hline \hline
1 & $\mathbb{P}^4[5]$ & ${1/5}$ & ${2/5}$ & ${3/5}$
& ${4/5}$ & -4 & 5 & -5 & 5 \\
2 & $\mathbb{WP}^{2,1,1,1,1}[6]$ & ${1/ 6}$ & ${1 / 3}$
& ${2/ 3}$ & ${5/ 6}$ & -3 & 4 & -3 & 3 \\
3 & $\mathbb{WP}^{4,1,1,1,1}[8]$ & ${1/ 8}$ & ${3 / 8}$
& ${5/ 8}$ & ${7/ 8}$ & -3 & 4 & -2 & 2 \\
4 & $\mathbb{WP}^{5,2,1,1,1}[10]$ & ${1/ 10}$ & ${3 / 10}$
& ${7/ 10}$ & ${9/ 10}$ & -2 & 3 & -1 & 1 \\
5 & $\mathbb{WP}^{2,1,1,1,1,1}[3,4]$ & ${1/ 4}$ & ${1 / 3}$
& ${2/ 3}$ & ${3/ 4}$ & -4 & 5 & -6 & 6 \\
6 & $\mathbb{WP}^{3,2,2,1,1,1}[4,6]$ & ${1/ 6}$ & ${1 / 4}$
& ${3/ 4}$ & ${5/ 6}$ & -2 & 3 & -2 & 2 \\
7 & * & ${1/ 12}$ & ${5 / 12}$ & ${7/ 12}$
& ${11/ 12}$ & -3 & 4 & -1 & 1 \\
\hline
8 & $\mathbb{P}^{5}[2,4]$ & ${1/ 4}$ & ${1/ 2}$
& ${1/ 2}$ & ${3 / 4}$ & -5 & 6 & -8 & 8 \\
9 & $\mathbb{P}^{6}[2,2,3]$ & ${1/ 3}$ & ${1/ 2}$
& ${1/ 2}$ & ${2 / 3}$ & -6 & 7 & -12 & 12 \\
10 & $\mathbb{WP}^{3,1,1,1,1,1}[2,6]$ & ${1/ 6}$ &
${1/ 2}$ & ${1/ 2}$ & ${5 / 6}$ & -4 & 5 & -4 & 4 \\
\hline
11 & $\mathbb{P}^5[3,3]$ & ${1/ 3}$ & ${1 / 3}$
& ${2/ 3}$ & ${2/ 3}$ & -5 & 6 & -9 & 9 \\
12 & $\mathbb{WP}^{2,2,1,1,1,1}[4,4]$ & ${1/ 4}$
& ${1 / 4}$ & ${3/ 4}$ & ${3/ 4}$ & -3 & 4 & -4 & 4\\
13 & $\mathbb{WP}^{3,3,2,2,1,1}[6,6]$ & ${1/ 6}$
& ${1 / 6}$ & ${5/ 6}$ & ${5/ 6}$ & -1 & 2 & -1 & 1\\
\hline
14 & $\mathbb{P}^7[2,2,2,2]$ & ${1/ 2}$ & ${1 / 2}$ &
${1/ 2}$ & ${1/ 2}$ & -7 & 8 & -16 & 16 \\
\hline
\end{tabular}
\vspace{8pt}

Table A.1: A summary of the various model parameters
\end{center}

The monodromies around $z=0$ and $\infty$ in the Meijer basis take the
form
\begin{equation}
T[0] = \begin{pmatrix}
1 & 0 & 0 & 0 \\
-1 & 1 & 0 & 0 \\
1 & -1 & 1 & 0 \\
0 & 0 & -1 & 1 \\
\end{pmatrix}, \ \ \
T[\infty] = \begin{pmatrix}
m_1 & m_2 & m_3 & m_4 \\
-1 & 1 & 0 & 0 \\
1 & -1 & 1 & 0 \\
0 & 0 & -1 & 1
\end{pmatrix}, \ \ \
T[1] = \left\{
\begin{array}{ll}
T[\infty] T[0]^{-1}, & \mbox{if Im$\,z < 0$} \\
T[0]^{-1} T[\infty], & \mbox{if Im$\,z > 0$}
\end{array}
\right.
\end{equation}

The intersection form in the Meijer basis and the transformation to
the symplectic basis are given by
\begin{equation}
Q_U = {1 \over \kappa^2 m_4^2} \begin{pmatrix}
0 & 0 & 0 & m_4 \\
0 & 0 & -m_4 & 0 \\
0 & m_4 & 0 & -m_2 \\
-m_4 & 0 & m_2 & 0
\end{pmatrix}, \ \ \
L = \kappa \begin{pmatrix}
0 & m_2 & 0 & m_4 \\
0 & C & -m_4 & 0 \\
0 & -1 & 0 & 0 \\
1 & 0 & 0 & 0
\end{pmatrix}.
\end{equation}

\section{A Brief Overview of Dimensional Reduction for Flux Compactifications}
\label{Reduction}

The low-energy effective description for type IIB string theory is
\begin{equation}\label{IIBAction}
S_{\textrm{IIB}} = {1 \over 2 \kappa_{10}^2} \int d^{10}x \sqrt{-g}
\left({\cal R} - {|\pd \tau|^2 \over 2 \tau_I^2} - {|G_{(3)}|^2 \over 12 \tau_I} - {\tilde{F}_{(5)}^2 \over 4 \cdot 5!}\right) + S_{\textrm{CS}} + S_{\textrm{loc}},
\end{equation} 
with
\begin{equation}
S_{\textrm{CS}} = {1 \over 8 i \kappa_{10}^2} \int {C_{(4)} \wedge G_{(3)} \wedge \overline{G}_{(3)} \over \tau_I},
\end{equation}
and $S_{\textrm{loc}}$ are contributions to the action from localized sources. We also have $\kappa_{10}^2 = (2\pi)^7 {\alpha'}^4$, and that the 5-form $\tilde{F}_{(5)}$ be self-dual. The NS-NS field strength $H_{(3)}$ and the R-R field strength $F_{(3)}$ have been combined into $G_{(3)} = F_{(3)} - \tau H_{(3)}$. The action is invariant under the $SL(2,\mathbb{Z})$ transformations
\begin{equation}
\tau \to {a \tau + b \over c\tau + d}, \ \ \ G_{(3)} \to {G_{(3)} \over c \tau + d}.
\end{equation}
The action is given in the 10D Einstein frame, which is related to the string frame by the metric rescaling $g_{\textrm{Einstein}} = e^{-\phi/2} g_{\textrm{String}}$, where $\phi$ is the dilaton.

\subsection{Conformal Calabi-Yau metric and the 4D Einstein frame}

The metric for compactification on a warped Calabi-Yau manifold is
\begin{equation}\label{CompactMetric}
ds_{10}^2 = e^{2 A(y)} \eta_{\mu\nu}\,dx^\mu dx^\nu + e^{-2 A(y)} \tilde{g}_{ij}\,dy^i dy^j.
\end{equation}
As discussed in \cite{GiddingsWarpDynamics}, there is some subtlety involved in bringing this metric into the 4D Einstein frame due to the fact that for non-trivial warping, the universal Kahler modulus, $\rho_I$ should actually be thought of as the zero-mode of the warp factor:
\begin{equation}
e^{-4 A(y)} = c + e^{-4 A_0(y)},
\end{equation}
where $c$ in the above is equivalent to $\rho_I$ (but is kept notationally distinct in keeping with the literature). As a result, a rescaling of the internal metric
\begin{equation}
\tilde{g}_{ij} \to \lambda \tilde{g}_{ij}
\end{equation} 
must be done in conjunction with a transformation of the warp factor
\begin{equation}
e^{2 A(y)} \to \lambda e^{2 A(y)},
\end{equation}
which ends up preserving the original internal term in the metric (\ref{CompactMetric}), while producing a factor of $\lambda$ in the spacetime term. To go to the 4D Einstein frame choose
\begin{equation}
\lambda = {V_{CY} \over V_W} = {V_{CY} \over V_W^0 + c V_{CY}},
\end{equation}
where
\begin{equation}
V_{CY} = \int_{\cal M} d^6 y \sqrt{\tilde{g}_6}, \ \ \ V_{W}^0 = \int_{\cal M} d^6 y \sqrt{\tilde{g}_6} e^{-4 A_0(y)},
\end{equation}
so in the 4D Einstein frame we have
\begin{equation}
ds_{10}^2 = e^{2 A(y)} {V_{CY} \over V_W} \eta_{\mu\nu}\,dx^\mu dx^\nu + e^{-2 A(y)} \tilde{g}_{ij}\,dy^i dy^j.
\end{equation}
Observe that in the limit where $c \gg e^{-4A_0}$ the above we have $V_W \to c V_{CY}$ and $e^{-4 A} \to c$, so the factor in front of the spacetime metric goes to the standard $e^{-6 u} = c^{-3/2}$ while the factor in front of the internal manifold term goes to $c^{1/2}$. This shows that in the limit of weak warping, the metric takes the standard 4D Einstein frame form for Calabi-Yau compactification.

In the above, we have treated the zero-mode $c$ and any other moduli as constant parameters. Promoting the moduli to spacetime fields involves introducing compensators and is rather technically involved. The analyses in \cite{DouglasWarped, DouglasKinetic} show that the Kahler potential receives warping corrections\footnote{The same techniques were applied to the study of D3 brane dynamics in warped compactifications in \cite{HengYu}.}. Complementary methods for computing warping effects on the Kahler potential have been described in \cite{Martucci}. The internal metric fluctuations corresponding to both complex and Kahler moduli should have a Kahler metric of the form
\begin{equation}
G_{IJ} = {1 \over 4 V_W} \int_{\cal M} d^6 y \sqrt{\tilde{g}_6} e^{-4 A(y)} \tilde{g}^{ik} \tilde{g}^{jl} \delta_I \tilde{g}_{ij} \delta_J \tilde{g}_{kl},
\end{equation}
where the deformations $\delta_I g_{ij}$ are given by
\begin{equation}
\delta_I \tilde{g}_{ij} = {\pd \tilde{g}_{ij} \over \pd u^I} + \delta_I \tilde{g}^*_{ij}.
\end{equation}
The term $\pd g_{ij}/\pd u^I$ is the standard metric fluctuation associated with the modulus $u^I$. The second term on the RHS is a correction from warping effects and is defined by a set of constraint equations
\begin{equation}
\delta_I A = {1\over 8} \tilde{g}^{ij} \delta_I \tilde{g}_{ij}, \ \ \ 
\tilde{\nabla}^k \left(\delta_I \tilde{g}_{kj} - \half \tilde{g}_{kj} \tilde{g}^{mn} \delta_I \tilde{g}_{mn} \right) = 4\tilde{g}^{ik} \pd_i A \delta_I \tilde{g}_{kj}.
\end{equation}

For the analysis near the conifold, strong warping forces us to take into account the above considerations. It turns out that the correct functional dependence on the complex structure modulus near the conifold can be derived through a somewhat less complicated procedure than that of \cite{DouglasKinetic}. In \cite{DouglasSUSYBreaking}, the authors assume a Kahler potential of the form
\begin{equation}
\int_{\cal M} d^6 y e^{-4 A(y)} \Omega \wedge \overline{\Omega}, 
\end{equation}
where $\Omega$ is the holomorphic 3-form of the unwarped Calabi-Yau. This simple form does {\em not} follow from the general considerations of warping above, and indeed, in the specific case of strong warping corrections to the conifold it does not produce the correct numerical coefficients. However, it does capture the correct functional dependence on the complex structure modulus which is sufficient for our purposes.

\subsection{Corrections to the universal Kahler modulus}

In \cite{DouglasUniversalKahler}, the authors considered the impact of warping on the universal Kahler modulus $\rho$ when it is promoted to a dynamical spacetime field. As noted above, $\rho$ is not an independent degree of freedom, its imaginary part is in fact identified with the zero-mode part of the warp factor itself. The real part is associated with an axion related to the 4-form potential
\begin{equation}
C_{(4)} = \half\,a\,\tilde{J} \wedge \tilde{J} + \cdots,
\end{equation}
where $\tilde{J}$ is the Kahler form associated with the metric $\tilde{g}_{ij}$. 

Surprisingly, the corrected Kahler potential for the universal Kahler modulus takes a simple, natural form
\begin{equation}
K_{\textrm{Kahler}} = -3 \log\left(2 {V_W \over V_{CY}}\right) = -3 \log \left(2 c + 2{V_W^0 \over V_{CY}}\right) 
= -3 \log\left(-i(\rho - \bar{\rho}) + 2{V_W^0 \over V_{CY}}\right).
\end{equation}
The above can be simplified by shifting the universal Kahler modulus, $\rho \to \rho - i(V^0_W / V_{CY})$. The shifted expression has the usual form and is what we work with in the main text of this paper.

\section{Numerical Computation of the Flux Potential}\label{sec-VacuumNumerics}

Numerical computation of the Kahler potential $K$, Kahler metric $K_{z\bar{z}}$, superpotential $W$, and the flux potential $V$ require the use of Meijer functions that solve the fourth-order Picard-Fuchs ODE. {\em Mathematica} has built-in Meijer functions, however these evaluate too slowly for analytical computations. Instead, we generate a table of look-up values for these functions on a grid running from $(-5.01,4.99)$ in both the Re($z$) and Im($z$) directions. The grid spacing is $0.05$ between each vertex.

The numerical Meijer functions for our models are defined using the built-in ones as follows
\begin{eqnarray}
U_0 &=& c\,{\tt MeijerG[\{\{1-\alpha_1,1-\alpha_2,1-\alpha_3,1-\alpha_4\},\{\}\},\{\{0\},\{0,0,0\}\},-z]}, \nn \\
U_1 &=& {c\over 2\pi i}\,{\tt MeijerG[\{\{1-\alpha_1,1-\alpha_2,1-\alpha_3,1-\alpha_4\},\{\}\},\{\{0,0\},\{0,0\}\},z]}, \nn \\
U_2^- &=& {c\over (2\pi i)^2}\,{\tt MeijerG[\{\{1-\alpha_1,1-\alpha_2,1-\alpha_3,1-\alpha_4\},\{\}\},\{\{0,0,0\},\{0\}\},-z]}, \nn \\
U_3 &=& {c\over (2\pi i)^3}\,{\tt MeijerG[\{\{1-\alpha_1,1-\alpha_2,1-\alpha_3,1-\alpha_4\},\{\}\},\{\{0,0,0,0\},\{\}\},z]}, \nn \\
\end{eqnarray}
where the $\alpha$-parameters are read off of table A.1 in appendix \ref{TheModels} and the constant is given by
\begin{equation*}
c = {1 \over \Gamma(\alpha_1) \Gamma(\alpha_2) \Gamma(\alpha_3) \Gamma(\alpha_4)}.
\end{equation*}
It is also useful to compute look-up tables for the derivatives of these functions:
\begin{eqnarray}
\pd_z U_0 &=& c\,{\tt MeijerG[\{\{-\alpha_1,-\alpha_2,-\alpha_3,-\alpha_4\},\{\}\},\{\{0\},\{-1,-1,-1\}\},-z]}, \nn \\
\pd_z U_1 &=& -{c\over 2\pi i}\,{\tt MeijerG[\{\{-\alpha_1,-\alpha_2,-\alpha_3,-\alpha_4\},\{\}\},\{\{0,-1\},\{-1,-1\}\},z]}, \nn \\
\pd_z U_2^- &=& {c\over (2\pi i)^2}\,{\tt MeijerG[\{\{-\alpha_1,-\alpha_2,-\alpha_3,-\alpha_4\},\{\}\},\{\{0,-1,-1\},\{-1\}\},-z]}, \nn \\
\pd_z U_3 &=& -{c\over (2\pi i)^3}\,{\tt MeijerG[\{\{-\alpha_1,-\alpha_2,-\alpha_3,-\alpha_4\},\{\}\},\{\{0,-1,-1,-1\},\{\}\},z]}. \nn\\
\end{eqnarray}

Initially, we arrange for the branch-cuts to lie along the real axis from $(-\infty,0]$ and $[1,\infty)$. To do this, we must define
\begin{equation}
U_2 = \left\{
\begin{array}{l l}
U_2^-, & \ \ \ \ \textrm{if Im($z$) $<$ 0}, \\ 
U_2^- - U_1, & \ \ \ \ \textrm{if Im($z$) $\geq$ 0}\\ 
\end{array} \right.
\end{equation}
and similarly for $\pd_z U_2$.

Let the look-up tables constructed from the above definitions be {\tt U0, U1, U2, U3, dU0, dU1, dU2}, and {\tt dU3}. These arrays contain just the values of the Meijer functions at the grid points. To form the interpolating function on the grid for say, $U_0$ one must form a table associating each entry in {\tt U0} to its corresponding grid point. One can then run {\em Mathematica}'s {\tt Interpolation} function on this table. Usually this is one of the final steps after computing the table of values for a function of interest such as the flux potential.

The arrays for the canonical periods $\Pi_j$ are computed by using the change of basis $L$ given in appendix \ref{TheModels} with $\kappa = 1$ and $C = 3$. This means that
\begin{eqnarray}
{\tt P3} &=& m_2 {\tt U1} + m_4 {\tt U3}, \nn \\
{\tt P2} &=& 3{\tt U1} - m_4{\tt U2}, \nn \\
{\tt P1} &=& -{\tt U1}, \nn \\
{\tt P0} &=& {\tt U0}.
\end{eqnarray}
The computation of all the other functions of interest in terms of the canonical periods now follow.

The final issue is that sometimes we must rotate the branch cut emanating from $0$ to $-\infty$ in order to locate minima that sit too close to this branch cut. We do this by using the monodromy transformation around the LCS point in the canonical basis. To define the new periods with the rotated branch cut, we essentially define the new functions to be the same as the old functions either above or below the new cut (depending on which way one is rotating), while defining the value beyond the cut as given by the monodromy matrix applied to the vector of periods.

\section{A Toy Model for Two-Field Tunneling}
\label{sec-toy}
 \subsection{General Formalism for Relaxation Method}
As mentioned in section \ref{sec-onefield}, the relaxation method allows us to find a path between two vacua which minimizes the tension,
\begin{equation}
\sigma = \int dz \left(\frac{G_{ij}}{2}\frac{d\phi_i}{dz}\frac{d\phi_j}{dz}+ V-V_1\right)~.
\end{equation}
Such a path will solve the equations of motion,
\begin{equation}
\frac{d}{dz}\left(G_{ij}\frac{d\phi_i}{dz}\right)=
\frac{\partial V}{\partial\phi_j}+\frac{1}{2}\frac{dG_{kl}}{d\phi_j}\frac{d\phi_k}{dz}\frac{d\phi_l}{dz}~,
\label{eq-multi}
\end{equation}
with boundary conditions
\begin{equation}
\phi_i(z=-\infty)\rightarrow\phi_i^{(1)},
\ \ \ \  \phi_i(z=\infty)\rightarrow\phi_i^{(2)}~. 
\label{eq-multibound}
\end{equation}

First we promote the equations of motion, Eq.~(\ref{eq-multi}), to $(1+1)$ dimensional PDEs,
\begin{equation}
\frac{\partial}{\partial t}\left(G_{ij}\frac{\partial \phi_i}{\partial t}\right)-
\frac{\partial}{\partial z}\left(G_{ij}\frac{\partial \phi_i}{\partial z}\right)=
-\frac{\partial V}{\partial\phi_j}+\frac{1}{2}
\frac{\partial G_{kl}}{\partial \phi_j}
\left(\frac{\partial\phi_k}{\partial t}\frac{\partial \phi_l}{\partial t}-\frac{\partial\phi_k}{\partial z}\frac{\partial \phi_l}{\partial z}\right)~,
\label{eq-static}
\end{equation}
which has conservation of energy (up to boundary terms)
\begin{equation}
\frac{G_{ij}}{2}
\left(\frac{\partial\phi_i}{\partial t}\frac{\partial\phi_j}{\partial t}
+\frac{\partial\phi_i}{\partial z}\frac{\partial\phi_j}{\partial z}\right)
+V=E=const.
\end{equation}

Note that a static solution of these PDEs will solve the equations of motion.  So the trick is to add a damping term,
\begin{equation}
\frac{\partial}{\partial t}\left(G_{ij}\frac{\partial \phi_i}{\partial t}\right)-
\frac{\partial}{\partial z}\left(G_{ij}\frac{\partial \phi_i}{\partial z}\right)
+\lambda(t) \frac{\partial\phi_j}{\partial t}=
-\frac{\partial V}{\partial\phi_j}+\frac{1}{2}
\frac{\partial G_{kl}}{\partial \phi_j}
\left(\frac{\partial\phi_k}{\partial t}\frac{\partial \phi_l}{\partial t}-\frac{\partial\phi_k}{\partial z}\frac{\partial \phi_l}{\partial z}\right)~.
\label{eq-relax}
\end{equation}
With $\lambda(t)>0$, as long as the solution is not static, the total energy will keep decreasing until it reaches a minima.

 \subsection{Application to a Two Fields Model}
 
Here we apply the relaxation method to a two field problem with standard kinetic terms.
\begin{equation}
\ddot{x}=\frac{\partial V}{\partial x}~, \ \ \ \ \ddot{y}=\frac{\partial V}{\partial y}~.\,
\end{equation}
The potential is designed to have both a run away direction in ``$-y$'', and a valley connecting two degenerate vacua at $(x_1,y_0)$, $(x_2,y_0)$ with vacuum energy $\Lambda$.
\begin{eqnarray}
V(x,y) &=&  V_{\rm global}(y)\left(1-\exp\left[-\left(\frac{y-y_{\rm valley}(x)}
{w_y}\right)^2\right]\left[1-V_{\rm local}(x)\right]\right)~, \\
V_{\rm global}(y) &=& \exp\left[y/2\right]~, \nonumber \\
V_{\rm local}(x)  &=& \Lambda e^{-y_0/2}+
(1-\Lambda e^{-y_0/2})\tanh^2\left[(x - x_1)(x - x_2)/(6w_x)\right]~, \nonumber \\
y_{\rm valley}(x) &=& y_0+k\sin[\pi(x - x_1)/(x_2 - x_1)]~. \nonumber
\end{eqnarray}
Basically, we drill two holes with widths controlled by $w_x,w_y$, and also a valley along $y_{\rm valley}(x)$, controlled by 1 parameter, $k$.

Our toy model shows that
\begin{itemize}
\item A complete valley is not necessary nor sufficient to exibit a tunneling path,\footnote{In\cite{Larfors}, the authors proposed that a tunneling path will pass through a saddle point between the vacua.  That claim is similar but slightly weaker than a valley-following path, since a valley guarantees a saddle point but not the other way around.  Our examples here shows that they are equally wrong.  Their claim is based on tuning the strength of gravity and assuming that a weak gravitational, Coleman-deLuccia instanton (which has a tunneling path) must continuously deform into a Hawking-Moss instanton \cite{HawMos82} (which just sits on the saddle point) in strong gravity.  It is already incorrect even with only one field, as the Coleman-deLuccia instanton can emerge from a different branch and is not connected to the Hawking-Moss instanton by continuous deformation \cite{Erick}.}
\item A global path has a simple analytical description.
\end{itemize}

\begin{figure}[h]
\begin{center}
\includegraphics[width=0.5\textwidth]{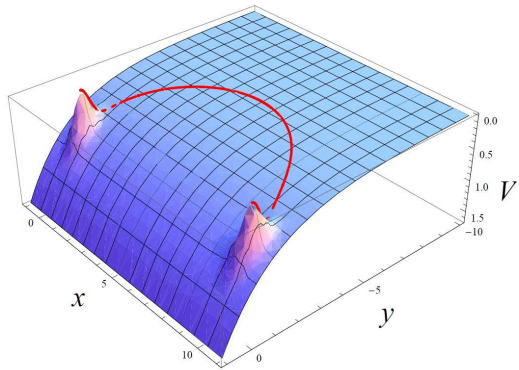}
\end{center}
\caption{The global path connecting two isolated vacua, showing in the inverse potential.  Here $w_x=w_y=0.5$, $x_1=0$, $x_2=10$, $y_0=-1$, $\Lambda=0$ and $k=1$ which is irrelevant.}
\label{fig-global}
\end{figure}

First we study the global path by choosing $w_x=w_y\ll|x_1-x_2|$, such that we have two sharp, isolated vacua with no valley between them, as in figure \ref{fig-global}.

A global path is the trajectory of a unit mass particle of total energy $-\Lambda$ on the inverse potential $-V(x,y)\sim-V_{\rm slope}(y)$.  It is as simple as a projectile motion, that $v_x$ stays constant while $y$ decreases to $y_m$ and comes back.
\begin{eqnarray}
v_x\delta t &=& |x_1-x_2|~, \nonumber \\
y_m         &=& 2\ln\left(v_x^2/2+\Lambda\right)~,\nonumber \\
\delta t    &=& 2\int_{y_m}^{y_0} \frac{dy}{\sqrt{2e^{y/2}-2\Lambda-v_x^2}}~, \nonumber
\end{eqnarray}
Combining these equations, we get
\begin{equation}
|x_1-x_2| = \frac{8 v_x}{\sqrt{2\Lambda+v_x^2}}
\tan^{-1}\frac{\sqrt{2e^{y_0/2}-2\Lambda-v_x^2}}{\sqrt{2\Lambda+v_x^2}}~.
\label{eq-dist}
\end{equation}
From (\ref{eq-dist}), we can see important properties of the global path.  For $\Lambda>0$, there is a maximum seperation $|x_1-x_2|<d_M$, beyond which no path can be found.  For separations within this range, there are two paths just like the usual projectile motion, a high path (smaller $v_x$) and a low path (bigger $v_x$).  According to the tension of a path given by
\begin{eqnarray}
\sigma(v_x) &=& \int (v_x^2+v_y^2)d\tau \nonumber \\
&=& 8\bigg(
\sqrt{2e^{-y_0/2}-2\Lambda-v_x^2}-\frac{2\Lambda}{\sqrt{2\Lambda+v_x^2}}
\tan^{-1}\frac{\sqrt{2e^{-y_0/2}-2\Lambda-v_x^2}}{\sqrt{2\Lambda+v_x^2}}\bigg)~,
\label{eq-tension2}
\end{eqnarray}
the low path has smaller tension, therefore is the solution we are looking for.

For $\Lambda=0$, the maximum separation occurs when $v_x\rightarrow0$.  This value $d_M=4\pi$ can be taken as the absolute maximum for all dS and Minkowski vacua.  For $\Lambda<0$, there is no maximum separation.

\begin{figure*}
\subfigure[]
{\includegraphics[width=.3\textwidth]{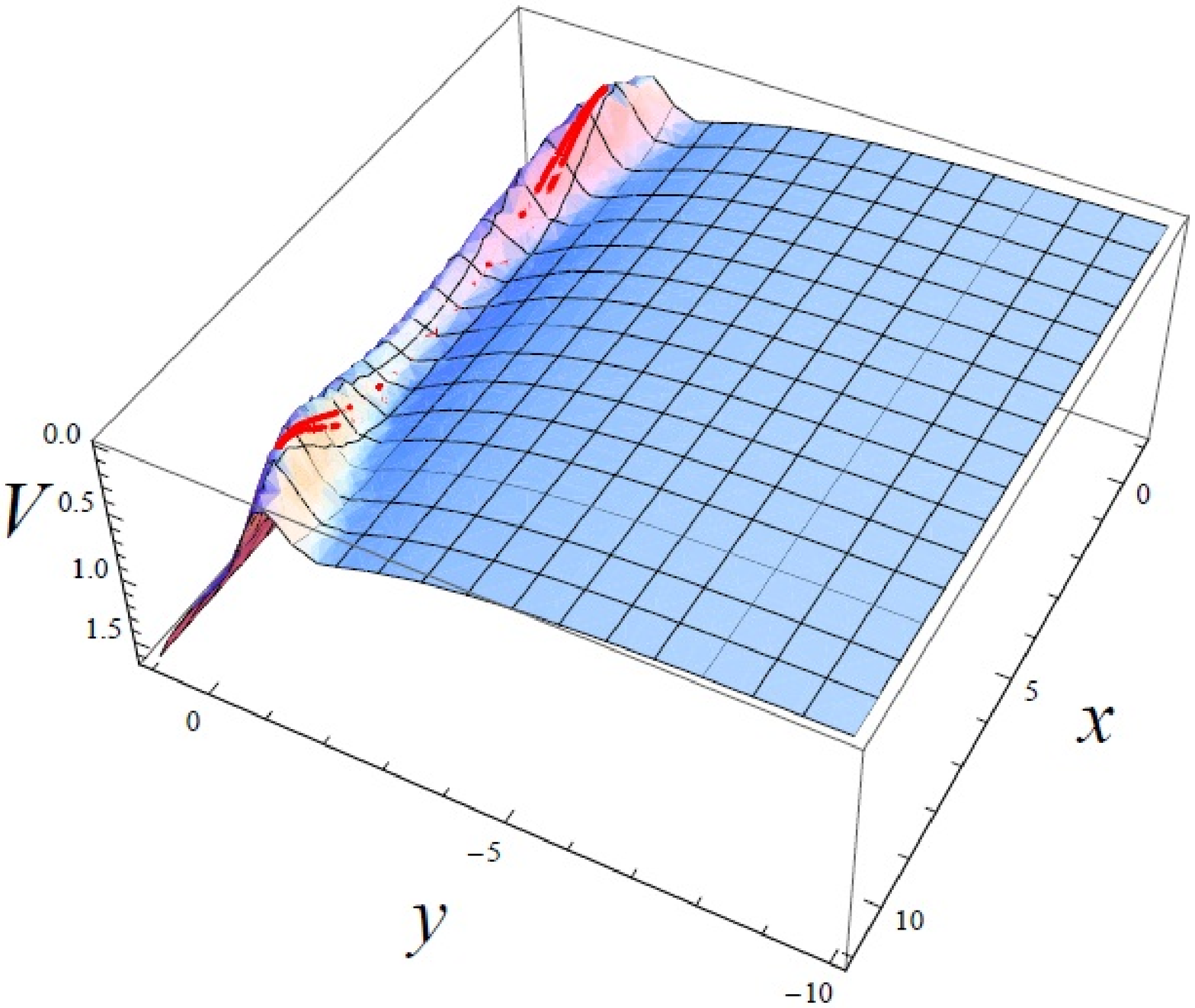}}
\subfigure[]
{\includegraphics[width=.3\textwidth]{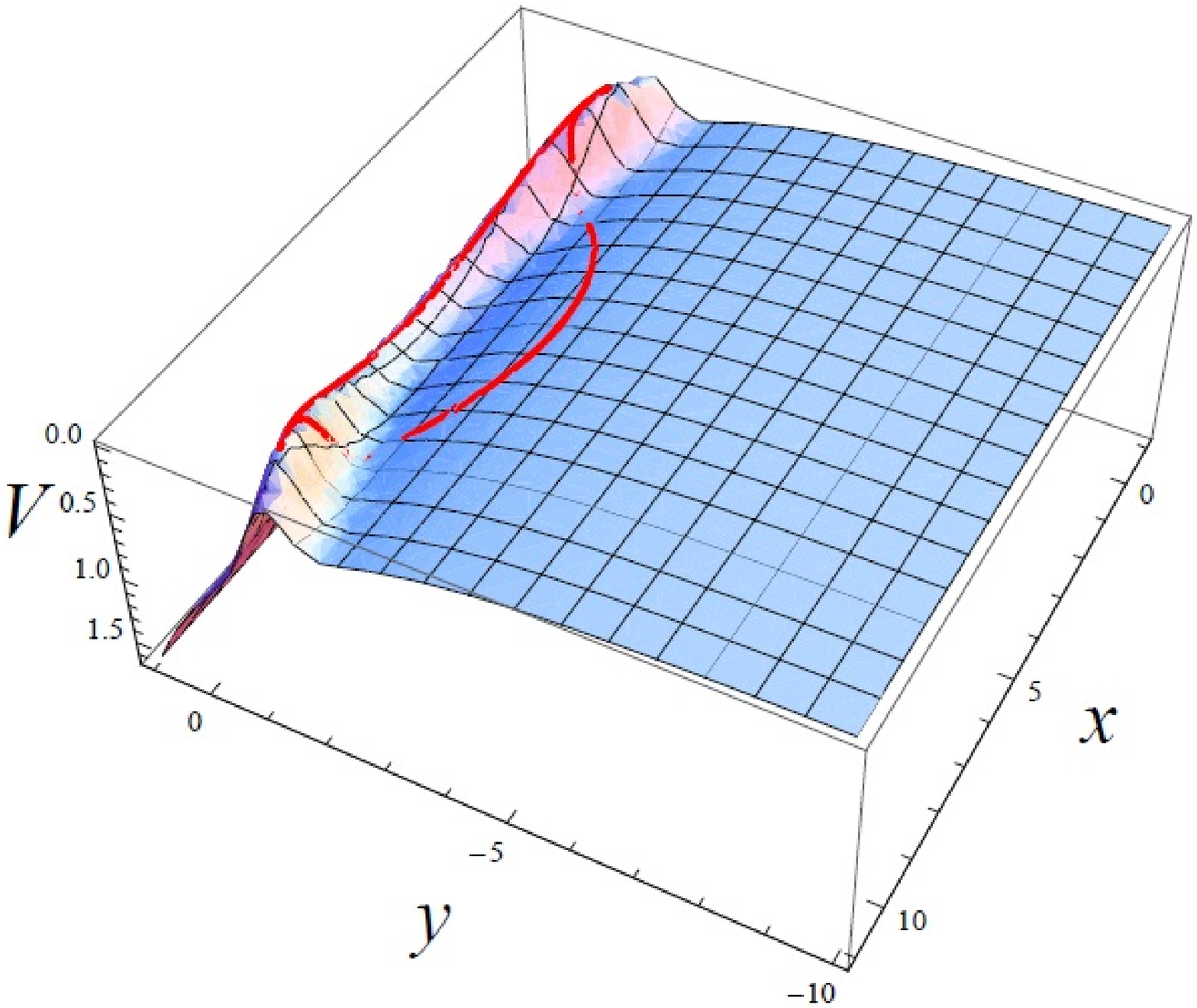}}
\subfigure[]
{\includegraphics[width=.3\textwidth]{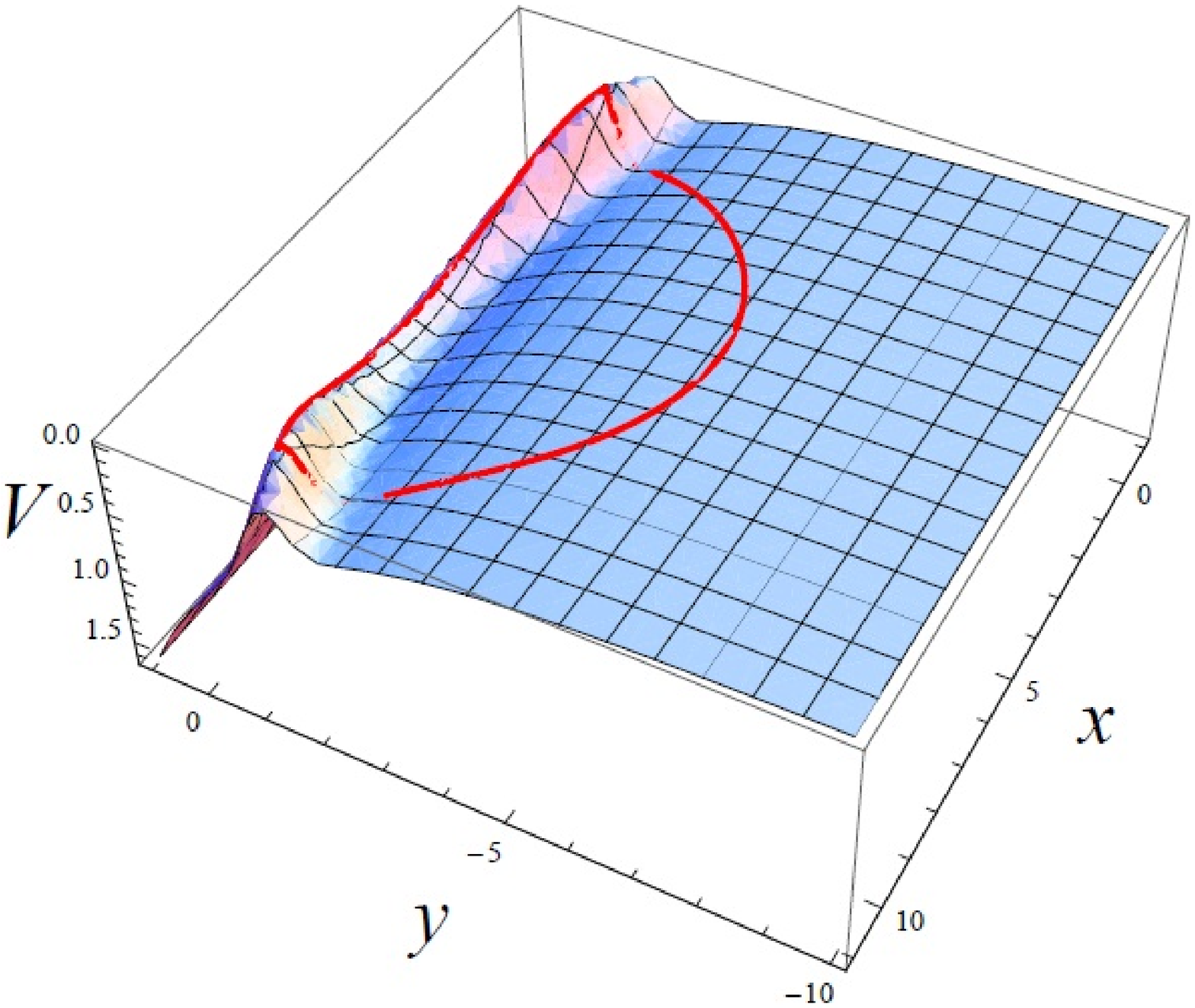}}
\caption{A potential with an obvious valley.  Here $w_y=0.5$, $w_x=4$, $x_1=0$, $x_2=10$, $y_0=-1$, $\Lambda=0$, and $k=1$.  With two different initial conditions, $a=0.3$ relaxes to the global path, and $a=0.5$ relaxes to the local path.}
\label{fig-both}
\end{figure*}

Now we turn our attention to local paths, namely the path which follows the valley. First we should increase $w_x$ so that the valley is more prominent.  Also, we should choose a reasonably large $k>0$, since negative $k$ makes it hard to distinguish the local path from the global path.  Then we can use the relaxation method with initial condition set by
\begin{equation}
y_{\rm initial}(x)=a y_{\rm valley}(x)~,
\label{eq-initial}
\end{equation}
where $a$ is a tunable parameter which should scan through a wide range of values if we are doing a general path search\footnote{In practice, the initial condition should be both $(x,y)$ as functions of $z$ between $z_{\rm left}$ and $z_{\rm right}$.  Here we choose $x$ to be linear in $z$.  One can choose a different initial parametrization but it will not significantly affect the result.  The path $y(x)$ is more important.}.  Obviously, if we want to relax into the local path, reasonable choices will have $a\sim1$.  In our first example, figure \ref{fig-both}, the valley is quite obvious and of course there is a saddle point right in the middle, we find both a local path and a global path.

The numerical data of the final paths allows us to estimate their tension.  In this example they are quite close and we can tune the parameters to make either one smaller.

\begin{figure*}
\subfigure[]
{\includegraphics[width=.3\textwidth]{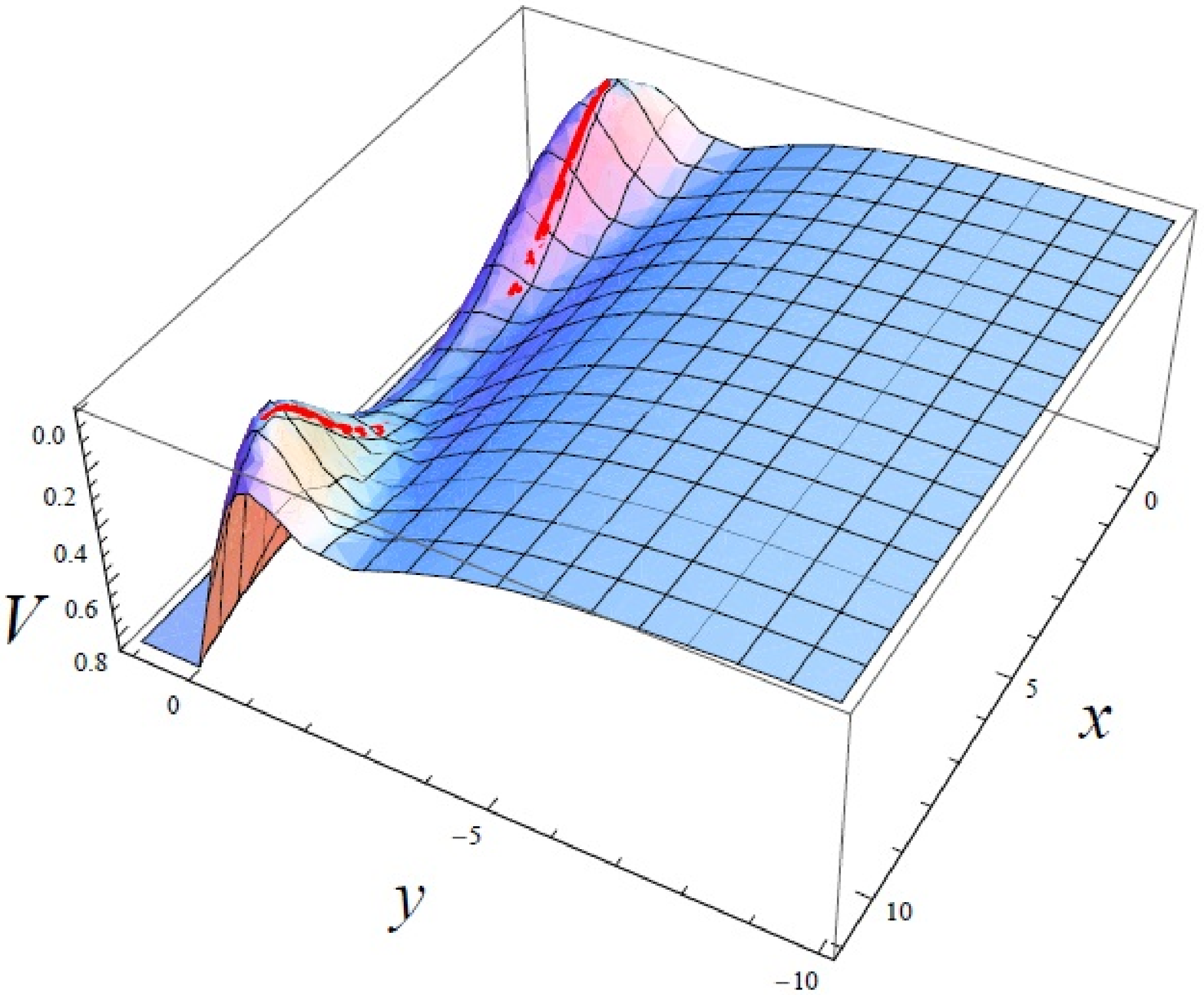}}
\subfigure[]
{\includegraphics[width=.3\textwidth]{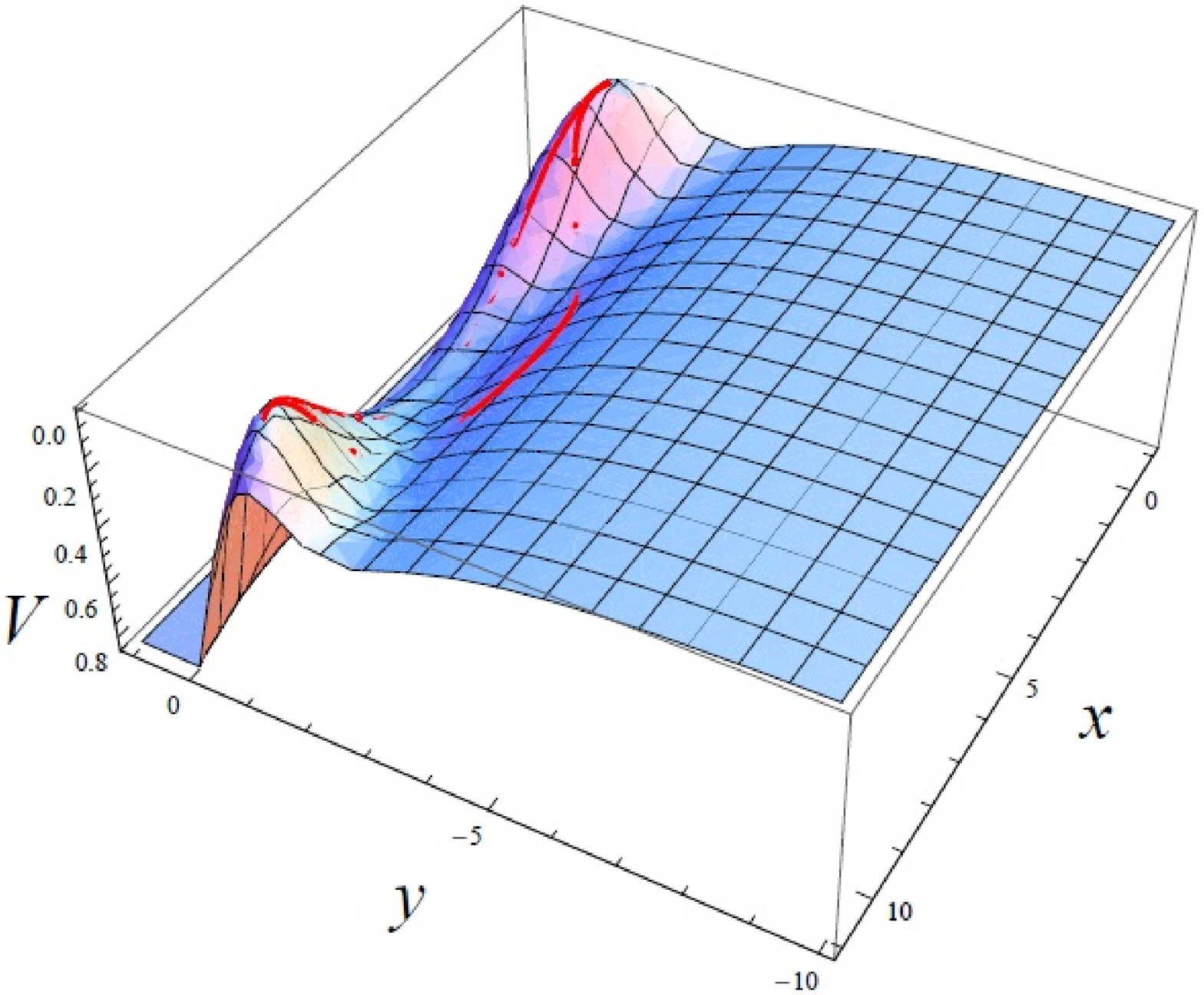}}
\subfigure[]
{\includegraphics[width=.3\textwidth]{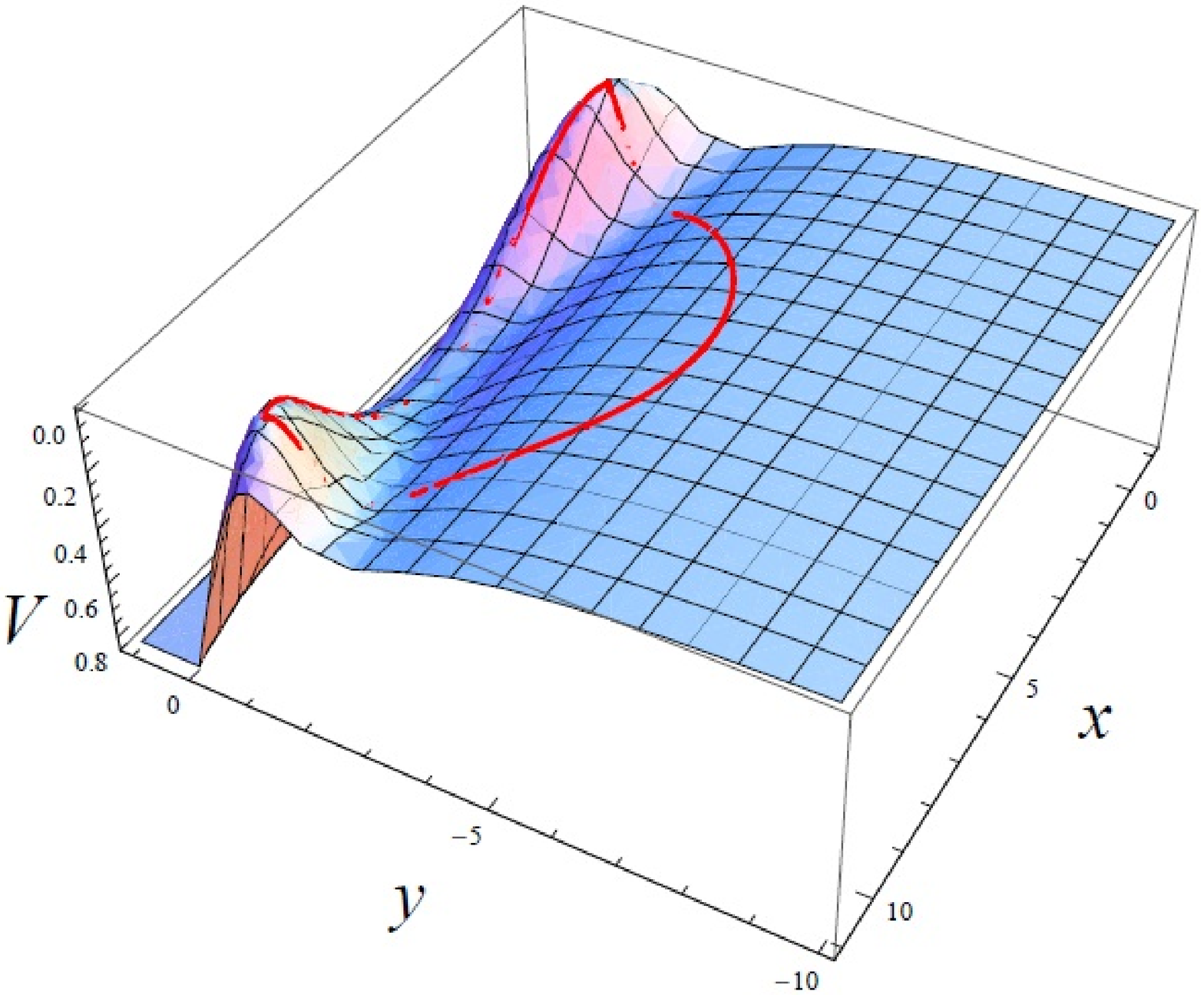}}
\caption{A local path without a saddle point.  Here $w_y=1$, $w_x=4$, $x_1=0$, $x_2=10$, $y_0=-1$, $\Lambda=0$, and $k=1$.  With two different initial conditions, $a=0.1$ relaxes to the global path, and $a=0.2$ relaxes to the local path.}
\label{fig-pns}
\end{figure*}

\begin{figure*}
\subfigure[]
{\includegraphics[width=.3\textwidth]{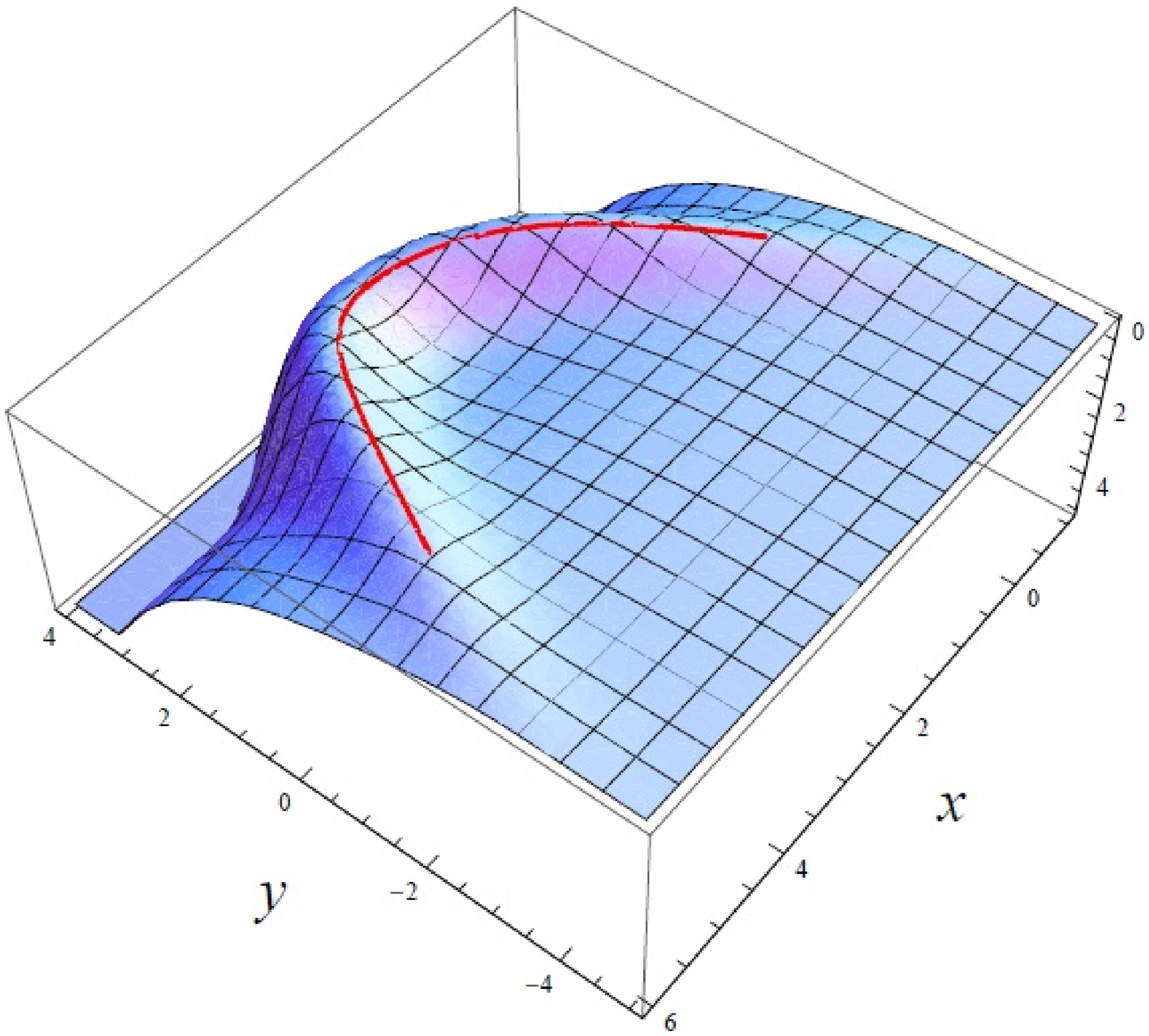}}
\subfigure[]
{\includegraphics[width=.3\textwidth]{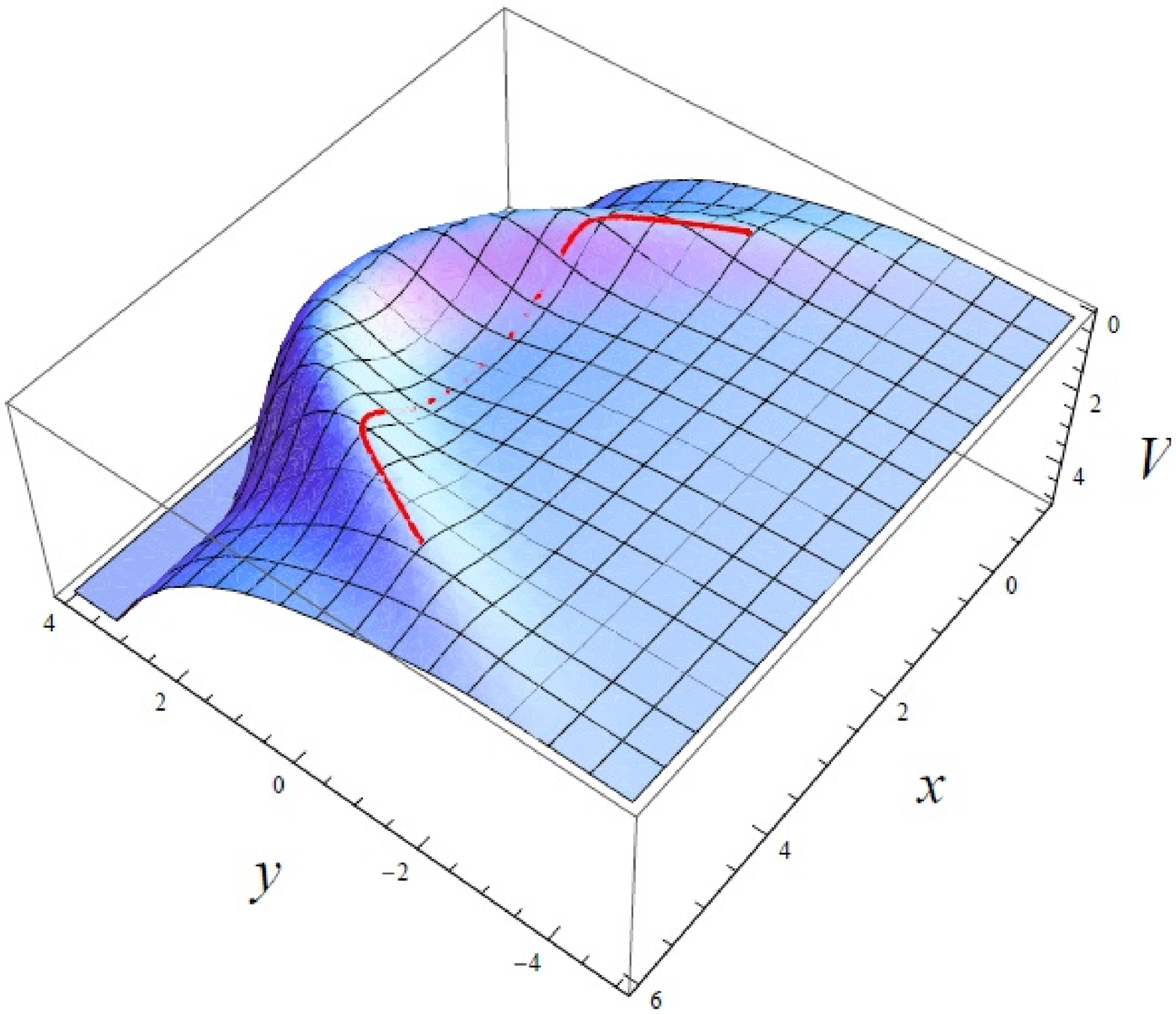}}
\subfigure[]
{\includegraphics[width=.3\textwidth]{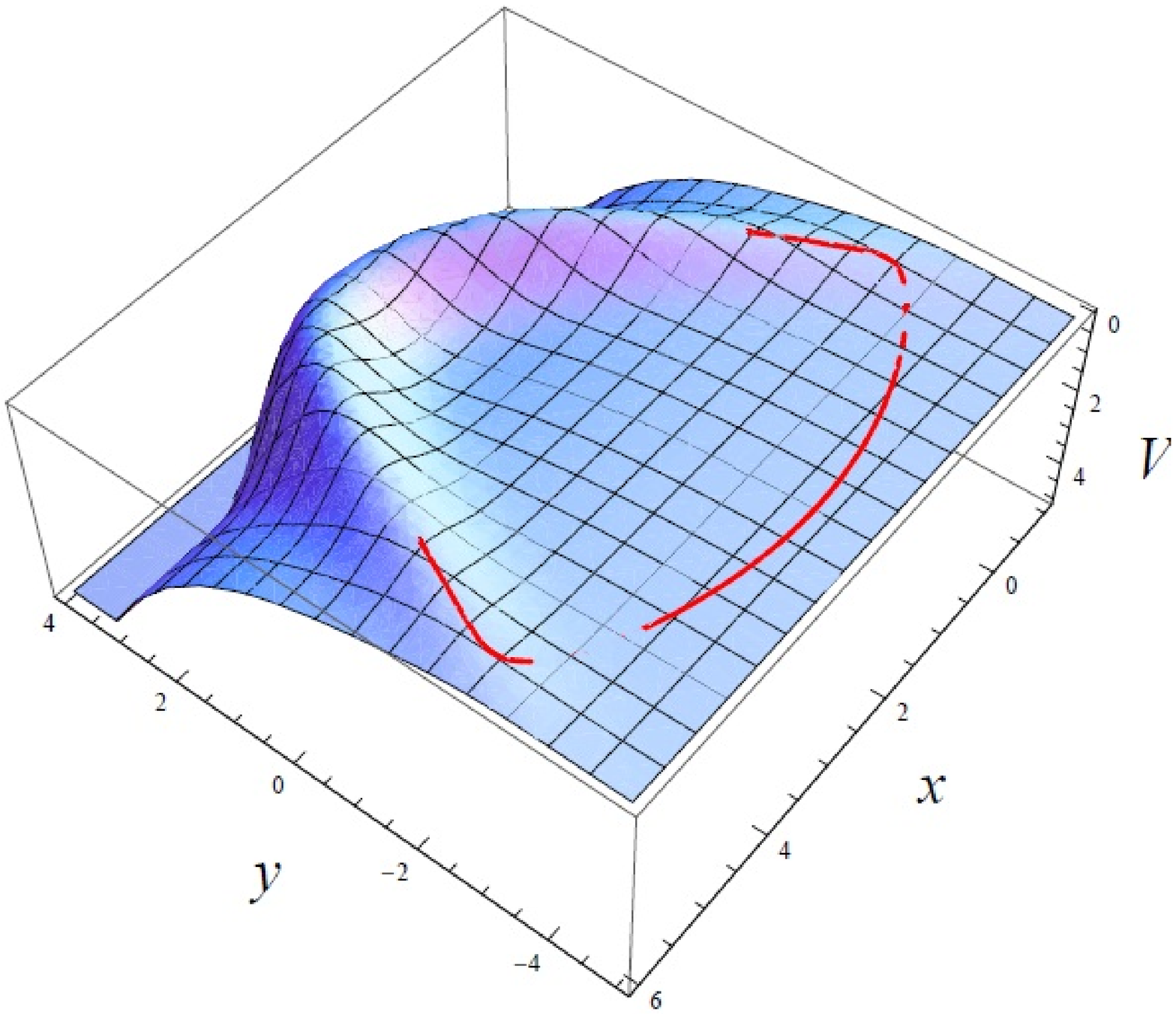}}
\caption{A saddle point without a local path.  Here $w_y=1.5$, $w_x=2$, $x_1=0$, $x_2=5$, $y_0=-1$, $\Lambda=0$, and $k=4$.  We tried several $a\sim1$ value in the neighborhood of the saddle point, and here shows $a=0.95$ which closely tracks the valley.  All of them left the valley and relaxed into the global path.}
\label{fig-snp}
\end{figure*}

In the next two examples, we adjust the valley and repeat the same process.  We can show that when there is no saddle point, there can still be a local path (figure \ref{fig-pns}).  Alternatively, in figure \ref{fig-snp} there is a saddle point but no local path.  Here in particular we increase the initial damping factor and look carefully for initial conditions near the saddle point, but no paths relax there.

It should not be surprising that the existence of a local path is not necessarily related to a saddle point (or a valley).  A path through the saddle point in general sees smaller values of the potential, but that is not the only factor.  While minimizing the action, we also need to consider the path length, which can be a comparable effect.  We can only hope that the local paths work in models where the path length does not vary too much.  Even so, we encourage people to always check whether such a path is real using methods such as numerical relaxation.

\section{Warping and the Conifold}\label{sec-ConifoldWarping}

A more careful analysis of the near conifold behavior of the flux potential would include the effects of warping due to fluxes---effects that are localized at the tip of the cone in the Calabi-Yau geometry when its moduli approach a conifold point. What follows is a slightly generalized version of the singular conifold calculations found in \cite{KS}.

\subsection{Geometry of the singular conifold}

The singular conifold has a Kahler metric given by
\begin{equation}
ds^2_{(6)} = d\rho^2 + \rho^2 d\Sigma^2,
\end{equation}
where
\begin{equation}
d\Sigma^2 = {1 \over 9}\left(2d\beta + \sum_{i=1}^2 \cos\theta_i d\phi_i\right)^2 + {1\over 6}\sum_{i=1}^2\left(d\theta_i^2 + \sin^2 \theta_i d\phi_i^2\right).
\end{equation}
is the metric on the base of the conifold, $T^{1,1}$. The base has the topology $S^3 \times S^2$. The coordinates $\theta_i$ go between $0,\pi$, while the $\phi_i$ and $\beta$ go between $0,2\pi$, thus these coordinates parametrize two $S^2$'s. The manifold $T^{1,1}$ can thus be thought of as a fiber bundle over $S^2 \times S^2$ with $S^1$ fibers. Alternatively, one can look at $T^{1,1}$ as a fiber bundle over $S^3$ with $S^2$ fibers. To trace out one of the $S^2$ fibers, hold $\beta$ fixed (say at $\beta = 0$), and then set $\theta_1 = \theta_2$ and $\phi_1 = -\phi_2$. The base is given by taking $\theta_2 = \phi_2 = 0$.

It's helpful to introduce the following two bases of one-forms:
\begin{eqnarray}
e^5 &=& 2 d\beta + \cos\theta_1 d\phi_1 + \cos\theta_2 d\phi_2,\nonumber \\
e^1 + i e^2 &=& i\left(d\theta_1 + i\sin \theta_1 d\phi_1\right),\nonumber \\
 e^3 + i e^4 &=& i e^{2i\beta} \left(d\theta_2 - i \sin\theta_2 d\phi_2\right),
\end{eqnarray}
and
\begin{equation}
g^{1,3} = {e^1 \mp e^3 \over \sqrt{2}}, \ \ g^{2,4} = {e^2 \mp e^4 \over \sqrt{2}}, \ \ g^5 = e^5.
\end{equation}
In this last basis the metric on $T^{1,1}$ is
\begin{equation}
d\Sigma^2 = {1\over 9}(g^5)^2 + {1\over 6} \sum_{i=1}^4 (g^i)^2.
\end{equation}

The volume of $T^{1,1}$ is given by the integral
\begin{equation}
\int_{T^{1,1}} {1 \over 6^2} g^1 \wedge g^2 \wedge g^3 \wedge g^4 \wedge {1 \over 3} g^5 =
{1 \over 3^3 \cdot 2^2} \int_{T^{1,1}} e^1 \wedge \cdots \wedge e^5.
\end{equation}
The integral over $T^{1,1}$ above becomes
\begin{equation}
\textrm{Vol}(T^{1,1}) = {2 \over 2^2 \cdot 3^3} \left(\int_0^{2\pi} d\phi\right)^3 \left(\int_0^\pi d(\cos \theta)\right)^2
= {2 \cdot (2\pi)^3 \cdot 2^2 \over 2^2 \cdot 3^3} = {16 \over 27}\pi^3.
\end{equation}

It is useful to define the forms
\begin{eqnarray}
\omega_{1i} &=& 2 d\beta + \cos \theta_i d\phi_i, \ \ \ \omega_{12} \wedge \omega_{11} 
= 2 \cos\theta_1 d\beta \wedge d\phi_1 - 2 \cos\theta_2 d\beta \wedge d\phi_2 - \cos \theta_1 \cos \theta_2 d\phi_1 \wedge d\phi_2 \nn \\ 
\omega_{2\,\,} &=& {1\over 2} \left( g^1 \wedge g^2 + g^3 \wedge g^4\right) = {1 \over 2}\left( e^1 \wedge e^2 + e^3 \wedge e^4\right) 
= {1\over 2} \left(\sin\theta_1 d\theta_1 \wedge d\phi_1 - \sin\theta_2 d\theta_2 \wedge d\phi_2\right), \nn\\
\omega_{3\,\,} &=& \omega_2 \wedge g^5 = {1\over 2}d\left(\omega_{12} \wedge \omega_{11}\right)
\end{eqnarray}
Given the relationship between $\omega_3$ and the $\omega_{1i}$'s one finds that
\begin{equation}
\omega_3 \wedge \omega_{12} \wedge \omega_{11} = 0,
\end{equation}
since $0 = d(\omega_{12} \wedge \omega_{11})^2 = 2\omega_3 \wedge \omega_{12} \wedge \omega_{11} + 2\omega_{12} \wedge \omega_{11} \wedge \omega_3$. We also have
\begin{equation}
\omega_2 \wedge \omega_{12} \wedge \omega_{11} 
= \cos \theta_2 d\beta \wedge d(\cos\theta_1) \wedge d\phi_1 \wedge d\phi_2 
- \cos \theta_1 d\beta \wedge d(\cos\theta_2) \wedge d\phi_1 \wedge d\phi_2 
\end{equation}

The forms $\omega_2$ and $\omega_3$ satisfy
\begin{equation}
\int_{S^2} \omega_2 = 4\pi, \ \ \ \int_{S^3} \omega_3 = 8\pi^2
\end{equation}
for the $S^2$ fibers and the $S^3$ base of $T^{1,1}$.

The 6D Hodge dual of $\omega_3$ (denoted by a $\star_6 \omega_3$ to keep it distinct from the 10D duality operator) is
\begin{equation}
\star_6 \omega_3 = {1\over 2} \left(\star_6 \left(g^1 \wedge g^2 \wedge g^5\right) + \star_6 \left(g^3 \wedge g^4 \wedge g^5\right) \right)
= {3 \over \rho} d\rho \wedge \omega_2,
\end{equation}
which can be easily seen by defining the forms $\lambda_i = \rho g^i / \sqrt{6}$, $\lambda_5 = \rho g^5 / 3$, and $\lambda_6 = d\rho$.

Clearly, $\star_6 \omega_3$ can also be expressed as an exact form
\begin{equation}
\star_6 \omega_3 = d\tilde{\omega}_2, \ \ \ \tilde{\omega}_2 = 3 \ln \rho \cdot \omega_2.
\end{equation}

Given a 3-form of the form
\begin{equation}
{\cal A} \omega_3 + {\cal B} \star_6 \omega_3,
\end{equation}
using $(\star_6)^2 = -1$, the condition for imaginary self-duality is that ${\cal B} = -i{\cal A}$. The 3-form is exact, with a potential:
\begin{equation}
\half{\cal A} \omega_{12} \wedge \omega_{11} + {\cal B} \tilde{\omega}_2.
\end{equation}
So for an imaginary self-dual 3-form, we have the form for the 3-form and its 2-form potential:
\begin{equation}
{\cal A}(1-i \star_6)\omega_3, \ \ \ {\cal A} \left(\half \omega_{12} \wedge \omega_{11} - i \tilde{\omega_2}\right).
\end{equation}

\subsection{Branes and fluxes on the singular conifold}

We imagine $N_{D3}$ D3-branes filling the large spatial dimensions. These branes are pointlike in the 6D conifold geometry and they sit at the tip of the conifold. These D3's act as magnetic sources for type IIB $F_{(5)}$ flux through the $T^{1,1}$ base of the conifold:
\begin{equation}
{1\over (4\pi^2 \alpha')^2} \int_{T^{1,1}} F_{(5)} = N_{D3}.
\end{equation}
This means that we can write
\begin{equation}
F_{(5)} = {\pi {\alpha'}^2 \over 2} N_{D3}\,\omega_2 \wedge \omega_3 = 27 \pi {\alpha'}^2 N_{D3} {\bf Vol}(T^{1,1}), 
\end{equation}
where ${\bf Vol}(T^{1,1})$ is the volume form on $T^{1,1}$.

In addition to the D3-branes, we consider $N$ D5-branes wrapped around $S^2$'s in $T^{1,1}$. Geometrically, the dual to the $S^2$ fiber in $T^{1,1}$ is the $S^3$ base, and thus, these D5's act as sources of magnetic R-R flux on the $S^3$:
\begin{equation}
{1\over 4\pi^2 \alpha'} \int_{S^3} F_{(3)} = N.
\end{equation}
This suggests that $F_{(3)}$ takes the form
\begin{equation}
F_{(3)} = {1\over 2} N \omega_3 + f \star_6 \omega_3,
\end{equation}
where $f$ is to be determined.

We also allow for some integer amount of NS-NS flux through the $S^3$:
\begin{equation}
{1\over 4\pi^2 \alpha'} \int_{S^3} H_{(3)} = M.
\end{equation}
In fact, we will always choose this to vanish (which is general since type IIB theory has an $SL(2,\mathbb{Z})$ symmetry that allows us to fix such a condition). For the sake of clarity, we will work with $M$ left arbitrary for now. Therefore, we can write
\begin{equation}
H_{(3)} = {1\over 2} M \omega_3 + h \star_6 \omega_3,
\end{equation}
with $h$ to be determined.

In type IIB supergravity, the R-R 3-form field strength $F_{(3)} = dC_{(2)}$ and the NS-NS 3-form field strength $H_{(3)} = dB_{(2)}$ are naturally combined into the imaginary self-dual 3-form
\begin{equation}
G_{(3)} = F_{(3)} - \tau H_{(3)} = \left(F_{(3)} - \tau_R H_{(3)}\right) - i \tau_I H_{(3)},
\end{equation}
where $\tau = C_{(0)} + i e^{-\phi}$ is the axio-dilaton and $g_s = e^\phi$ is the string coupling. Since $G_{(3)}$ is imaginary self-dual, it follows from before that it can be expressed in the form
\begin{equation}
{\cal A}\,\omega_3 - i {\cal A} \star_6 \omega_3.
\end{equation}
This implies that 
\begin{equation}
\half (N - \tau M) = i(f - \tau\, h),
\end{equation}
or taking real and imaginary parts
\begin{equation}
\half (N - \tau_R M) = \tau_I h, \ \ \ -\half \tau_I M = f - \tau_R h.
\end{equation}
So,
\begin{equation}
h = {1 \over 2\tau_I} (N - \tau_R M), \ \ \ f = {1 \over 2\tau_I} (N \tau_R - |\tau|^2 M).
\end{equation}
So we have
\begin{equation}
F_{(3)} = \half N\,\omega_3 + {1 \over 2\tau_I} (N \tau_R - |\tau|^2 M) \star_6 \omega_3, \ \ \
H_{(3)} = \half M\,\omega_3 + {1 \over 2\tau_I} (N - \tau_R M) \star_6 \omega_3.
\end{equation}
As mentioned above, we can always use $SL(2,\mathbb{Z})$ symmetry to set $M=0$, which simplifies the above expression
\begin{equation}
F_{(3)} = \half N\,\omega_3 + {\tau_R \over 2\tau_I}N \, \star_6 \omega_3, \ \ \
H_{(3)} = {1 \over 2\tau_I} N\, \star_6 \omega_3.
\end{equation}
Note that in much of the literature (including the famous paper by Klebanov and Strassler), $\tau_R$ is taken to vanish. The above reproduces those results.

The type IIB self-dual 5-form satisfies the Bianchi identity
\begin{equation}
d\tilde{F}_{(5)} = 2\kappa^2 T_3 \rho_3 + H_{(3)} \wedge F_{(3)},
\end{equation}
where $\kappa^2$ is the 10D Planck scale, $T_3$ is the tension of D3-branes, and $\rho_3$ is the D3-charge density from localized sources. We know that
\begin{equation}
H_{(3)} \wedge F_{(3)} = {N^2 \over 4 \tau_I} \star_6 \omega_3 \wedge \omega_3 = 27 \cdot {3N^2 \over 2 \tau_I} {d\rho \over \rho} \wedge {\bf Vol}(T^{1,1}).
\end{equation}
Define
\begin{equation}
{\cal F}_{(5)} = 27\pi {\alpha'}^2 N_{eff}(\rho) {\bf Vol}(T^{1,1}),
\end{equation}
where
\begin{equation}
N_{eff}(\rho) = N_{D3} + {3 N^2 \over 2\pi \tau_I} \ln {\rho \over \rho_0},
\end{equation}
where $\rho_0$ is a cut-off distance at which we choose to truncate the singular conifold geometry, which gives the hard-wall approximation to the deformed conifold in which the singularity at the tip is replaced by an $S^3$ of some minimal size.

The 10D self-dual 5-form field strength is then
\begin{equation}
\tilde{F}_{(5)} = (1+\star){\cal F}_{(5)}.
\end{equation}

\subsection{10D warped geometry and the warp factor}

The metric is taken to be
\begin{equation}
ds^2 = e^{2 A(y)} \eta_{\mu\nu} dx^\mu dx^\nu + e^{-2 A(y)} g_{ij}(y) dy^i dy^j,
\end{equation}
where $g_{ij}$ is the singular conifold metric and $e^{-4A(y)}$ is the warp factor. Recalling that the conifold metric can be written as
\begin{equation}
ds_6^2 = d\rho^2 + \rho^2 d\Sigma^2 = \sum_{i=1}^6 (\lambda_i)^2, \ \  \lambda_{1,\ldots,4} = {g^{1,\ldots,4} \over \sqrt{6}}\rho, \ \ \lambda_5 = {g^5 \over 3} \rho, \ \ \lambda_6 = d\rho,
\end{equation}
we can define the 10D vielbeins
\begin{equation}
E^\mu = e^A dx^\mu, \ \ \ F^i = e^{-A} \lambda_i. 
\end{equation}
Now,
\begin{equation}
\star (F^1 \wedge \cdots \wedge F^5) = e^{-5 A} \rho^5 \star {\bf Vol}(T^{1,1}) = F^6 \wedge E^0 \wedge \cdots \wedge E^3 = e^{3 A} d\rho \wedge dx^0 \wedge \cdots \wedge dx^3.\nn 
\end{equation}
So,
\begin{equation}
\star {\bf Vol}(T^{1,1}) = {e^{8 A} \over \rho^5} d\rho \wedge dx^0 \wedge \cdots \wedge dx^3.
\end{equation}
Furthermore, we know from 4D Poincar\'e invariance that
\begin{equation}
\star {\cal F}_{(5)} = d\alpha(y) \wedge dx^1 \wedge \cdots \wedge dx^3.
\end{equation}
Thus,
\begin{equation}
d\alpha = 27 \pi {\alpha'}^2 N_{eff}(\rho) e^{8 A} \rho^{-5} d\rho.
\end{equation}
The BPS constraints require that $\alpha = e^{4 A}$. Thus
\begin{equation}
{d\alpha \over \alpha^2} = 27 \pi {\alpha'}^2 {N_{eff}(\rho)\,d\rho \over \rho^{5}}.
\end{equation}
Integrating both sides yields
\begin{equation}\label{WarpFactor}
e^{-4 A} = c + {27 \pi {\alpha'}^2 \over 4 \rho^4} \left( {N_{D3} \over \tau_I} + {3 N^2 \ln(\rho/\rho_0) \over 2\pi\tau_I^2} + {3 N^2 \over 8\pi \tau_I^2}\right),
\end{equation}
where $c$ is a constant of integration. The constant actually plays a crucial role in understanding the dynamics of the universal Kahler modulus in warped flux compactifications since it represents the zero-mode of the warping and is, in fact, identified with the universal Kahler modulus field.

Note that a factor of $\tau_I^{-1}$ crops up in the final expression after integrating. This is due to a shift from the 10D string frame metric to the 10D Einstein frame metric.

\subsection{The functional form of warping corrections to the Kahler metric}

Following \cite{GiddingsWarpDynamics,DouglasSUSYBreaking}, we assume that the Kahler potential takes the form
\begin{equation}
K_{\textrm{cs}} = -\log \left(i \int_{\cal M} e^{-4 A} \Omega \wedge \overline{\Omega}\right).
\end{equation}
In general this is not the complete Kahler potential and additional corrections are needed (see \cite{DouglasWarped, DouglasKinetic}), however it was shown in \cite{DouglasKinetic} that the functional behavior of the warp corrections to the Kahler metric is the same in the case of the conifold whether or not one considers the additional corrections.

Splitting the integral into the bulk and conifold portions of the Calabi-Yau manifold gives
\begin{equation}
K_{\textrm{cs}} 
= -\log \left(i c \int_{{\cal M}_{\textrm{bulk}}} \Omega \wedge \overline{\Omega} + i \int_{{\cal M}_{\textrm{conifold}}} e^{-4 A} \Omega \wedge \overline{\Omega}\right),
\end{equation}
where we have used the fact that in the bulk contributions from $e^{-4 A_0}$ will be negligible compared to $c$, taken to be large. Defining
\begin{equation}
K_{\textrm{bulk}} = - \log \left(i c \int_{{\cal M}_{\textrm{bulk}}} \Omega \wedge \overline{\Omega}\right),
\end{equation}
since the volume of the bulk region is taken to be much larger than that of the conifold region, we can write the above as
\begin{equation}
K_{\textrm{cs}} \approx K_{\textrm{bulk}} + i e^{K_{\textrm{bulk}}} \int_{{\cal M}_{\textrm{conifold}}} e^{-4 A} \Omega \wedge \overline{\Omega} + \cdots.
\end{equation}
The Kahler metric near the conifold with the warping effects will thus be
\begin{equation}\label{ConifoldKahlerMetric}
(K_{\textrm{conifold}})_{\xi\bar{\xi}} \approx i e^{K_{\textrm{bulk}}} \int_{{\cal M}_{\textrm{conifold}}} e^{-4 A} \chi \wedge \overline{\chi},
\end{equation}
where $\chi$ is the (2,1)-form that corresponds to complex deformations of the conifold
\begin{equation}
\chi = {1 \over 8\pi^2} \left(\omega_3 - i \star_6 \omega_3 \right) = {1 \over 8\pi^2} \left(\omega_3 - 3 i {d\rho \over \rho} \wedge \omega_2 \right).
\end{equation}
We have
\begin{equation}
\chi \wedge \overline{\chi} = {i \over 32 \pi^4} \omega_3 \wedge \star_6 \omega_3 = -{81 i \over 16 \pi^4} {d\rho \over \rho} \wedge \mathbf{Vol}(T^{1,1}).  
\end{equation}
In the integral for the near conifold Kahler metric (\ref{ConifoldKahlerMetric}), the warp factor will only depend on $r$, so the volume integration over $T^{1,1}$ simply gives a factor of $16 \pi^3 / 27$. The integral we wish to calculate is then
\begin{equation}
{3 \over \pi} e^{K_{\textrm{bulk}}} \int_{|\xi|^{1/3}}^{\Lambda_0} {d\rho \over \rho} e^{-4 A(r)}.
\end{equation}
Plugging in the result for the warp factor (\ref{WarpFactor}), we find the strong warping correction to the standard result. Ignoring the details of the coefficients, we have
\begin{equation}
(K_{\textrm{conifold}})_{\xi \bar{\xi}} \sim c_1 \log {\Lambda_0^3 \over |\xi|} + {c_2 \over |\xi|^{4/3}},
\end{equation}
where we have substituted $\rho_0 = |\xi|^{1/3}$ as the hard-wall cut-off of the singular conifold and we assume that $|\xi|$ is small compared to the long-distance cut-off $\Lambda_0^3$. The ratio of the coefficients $c_1 / c_2$ is going to be of order $c$, which is taken to be large, but not strictly infinite. This means that strong warping will have an impact for small enough $|\xi|$. For the purposes of our simulations, we consider the coefficient in front of the warp correction term to be an adjustable parameter that should be set at some small order of magnitude.

Note that in our numerical analysis of near conifold tunneling behavior, we have ignored the additional $\tau_I$ dependence that comes from the warp factor. Including this effect will likely modify the $\tau$-field profile a little, but dramatically complicates the computation.\footnote{One might ask how the Kahler {\em potential} is modified given the warping corrections to the Kahler metric. The corrections imply a contribution to the Kahler potential that looks like $|\xi|^{2/3}$. Note that near the conifold point, this correction is completely negligible. We neglect the possible contributions of this term and first derivatives in our analysis and note that their inclusion will yield mild modifications (if any) to our results.}

\section{Near-conifold Potential and Numerical Data}\label{sec-NearConifoldPotential}

In general, a conifold locus in the moduli space represents Calabi-Yaus that develop various singular points due to the collapse of certain cycles. In the one-parameter examples, there is a single conifold point and a single cycle that degenerates while the periods of the other cycles become constant \cite{Candelas:1990rm}.

Due to the paired intersections of cycles in a Calabi-Yau, the collapsing cycle's partner develops interesting behavior in the moduli space (despite going to a constant at the conifold point). Call the collapsing cycle $\Asc$ and the intersecting cycle $\Bsc$. Making a closed loop in moduli space around the conifold point, one finds that there is an ambiguity involved in determining what happens to the $\Bsc$-cycle. Since the intersection of $\Asc$ with itself is zero, going around the conifold point can lead to some integer multiple of an $\Asc$-cycle adding on to the $\Bsc$-cycle. From the perspective of the periods, a loop around the conifold point sends $\Pi_\Bsc \to \Pi_\Bsc + \Pi_\Asc$. This implies that
\begin{equation}
\Pi_\Asc = \xi + \pi_\Asc(\xi), \ \ \ \Pi_\Bsc = {\xi \over 2\pi i} \log {\xi \over \Lambda_0^3} + \pi_\Bsc(\xi),
\end{equation}
where the functions $\pi_\Asc$ and $\pi_\Bsc$ are $O(\xi^2)$ and $O(1)$, respectively. The monodromy of the $\Bsc$-cycle period is captured by the behavior of $\log$ in the expression above. The constant $\Lambda_0^3$ arises from cutting off the conifold geometry and gluing it into the bulk Calabi-Yau at $r \sim \Lambda_0$ where $r$ is the radial coordinate for the singular conifold.

Given the above expressions, we can work out the behavior of the Kahler potential, Kahler metric, superpotential, and the flux potential in the near-conifold limit. We will first do this while ignoring corrections from strong warping.

\subsection{The Kahler potential and its derivatives}

The complex structure Kahler potential for the one-parameter models is
\begin{equation}
K_{cs} = -\log\left(i \left( \Pi_3 \overline{\Pi}_0 - \overline{\Pi}_3 \Pi_0 + \Pi_1 \overline{\Pi}_2 - \overline{\Pi}_1 \Pi_2 \right) \right).
\end{equation}
In our notation the collapsing cycle is given by $\Pi_\Asc = \Pi_3$, and its intersecting partner is $\Pi_\Bsc = \Pi_0$. Plugging in the near-conifold behavior of these cycles and sweeping up all of the $O(1)$ dependence into a function $k(\xi)$ gives
\begin{equation}
K_{cs} = \log\left( {|\xi|^2 \over 2\pi} \log {\Lambda_0^6 \over |\xi|^2} + k \right) \to -\log k,
\end{equation}
where the expression after the arrow indicates the limit of the Kahler potential when we neglect terms of order $O(\xi)$.

The derivative is then
\begin{equation}
K_\xi = e^{K_{cs}} \left( {\bar{\xi} \over 2\pi} \left( \log {\Lambda_0^6 \over |\xi|^2} - 1 \right) - k_\xi \right) \to - {k_\xi \over k}.
\end{equation}
And the Kahler metric is
\begin{equation}
K_{\xi\bar{\xi}} = |K_\xi|^2 + e^{K_{cs}} \left( {1\over 2\pi} \left( \log{\Lambda_0^6 \over |\xi|^2} - 2 \right) - k_{\xi\bar{\xi}} \right) 
\to {1\over 2\pi k} \log{\Lambda_0^6 \over |\xi|^2} + \kappa(\xi),
\end{equation}
where $\kappa(\xi) \sim O(1)$. The near conifold Kahler metric possesses a logarithmic singularity at the conifold point.

In order to include the effects of strong warping for very small $|\xi|$, we modify the expression for the Kahler metric above by introducing the warp correction term
\begin{equation}
K_{\xi\bar{\xi}} \approx  {1\over 2\pi k} \log{\Lambda_0^6 \over |\xi|^2} + {K_1 \over k} + {C_1 \over k|\xi|^{4/3}},
\end{equation}
where $C_1$ is taken to be very small, reflecting that we are working with a large (but finite) volume Calabi-Yau manifold. Note also that we have replaced $\kappa = K_1 / k$ in the above as it is more convenient to work with in the final expression for the flux potential.

\subsection{The superpotential and its derivatives}

The superpotential is as above
\begin{equation}
W = F \cdot \Pi - \tau H \cdot \Pi = A + \tau B.
\end{equation}
Recall that we can use the $SL(2,\mathbb{Z})$ invariance of type IIB superstrings to ensure that $H_3$ always vanishes. This means that while $A$ has non-trivial monodromy near the conifold point, $B$ does not since the flux multiplying $\Pi_0$ is set to zero.

The near-conifold behavior of these functions is easily computed
\begin{equation}
A = {F_\Asc \xi \over 2\pi i} \log {\xi\over\Lambda_0^3} + a(\xi) \to a(\xi), \ \ \ B = b(\xi),
\end{equation}
where $a$ and $b$ are $O(1)$ and depend on the choice of the fluxes associated to the other cycles (including the $\Bsc$-cycle). We also have
\begin{equation}
A_\xi = {F_\Asc \over 2\pi i} \left(\log {\xi \over \Lambda_0^3} + 1\right) + a_\xi \to {F_\Asc \over 2\pi i} \log {\xi \over \Lambda_0^3}.
\end{equation}
Thus, the derivatives $D_\xi W$ and $D_\tau W$ of the superpotential take the following form near the conifold
\begin{equation}
D_\xi W \approx {F_\Asc \over 2\pi i} \log {\xi \over \Lambda_0^3} + A_1 - \tau B_1,
\end{equation}
and
\begin{equation}
D_\tau W \approx \sqrt{k}(A_2 + \bar{\tau} B_2).
\end{equation}

\subsection{The flux potential near the conifold}

The leading behavior of the flux potential in $\xi$ is determined by the behavior of $A$ in the superpotential above. Recall that the flux potential is given by
\begin{equation}
V = {e^{K_{cs}} \over 16 \tau_I \rho_I^3} \left( K^{\xi \bar{\xi}} |D_\xi W|^2 + K^{\tau\bar{\tau}} |D_\tau W|^2 \right).
\end{equation}
Inserting the expressions for the Kahler potential, metric, and superpotential near the conifold, we have
\begin{equation}\label{FluxPotentialNearConifold}
V = {1 \over 16 \tau_I \rho_I^3} \left( \left({1\over 2\pi}\log {\Lambda_0^6 \over |\xi|^2} + K_1 + {C_1 \over |\xi|^{4/3}} \right)^{-1} 
\left|{F_\Asc \over 2\pi i} \log \xi + A_1 - \tau B_1 \right|^2 
 + |A_2+\bar{\tau}B_2|^2 \right),
\end{equation}
where $C_1$ is a small constant (it's order of magnitude mainly reflecting the large volume of the compactification manifold) and $\Lambda_0$ the cut-off characterizing where the singular conifold is glued into the bulk Calabi-Yau geometry.

\subsection{Mirror quintic near-conifold numerical data}\label{sec-NumericalParameterData}

Our numerical simulations have been carried out for vacua arising from flux compactification on the mirror quintic. We collect here some data to ease the replication of our results.

Given the fluxes $F = (3,-6,-9,-1)$ and $H = (-1,0,-7,0)$, the parameters in the near-conifold flux potential (\ref{FluxPotentialNearConifold}) are
\begin{eqnarray*}
K_1 &=& 0.524211, \\
A_1 &=& 13.1691 + 17.3632\,i,\\
B_1 &=& 0.209511+0.000277995\,i,\\
A_2 &=& -9.55217 + 7.75481\,i,\\
B_2 &=& -2.26182\,i.
\end{eqnarray*}
The choice of $F$ flux implies that $F_\Asc \equiv F_3 = -1$.

The mirror quintic period data near the conifold is approximated as follows: for the period functions that are regular near the conifold, we used {\em Mathematica} to compute them in terms of Meijer G functions and find their expansions to first order around the conifold point. The period $\Pi_0$ picks up a monodromy on sending $\xi \to e^{2\pi i} \xi$. We captured this behavior, the $O(1)$ and $O(z)$ behavior by fitting a function of the appropriate form to a numerically generated period function. The fit is good for $|\xi| \sim 0.04$ for $\Lambda_0^3 \sim 1$. Note that the variable $z$ is set up so that $z=1$ is the conifold point, $z=0$ is called the large-complex-structure (LCS) point. 
\begin{eqnarray*}
\Pi_3 &\to&  \xi \equiv -0.355878\,(z-1) i, \\
\Pi_2 &\to&  6.19501 - 7.11466\,i - (2.33032 + 2.85683\,i)\xi, \\
\Pi_1 &\to&  1.29357\,i + 0.423645\,\xi, \\
\Pi_0 &\to&  {\xi \over 2\pi i} \log(-i\xi) + 1.07128 - 0.0630147\,i\,\xi.
\end{eqnarray*}

\end{appendix}

%*******************************************************************

\end{document}